\documentclass[preprint2]{aastex}

\usepackage{subfigure}
\usepackage{natbib}%
\usepackage{graphicx}
\usepackage{multirow}
\usepackage{epstopdf}
\usepackage{amsmath}
\usepackage{url}
\usepackage{hyperref}
\shorttitle{GBM Spectral Catalog}
\shortauthors{Goldstein, et al.}

\begin{document}

%%Title
\title{The \emph{Fermi} GBM Gamma-Ray Burst Spectral Catalog: \\
         The First Two Years }

%%Authors
\author{Adam~Goldstein\altaffilmark{1}, 
J.~Michael~Burgess\altaffilmark{1}, 
Robert~D.~Preece\altaffilmark{1},
Michael~S.~Briggs\altaffilmark{1},  
Sylvain Guiriec\altaffilmark{1}, 
Alexander J.~van der Horst\altaffilmark{2},
Valerie Connaughton\altaffilmark{1}, 
Colleen A. Wilson-Hodge\altaffilmark{3},
William S.~Paciesas\altaffilmark{1}, 
Charles A.~Meegan\altaffilmark{2}, 
Andreas~von~Kienlin\altaffilmark{4}, 
P.~N.~Bhat\altaffilmark{1}, 
Elisabetta~Bissaldi\altaffilmark{4}, 
Vandiver~Chaplin\altaffilmark{1}, 
Roland~Diehl\altaffilmark{4}, 
Gerald~J.~Fishman\altaffilmark{3}, 
Gerard~Fitzpatrick\altaffilmark{5}, 
Suzanne~Foley\altaffilmark{4}, 
Melissa~Gibby\altaffilmark{6}, 
Misty~Giles\altaffilmark{6}, 
Jochen~Greiner\altaffilmark{4}, 
David~Gruber\altaffilmark{4}, 
R.~Marc~Kippen\altaffilmark{7},  
Chryssa~Kouveliotou\altaffilmark{3}, 
Sheila~McBreen\altaffilmark{5},
Sin$\acute{e}$ad~McGlynn\altaffilmark{8}$^,$\altaffilmark{4}
Arne~Rau\altaffilmark{4}, and
Dave~Tierney\altaffilmark{5}}

%%Affiliations
\altaffiltext{1}{University of Alabama in Huntsville, 320 Sparkman Drive, Huntsville, AL 35805, USA}
\altaffiltext{2}{Universities Space Research Association, 320 Sparkman Drive, Huntsville, AL 35805, USA}
\altaffiltext{3}{Space Science Office, VP62, NASA/Marshall Space Flight Center, Huntsville, AL 35812, USA}
\altaffiltext{4}{Max-Planck-Institut f$\rm \ddot{u}$r extraterrestrische Physik (Giessenbachstrasse 1, 85748 Garching, 
Germany)}
\altaffiltext{5}{School of Physics, University College Dublin, Belfield, Stillorgan Road, Dublin 4, Ireland}
\altaffiltext{6}{Jacobs Technology}
\altaffiltext{7}{Los Alamos National Laboratory, PO Box 1663, Los Alamos, NM 87545, USA}
\altaffiltext{8}{Exzellenzcluster Universe, Technische Universit\"{a}t M\"{u}nchen, Boltzmannstrasse 2, 85748 Garching, Germany}

%%Abstract
\begin{abstract}
We present systematic spectral analyses of GRBs detected by the \emph{Fermi} Gamma-Ray Burst Monitor (GBM) during its 
first two years of operation.  This catalog contains two types of spectra extracted from 487 GRBs, and by fitting four different 
spectral models, this results in a compendium of over 3800 spectra.  The models were selected based on their empirical 
importance to the spectral shape of many GRBs, and the analysis performed was devised to be as thorough and objective as 
possible.  We describe in detail our procedure and criteria for the analyses, and present the bulk results in the form of parameter 
distributions.  This catalog should be considered an official product from the \emph{Fermi} GBM Science Team, and the data 
files containing the complete results are available from the High-Energy Astrophysics Science Archive Research Center 
(HEASARC).

\end{abstract}

%%Introduction
\section{Introduction}
Compendia of physical observations of an enigmatic phenomenon such as Gamma-Ray Bursts (GRBs) should be as complete 
and uniform as possible, so that useful inferences can be drawn from the assembled data. Here we have followed in a line of 
successful GRB Catalogs from the Burst And Transient Source Experiment (BATSE) to present the community with 
representative spectral fits for a complete sample of \emph{Fermi} Gamma-Ray Burst Monitor (GBM) bursts, observed during 
the first two years of operation (July 14, 2008 to July 13, 2010).  The methodology of producing this catalog strives to balance 
completeness, uniformity, and accuracy.  The selection criteria, techniques, spectral models used for fitting, etc., all derive from a 
previous work, based upon BATSE data \citep{Mallozzi95}, that is being prepared for publication as a complete sample from that 
earlier mission. It is our hope that the two data sets, the current GBM sample and the complete BATSE sample, can serve as a 
whole for comparison and contrast.  Further, we will continue to add to this series as more observations are accumulated by 
GBM.

Many correlative studies in the field of GRB global properties studies rely on only two very distinct  representative spectra \citep
{Amati,Ghirlanda,Yonetoku}. These are the {\em time-integrated} and {\em peak flux} spectra, representing (respectively) the 
average emission and the most luminous. Defining these uniformly is not trivial and trade-offs inevitably must be made. We have 
chosen to accumulate every spectrum that meets a minimum flux significance into our average spectrum, which leaves out 
quiescent periods that will dilute the signal. For the peak flux spectra we have chosen a slice of the burst representing the peak 
flux on a particular time scale.  Because of the large dynamic range of burst durations (ms to thousands of s) we have chosen a 
smaller accumulation interval for those bursts with durations shorter than 2 s compared to those longer than 2 s. In this way, we 
can still make a distinction between the fluence and the peak flux spectra for the traditionally short-class GRBs.

The set of functions for fitting spectra was chosen for uniformity with the existing set of BATSE GRB Spectral Catalogs \citep
{Band93,Kaneko06,Goldstein11}, as well as with the (forthcoming) GBM Time Resolved Spectroscopy Catalog of Bright Bursts. 
In most cases where the statistics of the spectra are sufficient, the data can support a number of free parameters that is rarely 
greater than four. Hence, we have selected spectral functions that simply exhibit two, three, and four free parameters. Our 
choice of the best model for each GRB, based upon the fit statistics, will be discussed below; however, for many cases it is clear 
when a simple model fails to fit the data,  or when the fit does not significantly improve with a more complicated model. 

We start with a brief description of the GBM detectors and calibration in Section 2 and refer to \citet{Meegan} and \citet{Bissaldi} 
for a more thorough and complete description of the instrument and calibration respectively.  This is followed, in Section 3, by a 
description of the methodology used in the production of this catalog, including  detector selection, data types  used, energy 
selection and background fitting, and the source selection.  We then offer a description of the spectral models used in this 
catalog in Section 4.  In Section 5, we present the spectral analysis methods and results.  A description of the Catalog 
file format can be found in the Appendix.

%%Detectors & Calibration
\section{GBM Detectors and Calibration}
GBM is one of two instruments on-board the \emph{Fermi} Gamma-Ray Space Telescope, which was launched and 
placed into orbit on June 11, 2008.  GBM is a 14--detector instrument designed to study the gamma-ray sky in the energy band 
of $\sim$8 keV--40 MeV.  Twelve of the detectors are sodium iodide (NaI) scintillation detectors, placed in groups of three on the 
side edges  of the spacecraft and pointed at various angles from the \emph{Fermi} Large Area Telescope (LAT) boresight.  The 
pointing angles were optimized through simulations to adequately survey the entire unocculted sky at any time during the orbit, 
while also biasing coverage toward the LAT pointing direction.  The other two detectors are each composed of a bismuth 
germanate (BGO) crystal and are placed on either side of the spacecraft pointing at right angles to the LAT boresight.  Each BGO 
crystal is coupled to two photomultiplier tubes, which collect the scintillation photons and convert them into electronic signals.  
The NaI detectors cover an energy range from 8 keV--1 MeV, and the BGO detectors cover a range of 200 keV--40 MeV.  
Because of the all-sky coverage of the NaI detectors, and the fact that the diameter of each NaI crystal (12.7~cm) is ten times its 
thickness,  the approximate location of a burst can be determined by comparing the relative count rates in the detectors that 
observed the bursts.  The source location is then calculated in spacecraft coordinates and used in the production of the detector 
response matrices (DRMs).  The DRMs are mathematical models of the detectors' response used to map the observed counts 
into photons of known energy.  Each detector's response is dependent on incident photon energy, the measured detector output 
energy, and the detector--source angle and the earth--source--spacecraft geometry.

A key issue for accurate spectroscopy is the fidelity of the detector calibration model, which matches detector pulse height 
analyzer (PHA) channels to physical energies. The individual GBM detectors were extensively calibrated before launch using 
several nuclear line sources \citep{Bissaldi}. In addition to line sources, the BESSY accelerator in Berlin, Germany was used to 
determine the low-energy light output of four of the flight-qualified NaI detectors. Finally, a source survey calibration was 
performed on the entire assembled Fermi Observatory, just prior to observatory-level thermal vacuum testing. This last 
calibration consisted of several radioactive sources being individually placed at various angles relative to the observatory in 
order to verify and validate the spacecraft and instrument scattering models, used in the DRMs \citep{Hoover2008}.

Given that the lowest energy of the radioactive sources was 14.4 keV and that most of the detectors were not calibrated with a 
continuum energy source such as BESSY, it is useful to validate the GBM calibration using on-orbit observations of known 
astrophysical sources. A good spectroscopy standard is the Crab Nebula, which has been studied since the infancy of 
gamma-ray astronomy. However, GBM is not a pointed instrument, so we have made use of the Earth occultation technique, as 
pioneered with BATSE \citep{Harmon2002}. We have obtained excellent fits that are in good agreement in every detector with 
the canonical Crab Nebula spectrum \citep[for details, see][Wilson-Hodge et al. in prep.]{Case2011}. Notably, there are no 
systematic deviations in the residuals to the continuum fits that would indicate issues with the calibration.   The Earth occultation 
technique uses 16 re-binned broad energy channels in the CSPEC data type \citep{Meegan}. 

\begin{figure}[p!]
	\begin{center}
		\includegraphics[scale=0.45]{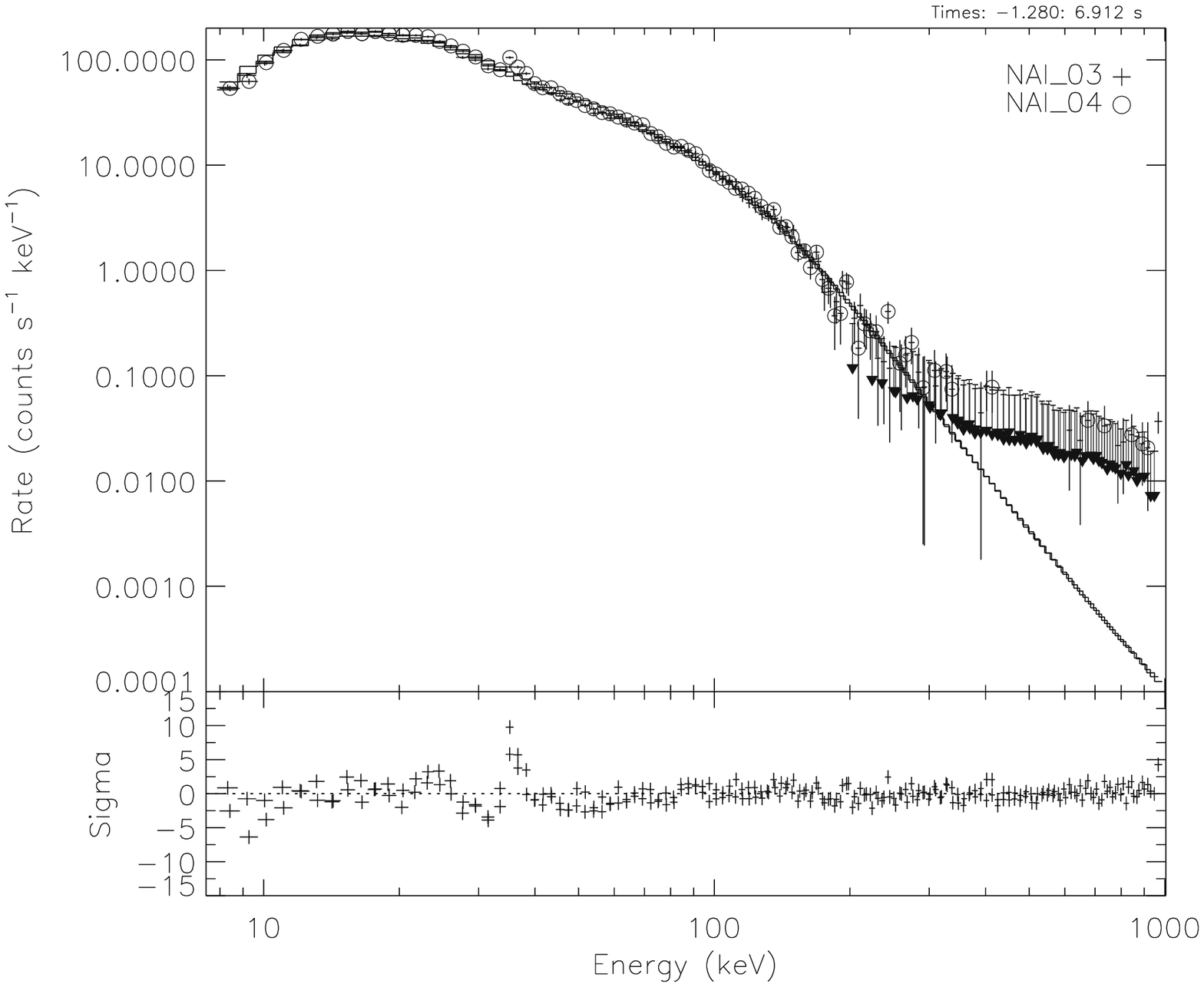}
	\end{center}
\caption{Spectrum of GRB 081009 using NaI 3 \& 4 with a Band function fit.  Note the excess in residuals at the K-edge.  The 
current calibration is inadequate in removing this instrumental effect for many bright bursts.
\label{kedge}}
\end{figure}

When some very bright sources have been fitted using the 128 energy channel CSPEC data type, a kink in the spectrum at $\sim
$33 keV can be observed. This takes a `p-cygni' type form, where some channels (at lower energies) lie below the fitted model 
and others (at higher energies) lie above it. The total extent in channel space is roughly 4--6 channels, while in energy space it is 
about 5--10 keV. Since the excess residuals have equal weight below and above the model fit and do not change the fitted 
parameters appreciably when the affected channels are excluded, we have chosen to ignore this effect in the current work. It is 
likely that the non-linear jump in the detector light output at the Na K-edge has not been modeled completely (see the NaI 
K-edge discussion in \citet{Bissaldi}), which we will address in future calibration work.  As an example of non-linearity of the 
response, see Figure \ref{kedge}, which is the spectrum of the bright GRB 081009 using two NaI detectors.

Solar flares provide an additional means of calibration of the GBM detectors, particularly the high-energy BGO detectors .  For 
example, the positron annihilation line (511~keV), neutron capture line (2.2~MeV), and various nuclear lines (e.g.,~4.4 and 
6.1~MeV) were observed with BGO detector 0 in the solar flare of 2010 June 12 (Abdo et al. in prep.), demonstrating the 
calibration of that detector.   In that flare there was a small gain shift (1\%), showing limitations of the gain adjustment software 
and the 511~keV line had an additional small offset (1\%), possibly showing limitations of the energy calibration of the BGO 
detectors.

In addition to persistent, known sources, calibration of GBM can be studied with other transient events such as Soft Gamma 
Repeaters (SGRs). SGR bursts are short, with typical durations of several tens to a few hundred milliseconds. Their $\nu F_
{\nu}$ energy spectra typically peak around 40 keV with a very steep decline in flux above the peak.  Analysis of all the SGR 
bursts detected by GBM is currently underway (see Lin et al. 2011; van der Horst et al., in preparation; von Kienlin et al., in 
preparation). Typically the models and fit parameters are consistent with those found from the observations of other instruments 
(for reviews, see \citet{WoodsThompson, Mereghetti}).  We fit the integrated count spectrum of one of the brightest bursts of 
SGR\,$1550-5418$ observed by GBM.  The spectrum is best fit with a model that agrees within the parameter uncertainties with 
the \citet{Israel} results obtained for SGR\,1900+14 with {\it Swift}). This demonstrates the ability of the low-energy calibration to fit 
the overall spectrum well, except for the area around the NaI K-edge where we notice that the calibration is not adequate in 
removing instrumental effects, as discussed above.

Concerning the calibration of GBM with respect to GRBs, a number of efforts have been pursued to compare the GRB spectra 
attained by GBM to spectra obtained by other instruments observing the same bursts in approximately the same energy band.  
One of the first GRBs detected by GBM (GRB 080723B) was also observed by two instruments onboard the INTEGRAL 
spacecraft.  A study \citep{Andreas} was performed to compare GBM's spectral analysis results with the ISGRI instrument \citep
[15 keV--1 MeV;][]{Lebrun} and the SPI instrument \citep[18 keV--8 MeV;][]{Vedrenne}.  This burst was selected because it was 
bright, spectrally hard, and seen in most GBM detectors.  This analysis was performed before the first in-flight calibration was 
released, and before an accurate modeling of the spacecraft solar panels, which could block or attenuate the observation from 
some detectors.  A later study \citep{Tierney} was performed on four other GRBs also detected by ISGRI, taking into account 
blocked detectors and a new in-flight calibration.  These studies have confirmed that GRB spectra as measured by GBM are 
consistent with other well-established and calibrated instruments. 

%%Method
\section{Method}
During the first two years of operation, GBM triggered on 491 GRBs, 487 of which are presented in this catalog.  The remaining 
bursts are excluded due to a low accumulation of counts or a lack of spectral/temporal coverage.  In order to provide the 
most useful analysis to the community, we have attempted to make the method as systematic and uniform as possible. When 
we deviate from uniformity we indicate the circumstances clearly. Here we detail the detector and data selection as well as the 
process used to fit the data. Many of the criteria are adopted from the GBM Burst Catalog, (Paciesas et al., in prep.) and we 
have attempted to maintain this in all aspects. However, due to the nature of spectral analysis we demand stricter criteria to 
ensure that we have adequate signal in all energy channels. This also effectively reduces the GRB sample from that used in the 
burst catalog.  In addition, this catalog includes 4 GRBs not included in the GBM Burst Catalog.  GBM trigger number 
081007.224 contains an observation of a GRB triggered by \emph{Swift} \citep[GRB 081007A;][]{081007} at $\sim$121 s after 
GBM.  The other three (GBM triggers 091013.989, 091022.752, \& 091208.623) where originally classified as unknown triggers 
and were not included in the Burst Catalog, however their spectra are consistent with the GRB spectra observed by GBM.  Two 
other triggers were classified as unknown, and the spectral analysis of those triggers reveal that they are too weak to determine if 
they have GRB-like spectra.

%%Detector Selection
\subsection{Detector Selection}
\begin{figure}[p!]
	\begin{center}
		\includegraphics[scale=0.45]{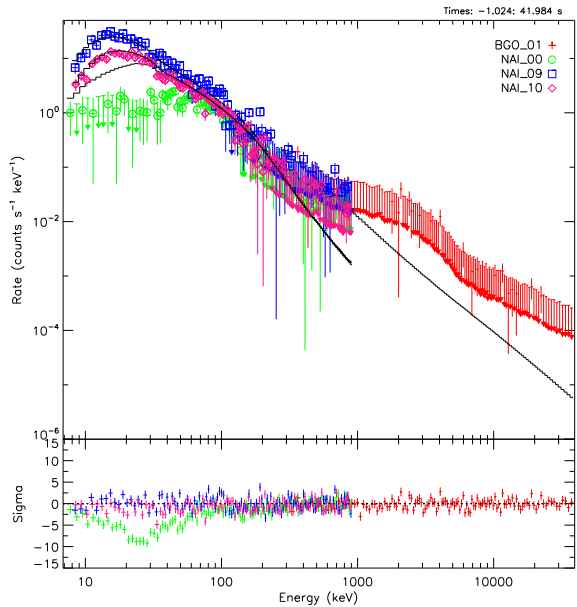}
	\end{center}
\caption{Plot of the counts spectrum of GRB 090131 using NaI detectors 0, 9 and 10, and BGO detector 1.  Close inspection of 
the $\sim$8-60 keV range reveals that NaI 0 shows a significant negative trend in residuals and a large deficit in count rate at 
low energy when compared to the other detectors, indicating a blockage from what is likely the LAT radiator, which is not 
properly modeled in the DRM. \label{blockage}}
\end{figure}

GBM employs 14 detectors, situated at different angles to promote monitoring of the full unocculted sky: twelve NaI detectors 
with an energy coverage of 8 keV to 1 MeV and two BGO detectors with an energy coverage of 200 keV to 40 MeV.  For this 
catalog, we use the data produced by both NaI and BGO detectors.  GBM was designed to observe the entire unocculted sky at 
all times in order to increase the number of GRBs detected. However, when GBM triggers on a GRB, not all detectors have an 
optimal pointing for reconstructing the observed GRB spectrum. Therefore, we begin with all 12 NaI detectors and select a 
subset of these detectors with optimum viewing angles.  To ensure sufficient detector response, we determined that using the 
detectors with viewing angles less than $60^{\circ}$ from the location of a burst maximize the effective area and provide the best 
signal for spectral analysis, since the detector angle is strongly correlated with the count rate in the detector. In addition, we 
select the BGO detector that is closest to the NaI detectors in the subset. With this subset we run a program designed to 
determine any spacecraft blockage that would interfere with the signal. In most cases this provides a smaller subset of detectors 
that should be free of any blockage. However, due to small inaccuracies in the detector mass model or location uncertainties, the 
blockage code does not always return a subset of detectors that is free from blockage. This is evident when the low-energy data 
deviate strongly from the fit model, as depicted in Figure \ref{blockage}.  This can affect particular detectors when the burst is 
partially occulted by the spacecraft.  When this occurs we remove these detectors from the selected sample.  Finally, in the event 
that more than 3 NaI detectors exist in our subsample for a particular burst, we discard all but the 3 detectors with the smallest 
source angles.  This helps to prevent a fitting bias toward lower energies for bursts with more than 3 sufficient detectors, since 
the high significance of data in the NaI detectors will influence the spectral fitting procedure more than the lesser amount of data 
in the BGO detectors.  With this final detector set for each burst, we have an adequate and reliable dataset for performing 
spectral analysis.  

%%Data Types
\subsection{Data Types}
The primary data type used in this catalog was the time-tagged event (TTE) data, which provide absolute time resolution to 2 $
\mu s$ based on GPS, and high energy resolution of 128 channels.  For the purpose of this catalog, we choose a standard time 
binning of 1024 ms for bursts greater than 2 s in duration as defined by the burst $\rm t_{90}$ (see Paciesas et al. 2011) and 
64 ms for bursts of duration 2 s and shorter.  The time history of TTE typically starts at $\sim$30 s before trigger and extends to $
\sim$300 s after trigger.  This timespan is adequate for the analysis of most GRBs.  For GRBs that have evident precursors or 
emissions that last more than 300 s after trigger, we use the CSPEC data, which extend $\sim$4000 s before and after the 
burst. This datatype also has a continuous 128 channel energy resolution which is normally accumulated at 4.096 s temporal 
resolution that changes to 1.024 s resolution for $\sim$600 s starting at trigger time.  CSPEC data was used for 22 GRBs in this 
catalog.

%% Energy Selection & Background Fitting
\subsection{Energy Selection and Background Fitting}
With the optimum subset of detectors selected, the best time and energy selections are chosen to fit the data.  The available 
reliable energy channels in the NaI detectors lie between $\sim$8 keV and $\sim$1 MeV. This selection excludes overflow 
channels and those channels where the instrument response is poor and the background is high.  Since the issue with the 
sodium K-edge noticeably affects only the brightest GBM bursts, which comprise a small percentage of this catalog, we have 
chosen to keep the channels affected by the K-edge in the analysis of all GRBs.  The fit parameters for bursts that are affected by 
non-linearity around the K-edge do not noticeably change; the only evident change in the results is the fit statistics and hence the 
goodness-of-fit.  We perform a similar selection to the BGO detector for each burst, selecting channels between $\sim$300 keV 
and $\sim$38 MeV.  With the resulting time series we select enough pre- and post-burst background to sufficiently model the 
background and fit a single energy dependent polynomial (choosing up to $4^{\rm th}$ order) to the background. For each 
detector the time selection and polynomial order are varied until the $\chi^2$ statistic map over all energy channels is 
minimized, resulting in an adequate background fit. 

%% Source Selection
\subsection{Source Selection}
\begin{figure}
	\begin{center}
		\subfigure[]{\label{acumtime}\includegraphics[scale=0.35]{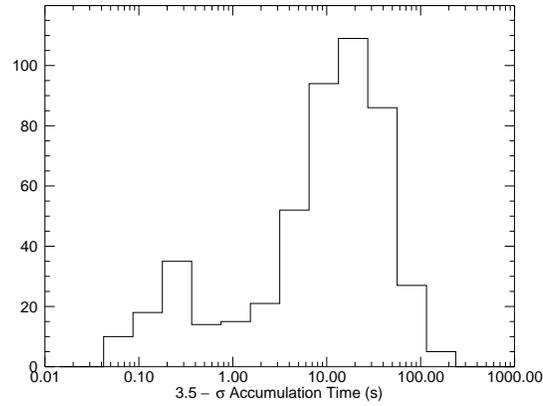}}\\
		\subfigure[]{\label{timepfluence}\includegraphics[scale=0.35]{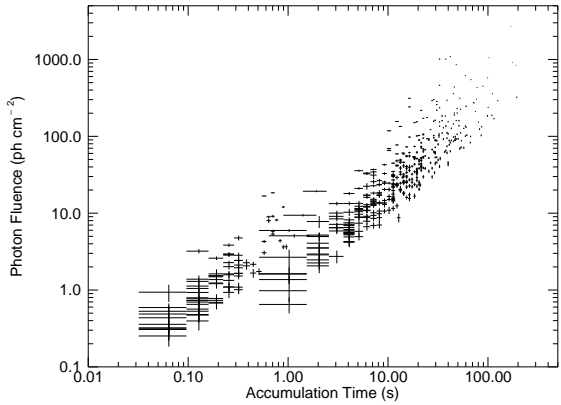}}
		\subfigure[]{\label{timepflux}\includegraphics[scale=0.35]{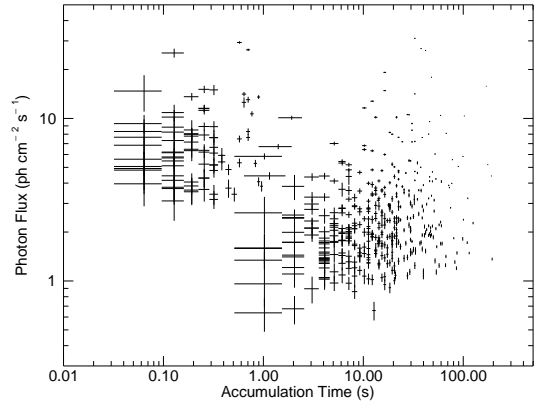}}
	\end{center}
\caption{\ref{acumtime} is the distribution of the accumulation time based on the 3.5$\sigma$ signal-to-noise selections.  Note 
the similarity to the traditional $\rm t_{90}$ distribution, with the minimum near 1 second.  No other estimation of the duration was 
factored into the production of the accumulation time. \ref{timepfluence} and \ref{timepflux} show the comparison of the model 
photon fluence and photon flux to the accumulation time respectively.  The fluxes and fluences shown in these figures are from 
the estimated BEST model fits. \label{allTime}}
\end{figure}

Once the background count rates are determined for each detector for a given burst, we sum the count rates and background 
models over all NaI detectors to produce a single lightcurve from which we make objective time selections.  We subtract the 
background, convert the rates to counts, and calculate the signal-to-noise ratio (SNR) for each time history bin.  Only the time 
bins that have a SNR greater or equal to 3.5 sigma are selected as signal.  The time selections produced from this criterion are 
then applied to all detectors for a given burst.  This criterion ensures that there is adequate signal to successfully perform a 
spectral fit and constrain the parameters of the fit. This does however eliminate some faint bursts from the catalog sample (i.e., 
those with no time bins with signal above 3.5 sigma).  While this strict cut was performed to provide an objective catalog, it is 
possible that not all signal from a burst was selected.  However, most of the signal below 3.5 sigma is likely indiscernible from 
the background fluctuations, so a spectral analysis including those bins would likely only increase the uncertainty in the 
measurements.  This selection is what we refer to as the ``fluence'' selection, since it is a time-integrated selection, and the 
fluence derived is representative of the fluence over the total duration of the burst.  The other selection performed is a 1.024 ms 
peak photon flux selection for long bursts and 64 ms peak count rate flux selection for short bursts as defined by their 
$\rm t_{90}$ duration (Paciesas et al. in prep.).  This selection is made by adding the count rates in the NaI detectors again and 
selecting the single bin of signal with the highest background-subtracted count rate.  This selection is a snapshot of the 
energetics at the most intense part of the burst.  Figure \ref{allTime} shows the distribution of accumulation time based on the 
signal-to-noise selection criteria.  The accumulation time reported is similar to the observed emission time of the burst, excluding 
quiescent periods, \citep[e.g.][]{Mitrofanov}.  Figure \ref{allTime} also includes the comparisons of the model photon fluence and 
photon flux compared to the accumulation time.  Note that both comparisons contain two distinct regions associated with short 
and long GRBs.  While there appears to be a very clear correlation between the photon fluence and the accumulation time, there 
is little correlation between the burst-averaged photon flux and the accumulation time.  However, the latter depicts a relationship 
similar to the hardness--duration relationship explored in \citet{Kouveliotou}.

%% Models
\section{Models}
We chose four spectral models to fit the spectra of GRBs in our selection sample. These models include a single power law 
(PL), Band's GRB function (BAND), an exponential cut-off power-law (COMP), and smoothly broken power law (SBPL). All 
models are formulated in units of photon flux with energy (\emph{E}) in keV and multiplied  by a normalization constant \emph
{A} ($ \rm ph \ s^{-1} \ cm^{-2} \ keV^{-1}$). Below we detail each model and its features.

%% PL
\subsection{Power-Law Model}
An obvious first model choice, ubiquitous in astrophysical spectra, is the single power law with two free parameters,
\begin{equation}
f_{ PL } ( E ) = A \left( \frac{ E }{ E_{piv} } \right )^{ \lambda }
\end{equation}
where \emph{A} is the amplitude and $\lambda$ is the spectral index. The pivot energy ($E_{piv}$) normalizes the model to the 
energy range under inspection and helps reduce cross-correlation of other parameters.  In all cases in this catalog, $E_{piv}$ is 
held fixed at 100 keV.  While most GRBs exhibit a spectral break in the GBM passband, some weak GRBs are too weak to 
adequately constrain this break in the fits and therefore we chose to fit these with the PL model.

%% BAND
\subsection{Band's GRB function}
Band's GRB function \citep{Band93} has become the standard for fitting GRB spectra, and therefore we include it in our analysis:

\begin{equation}
\begin{split}
&  f_{ BAND } ( E ) = \\
& A \begin{cases} 
\biggl(\frac{E}{100 \ \rm keV }\biggr)^{\alpha} \exp \biggl[- \frac{ (\alpha +2) E}{ E_{peak} } \biggr], \ E \geq \frac{ (\alpha - \beta) \ 
E_{peak} } { \alpha +2} \\
\biggl( \frac{E}{ 100 \ \rm keV } \biggr)^{ \beta } \exp (\beta -\alpha) \biggl[ \frac{(\alpha-\beta ) E_{peak}}{100 \ \rm keV \ (\alpha 
+2)} \biggr]^{\alpha-\beta }, \\ E < \frac{(\alpha -\beta ) \ E_{peak}}{\alpha +2}
\end{cases}
\end{split}
\end{equation}
The four free parameters are the amplitude, \emph{A}, the low and high energy spectral indices, $\alpha$ and $\beta$ 
respectively, and the $\nu F_{\nu}$ peak energy, $E_{peak}$. This function is essentially a smoothly broken power law with a 
curvature defined by its spectral indices. The low-energy index spectrum is a power law only asymptotically. 

%% COMP
\subsection{Comptonized Model}
This model considered for GRB spectra is an exponentially cutoff power-law, which is a subset of the Band function in the limit 
that $\beta \to -\infty$:
\begin{equation}
f_{COMP}(E) = A \ \Bigl(\frac{E}{E_{piv}}\Bigr) ^{\alpha} \exp \Biggl[ -\frac{(\alpha+2) \ E}{E_{peak}}  \Biggr]
\end{equation}	  
The three free parameters are the amplitude \emph{A}, the low energy spectral index $\alpha$ and $E_{peak}$. $E_{piv}$ is 
again fixed to 100 keV, as is the case for the power law model. With the extended high-energy response of GBM compared 
to BATSE, we rarely suffer from an inability to measure the high-energy spectra of medium to strong intensity GRBs; however, 
we still find that the Comptonized model adequately describes many GRB spectra.  

%% SBPL
\subsection{Smoothly Broken Power-Law}
The final model that we consider in this catalog is a broken power-law characterized by one break with flexible curvature able 
to fit spectra with both sharp and smooth transitions between the low and high energy power laws.  This model, first published in 
\citet{Ryde}, where the logarithmic derivative of the photon flux is a continuous hyperbolic tangent, has been re-parametrized 
\citep{Kaneko06} as such:
\begin{equation}
f_{SBPL}(E)=A \biggl(\frac{E}{E_{piv}} \biggr)^b  \ 10^{(a - a_{piv})}
\end{equation}
where
\begin{equation}
	\begin{split}
&a=m\Delta \ln \biggl(\frac{e^q+e^{-q}}{2}\biggr), \\
&\\
&a_{piv}=m\Delta \ln \biggl(\frac{e^{q_{piv}}+e^{-q_{piv}}}{2} \bigg), \\
&\\
&q=\frac{\log (E/E_b)}{\Delta}, \quad q_{piv}=\frac{\log(E_{piv}/E)}{\Delta},\\
&\\
&m=\frac{\lambda_2-\lambda_1}{2}, \quad b=\frac{\lambda_1+\lambda_2}{2}.
	\end{split}
\end{equation}
In the above relations, the low- and high-energy power law indices are $\lambda_1$ and $\lambda_2$ respectively, $E_b$ is 
the break energy in keV, and $\Delta$ is the break scale in decades of energy.  The break scale is independent and not 
coupled to the power law indices as it is with the Band function, and as such represents an additional degree of freedom.  
However, \citet{Kaneko06} found that an appropriate value for $\Delta$ for GRB spectra is 0.3, therefore we fix $\Delta$ at this 
value.  Studies on the behavior of $\Delta$ pertaining to GBM GRBs may be possible after more bright bursts have been 
detected; therefore this is left to possible future catalogs.

%%Data Analysis & Results
\section{Data Analysis \& Results}
To study the spectra resulting from the GBM detectors, a method must be established to associate the energy deposited in the 
detectors to the energy of the detected photons.  This association is dependent on effective area and the angle of the detector 
to the incoming photons.  To do this, detector response matrices (DRMs) are used to convert the photon energies into detector 
channel energies.  The DRM is a mathematical model of the deposition of photon energy in the crystal---a photon that interacts 
by the photoelectric effect will deposit 100\% of its energy, subject to resolution broadening, while a photon that interacts by a 
single Compton scatter may deposit only a portion of its energy.  The exception to this is when a photon carrying the energy of 
the iodine K--shell escapes the crystal.  The energy calibration determines the energy boundaries of the energy deposition 
channels.  The DRMs record this energy response of the detectors at various angles, which was determined through extensive 
simulations using GRESS \citep{Hoover2010} and validated by the source survey.  The response matrices for all GRBs in the 
catalog were made using GBMRSP v1.9 of the response generator and version 2 of the GBM DRM database, and all responses 
employ atmospheric response modeling to correct for possible atmospheric interference.  In particular, we use RSP2 files, which 
contain multiple DRMs based on the amount of slew the spacecraft experiences during the burst.  A new DRM is calculated for 
every $2^\circ$ of slew, changing the effective area of each detector based on its angle to the source.  These DRMs are then all 
stored in a single RSP2 file for each detector.  During the fitting process, each DRM is weighted by the counts fluence through 
the detector during each $2^\circ$ slew segment.

The spectral analysis of all bursts was performed using RMfit, version 3.4rc1.  RMfit employs a modified, forward-folding 
Levenberg-Marquardt algorithm for spectral fitting.  The Castor C-Statistic, which is a modified log likelihood statistic based on 
the Cash parametrization \citep{Cash} is used in the model-fitting process as a figure of merit to be minimized.  This statistic is 
preferable over the more traditional $\chi^2$ statistic minimization because of the non-Gaussian counting statistics present 
when dividing the energy spectra of GBM GRBs into 128 channels.  Although it is advantageous to perform the spectral fitting 
using C-Stat, this statistics provides no estimation of the goodness-of-fit, since there exists no standard probability distribution 
for likelihood statistics.  For this reason, we also calculate $\chi^2$ for each spectral fit that was performed through minimizing 
C-Stat.  This allows an estimation of the goodness-of-fit of a function to the data even though $\chi^2$ was not minimized.  This 
also allows for easy comparison between nested models.

\begin{figure}
	\begin{center}
		\includegraphics[scale=0.7]{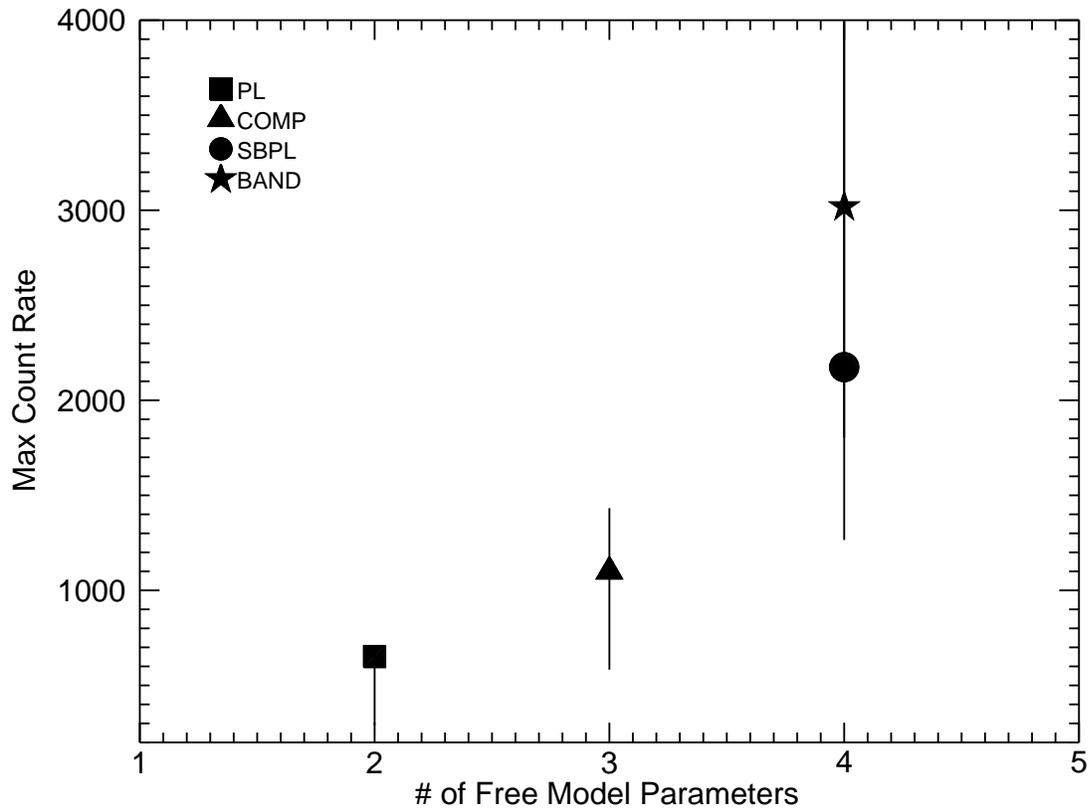}
	\end{center}
\caption{Plot of the average maximum background-subtracted count rates (as a proxy for observed intensity) versus the number 
of degrees of freedom of the best fit model.  The count rates in the NaI detectors for each burst was summed, the background 
subtracted, and the maximum count rate was computed.  The best fit model was determined for each burst, and a geometric 
average was calculated for the maximum count rates of the bursts for each best fit model.  The error bars shown are the 1$
\sigma$ standard deviations of the distributions of maximum count rates for each best fit model.
 \label{countrate}}
\end{figure}

We fit  the four functions described in Section 4 to the spectrum of each burst.  The BAND and COMP functions are 
parametrized with $E_{peak}$, the peak in the power density spectrum, while the SBPL is parametrized with the break energy, 
$E_{break}$.  We choose to fit these four different functions because the measurable spectrum of GRBs is dependent on 
intensity, as is shown in Figure \ref{countrate}.  Observably less intense bursts provide less data to support a large number of 
parameters.  This may appear obvious, but it allows us to determine why in many situations a particular empirical function 
provides a poor fit, while in other cases it provides an accurate fit.  For example, the energy spectra of GRBs are normally well fit 
by two smoothly joined power laws.  For particularly bright GRBs, the BAND and SBPL functions are typically an accurate 
description of the spectrum, while for weaker bursts the COMP function is most acceptable.  Bursts that have signal significance 
on the order of the background fluctuations do not have a detectable distinctive break in their spectrum and so the power law is 
the most acceptable function.  Although for weaker GRBs a model with more parameters is not statistically preferred, it is 
instructive to study the parameters of even the weaker bursts.  In addition, the actual physical GRB processes can have an effect 
on the spectra and different empirical models may fit certain bursts better than others.  The spectral results, including the best fit 
spectral parameters and the photon model, are stored in files following the FITS standard (see Appendix for file description) and 
are hosted as a public data archive on HEASARC\footnote{\url{http://heasarc.gsfc.nasa.gov/W3Browse/fermi/fermigbrst.html}}.

When inspecting the distribution of the parameters for the fitted models, we first define a data cut based on the goodness-of-fit.  
We require the $\chi^2$ statistic for the fit to be within the  3$\sigma$ expected region for the $\chi^2$ distribution of the given 
degrees of freedom, and we define a subset of each parameter distribution of this data cut as GOOD if the parameter error is 
within certain limits.  Following \citet{Kaneko06}, for the low-energy power law indices, we consider GOOD values to have errors 
less than 0.4, and for high-energy power law indices we consider GOOD values to have errors less than 1.0.  For all other 
parameters we consider a GOOD value to have a relative error of 0.4 or better.  The motivation for this is to show 
well--constrained parameter values, rather than basing interpretations on parameters that are poorly constrained.

In addition, we define a BEST sample where we compare the goodness-of-fit of all spectral models for each burst and select 
the most preferred model based on the difference in $\chi^2$ per degree of freedom.  The criterion for accepting a model with 
a single additional parameter is a change in $\chi^2$ of at least 6 since the probability for achieving this difference is $\sim
$0.01.  The parameter distributions are then populated with the spectral parameters from the BEST spectral fits.

%% Fluence Spectra
\subsection{Fluence Spectra}
\begin{figure}
	\begin{center}
		\subfigure[]{\label{indexlof}\includegraphics[scale=0.35]{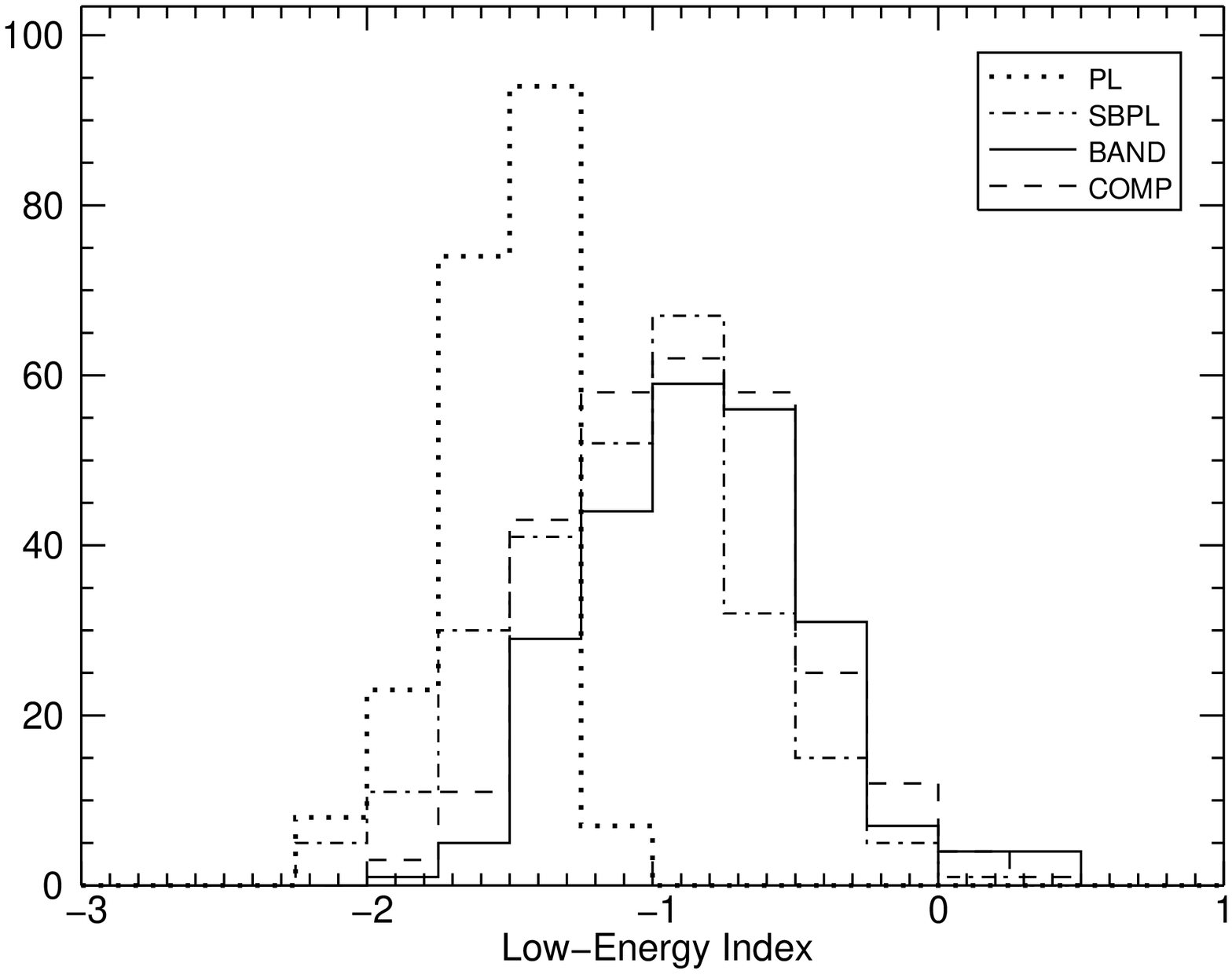}}
		\subfigure[]{\label{alphasbplf}\includegraphics[scale=0.35]{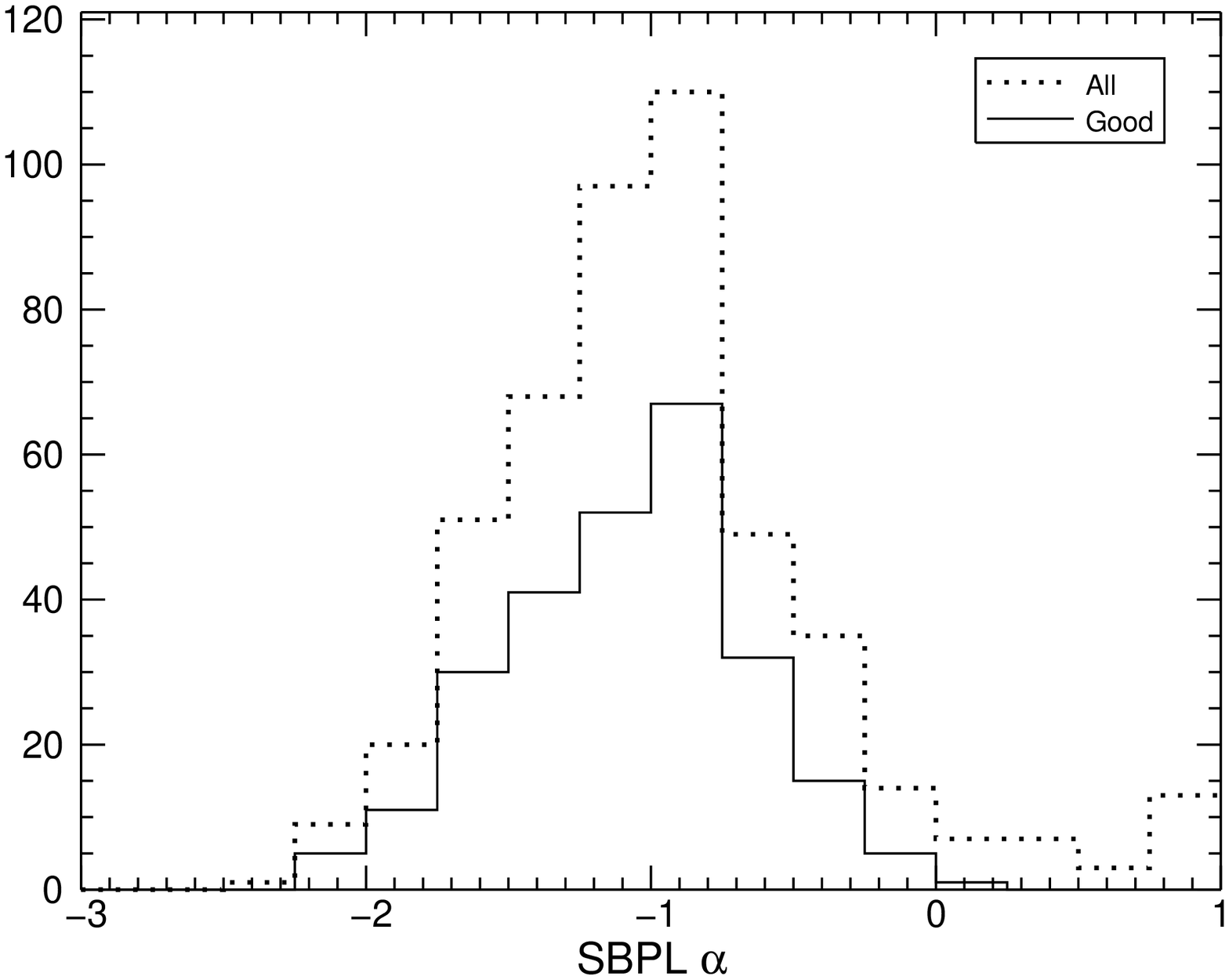}}\\
		\subfigure[]{\label{alphabandf}\includegraphics[scale=0.35]{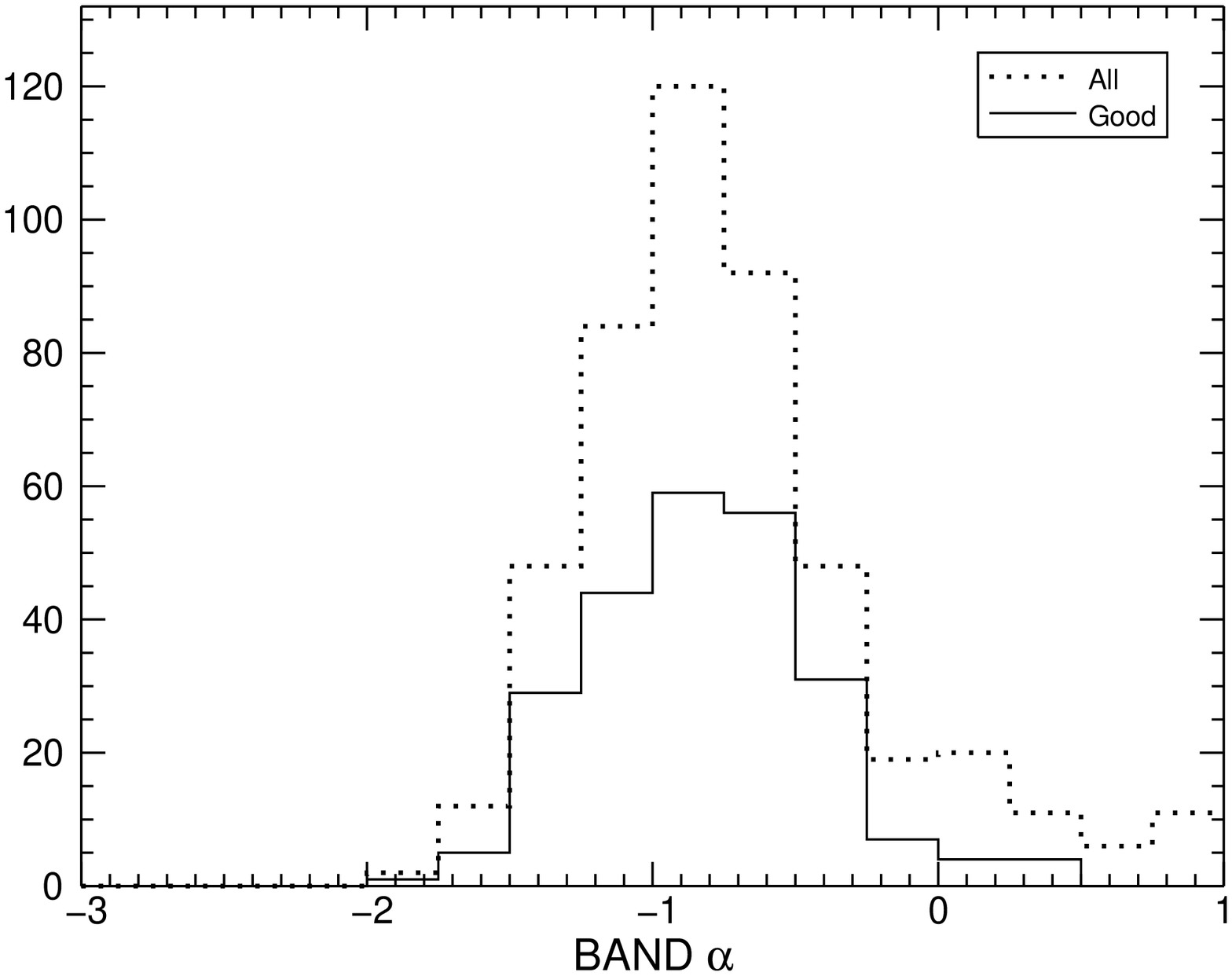}}
		\subfigure[]{\label{alphacompf}\includegraphics[scale=0.35]{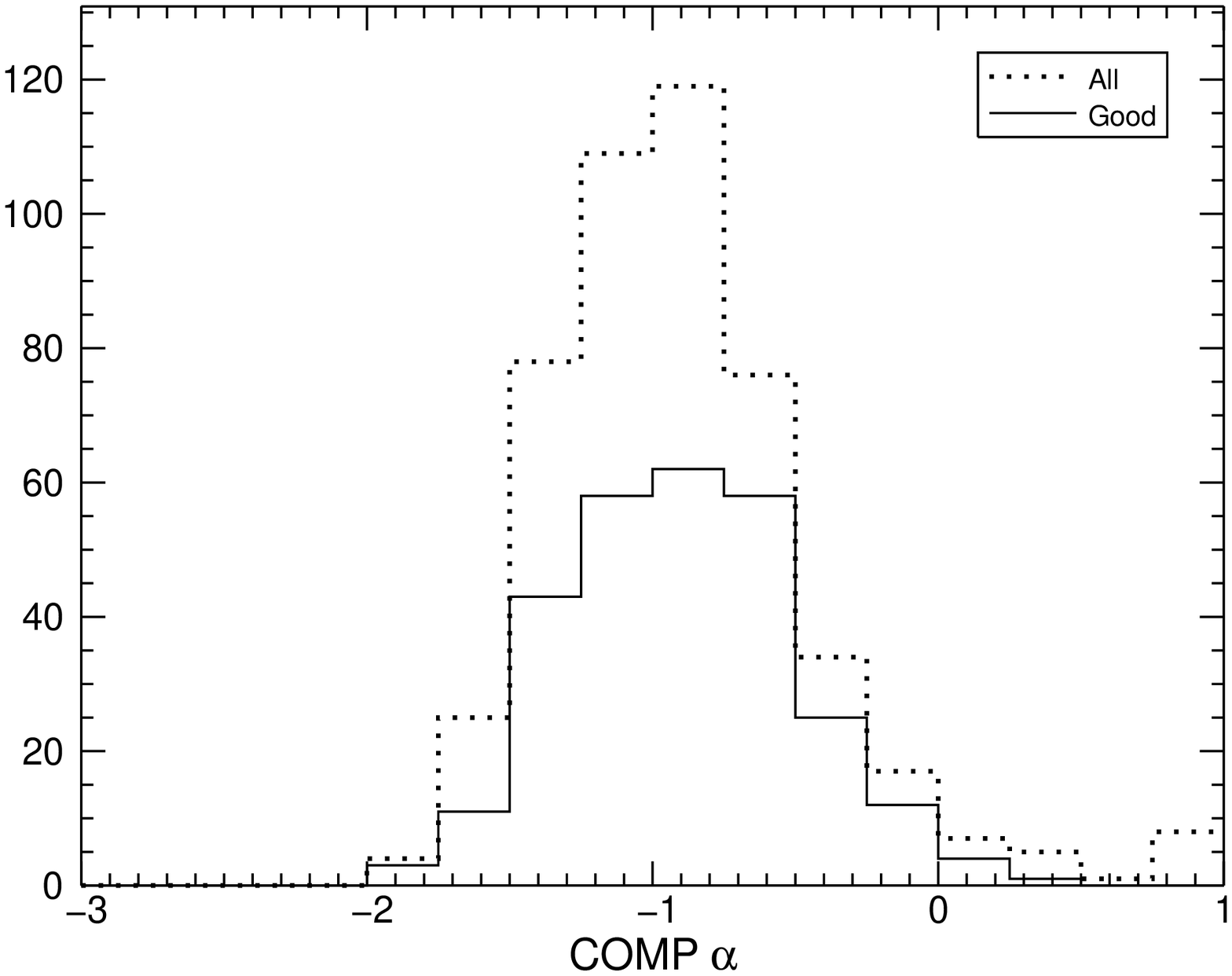}}
	\end{center}
\caption{Distributions of the low-energy spectral indices from fluence spectral fits.  \ref{indexlof} shows the distributions of 
GOOD parameters and compares to the distribution of PL indices.  \ref{alphasbplf}--\ref{alphacompf} display the comparison 
between the distribution of GOOD parameters and all parameters with no data cuts.  The last bin includes values greater than 
1. \label{loindexf}}
\end{figure}

\begin{figure}
	\begin{center}
		\subfigure[]{\label{indexhif}\includegraphics[scale=0.35]{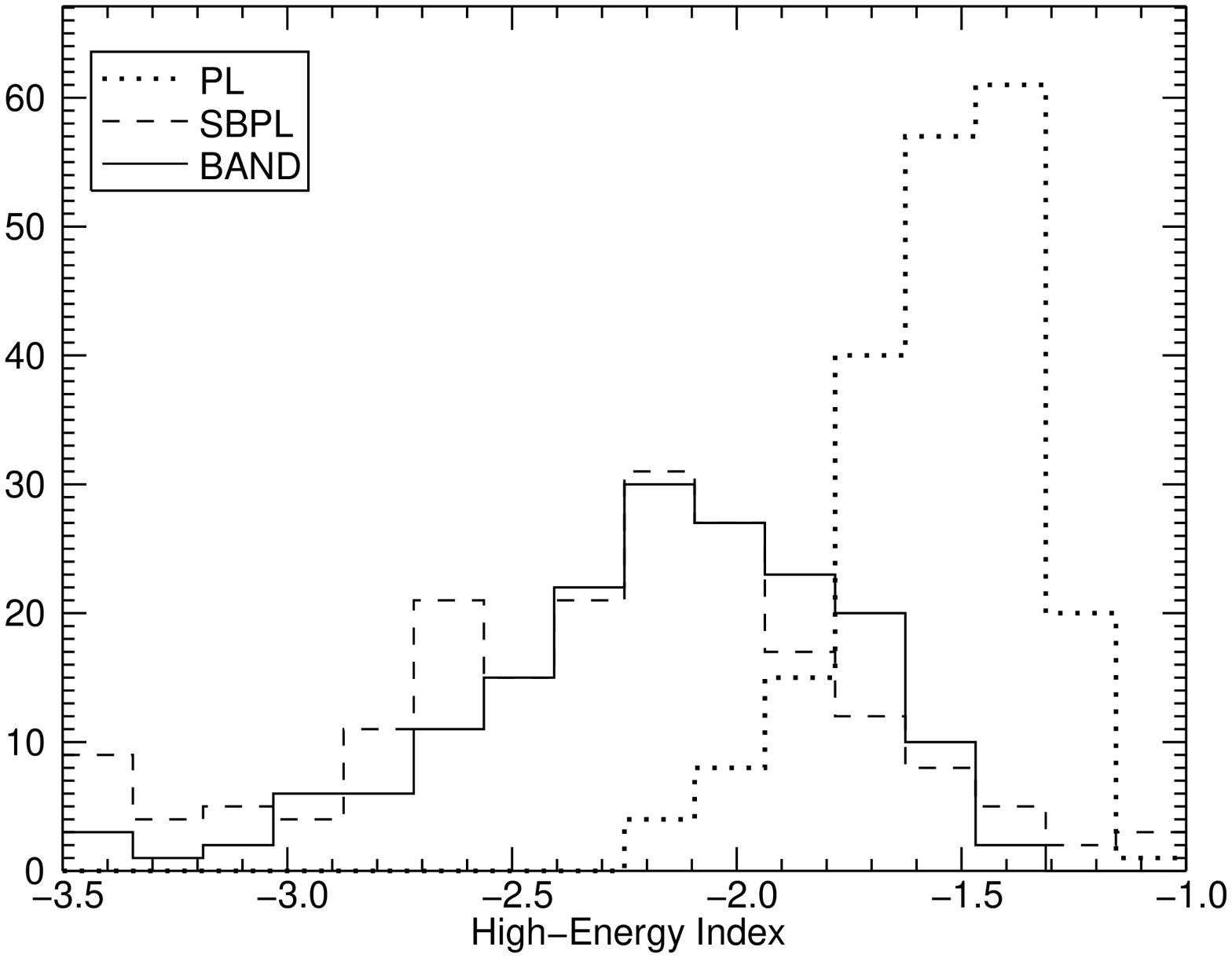}}
		\subfigure[]{\label{betasbplf}\includegraphics[scale=0.35]{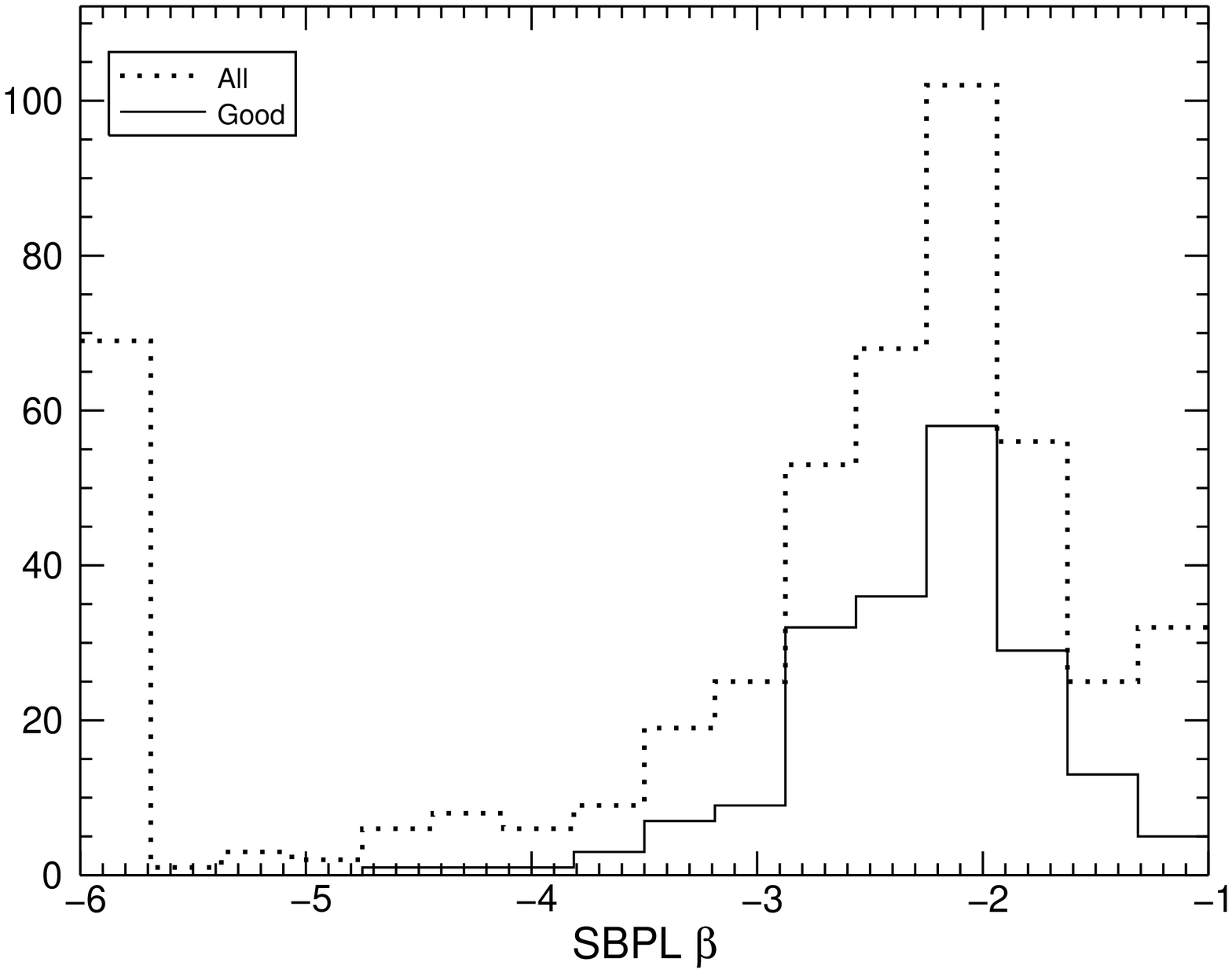}}\\
		\subfigure[]{\label{betabandf}\includegraphics[scale=0.35]{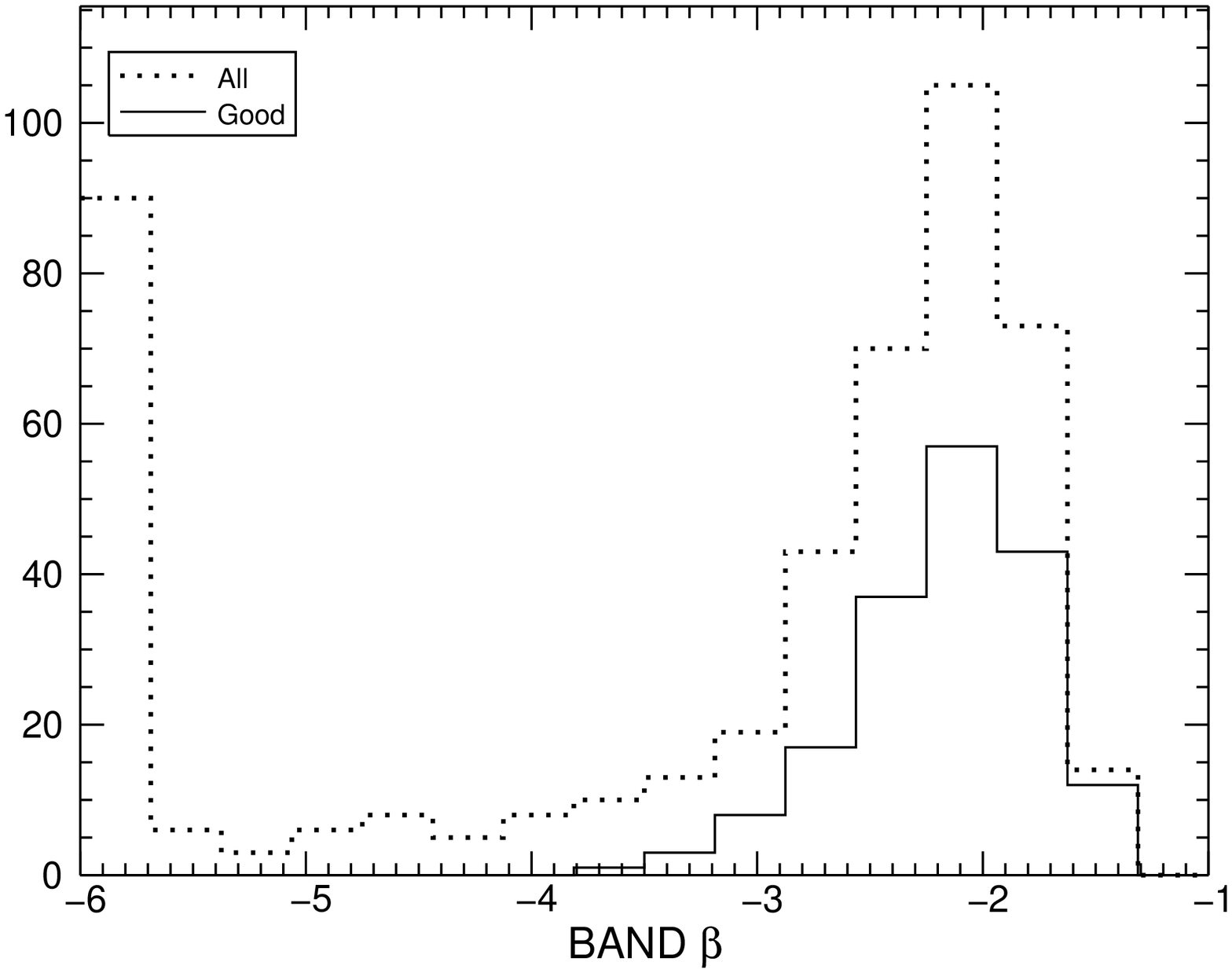}}
		\subfigure[]{\label{deltasf}\includegraphics[scale=0.35]{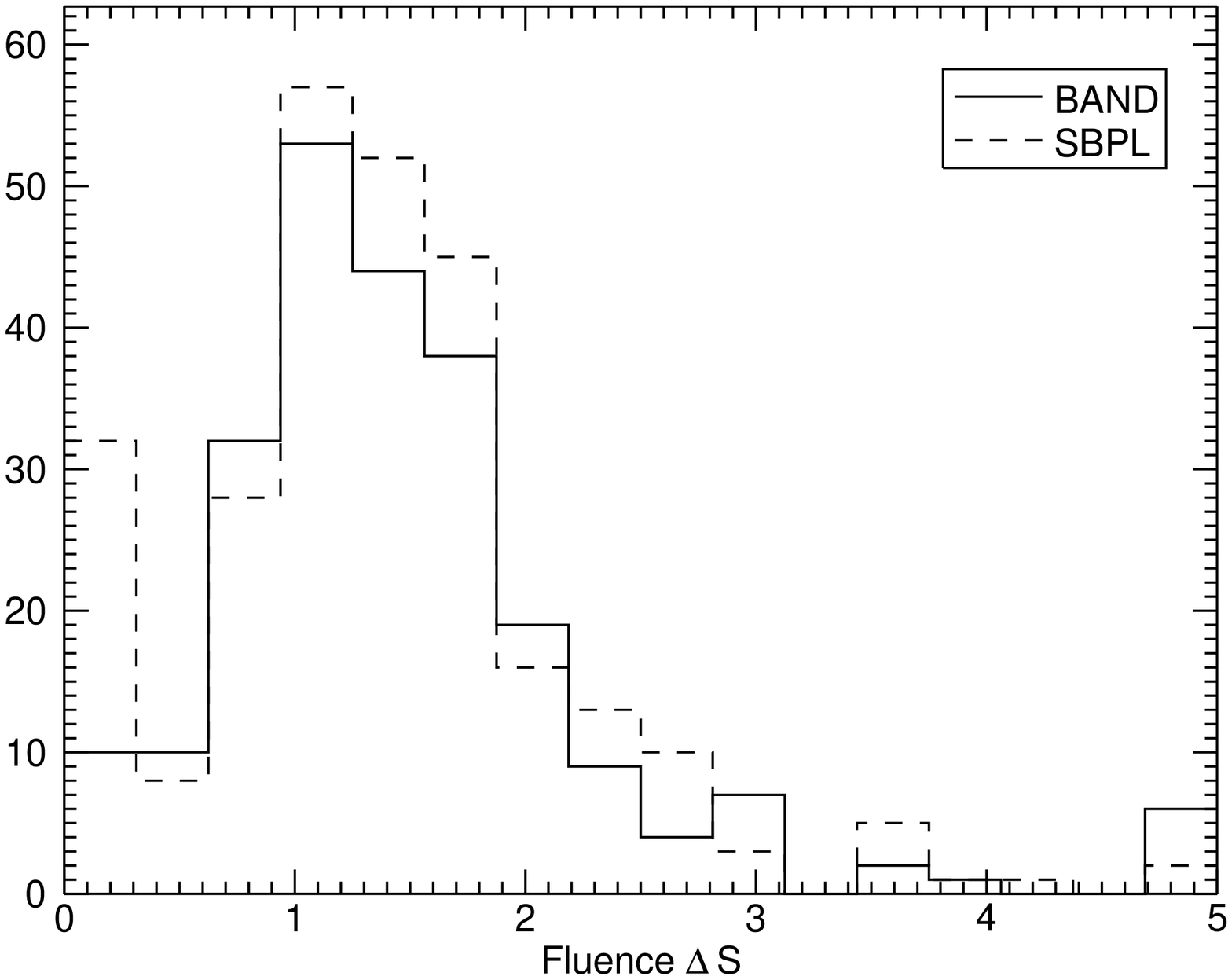}}
	\end{center}
\caption{\ref{indexhif} - \ref{betabandf} are distributions of the high-energy spectral indices from fluence spectral fits.  \ref
{indexhif} shows the distributions of GOOD parameters and compares to the distribution of PL indices.  \ref{betasbplf} and \ref
{betabandf}  display the comparison between the distribution of GOOD parameters and all parameters with no data cuts.  The 
first bins include values less than -6 and the last bins includes values greater than -1.  \ref{deltasf} shows the distribution of the 
difference between the low- and high-energy indices. The first bin contains values less than 0, indicating that the centroid value 
of alpha is steeper than the centroid value of beta. \label{hiindexf}}
\end{figure}

The time-integrated fluence spectral distributions admit results that are averaged over the duration of the observed emission.  It 
should be noted that the following distributions do not take into account any spectral evolution that may exist within bursts.  The 
low-energy indices, as shown in Figure \ref{loindexf}, distribute about a $-1$ power law typical of most GRBs.  Up to 33\% of 
the GOOD low-energy indices violate the $-2/3$ synchrotron ``line-of-death'', while an additional 62\% of the indices violate the 
$-3/2$ synchrotron cooling limit.  The high-energy indices in Figure \ref{hiindexf} peak at a slope slightly steeper than $-2$ and 
have a long tail toward steeper indices.  Note that the large number of unconstrained (or very steep) high-energy indices in the 
distribution of all high-energy index values indicates that a large number of GRBs are  better fit by the COMP model, which is 
equivalent to a BAND function with a high-energy index of $- \infty$.  The comparison of the simple power law index to the low- 
and high-energy indices makes evident that the simple power law index is averaged over the break energy, resulting in a 
index that is on average steeper than the low-energy index yet shallower than the high-energy index.  We also show in Figure 
\ref{deltasf} the difference between the time-integrated low- and high-energy spectral indices, $\Delta S =(\alpha-\beta)$.  This 
quantity is useful since the synchrotron shock model makes predictions of this value in a number of cases \citep{Preece}.  The 
fluence spectra distribution of $\Delta S$ bears resemblance to the time-resolved results in \citet{Preece}, as well as 
time--integrated results in \citet{Kaneko06}. 

\begin{figure}
	\begin{center}
		\subfigure[]{\label{ebreaksbplf}\includegraphics[scale=0.35]{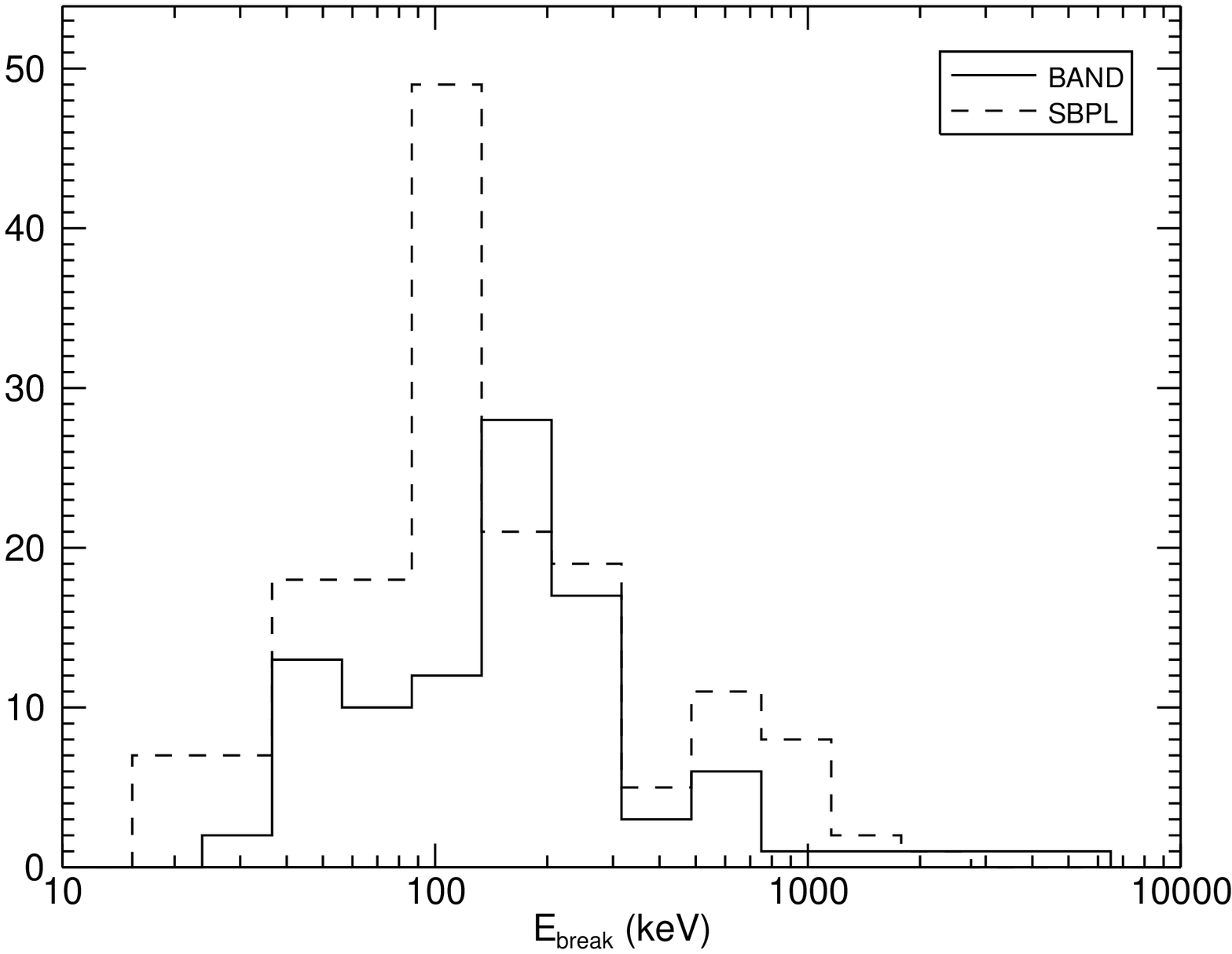}}
		\subfigure[]{\label{epeakf}\includegraphics[scale=0.35]{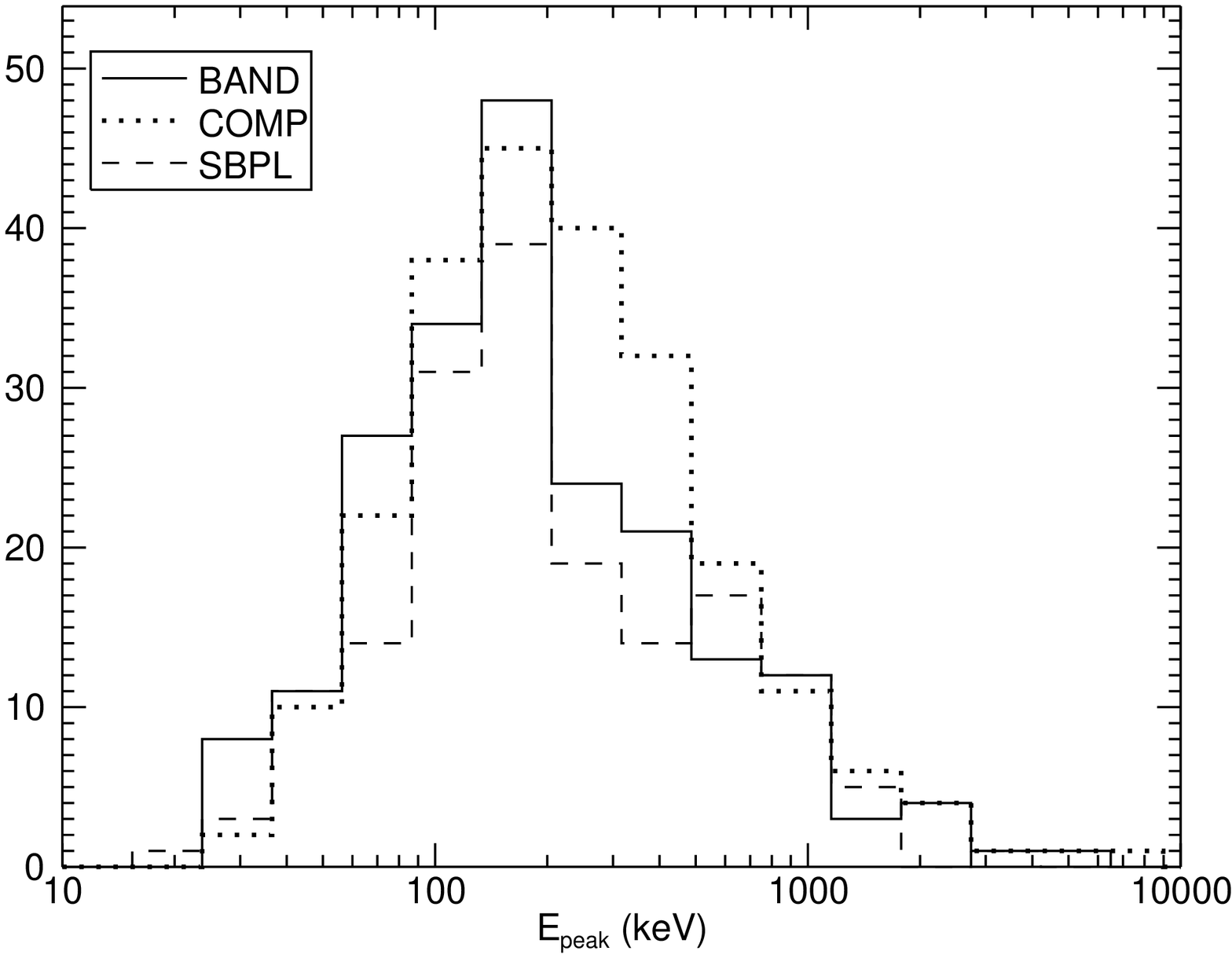}}\\
		\subfigure[]{\label{epeakbandf}\includegraphics[scale=0.35]{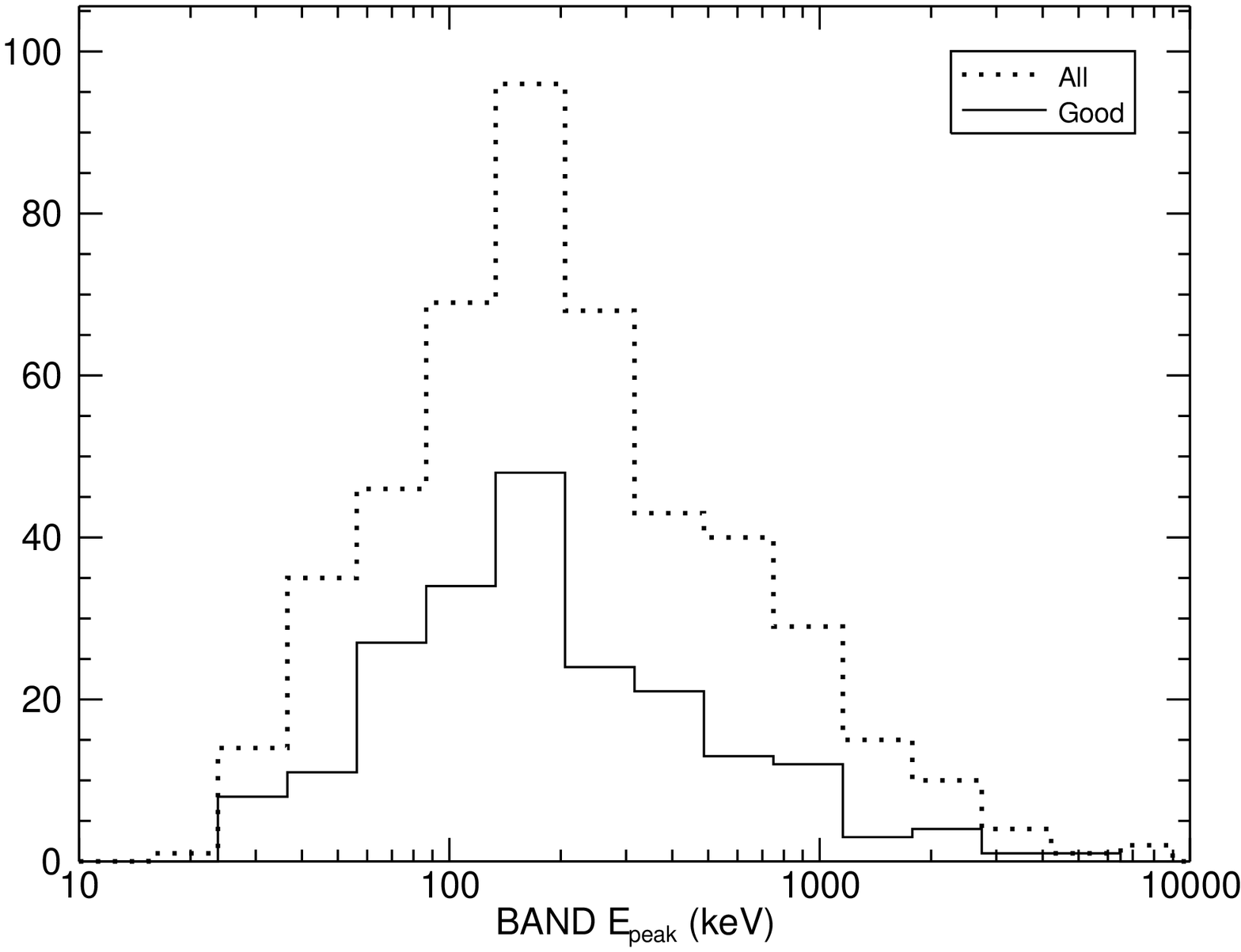}}
		\subfigure[]{\label{epeakcompf}\includegraphics[scale=0.35]{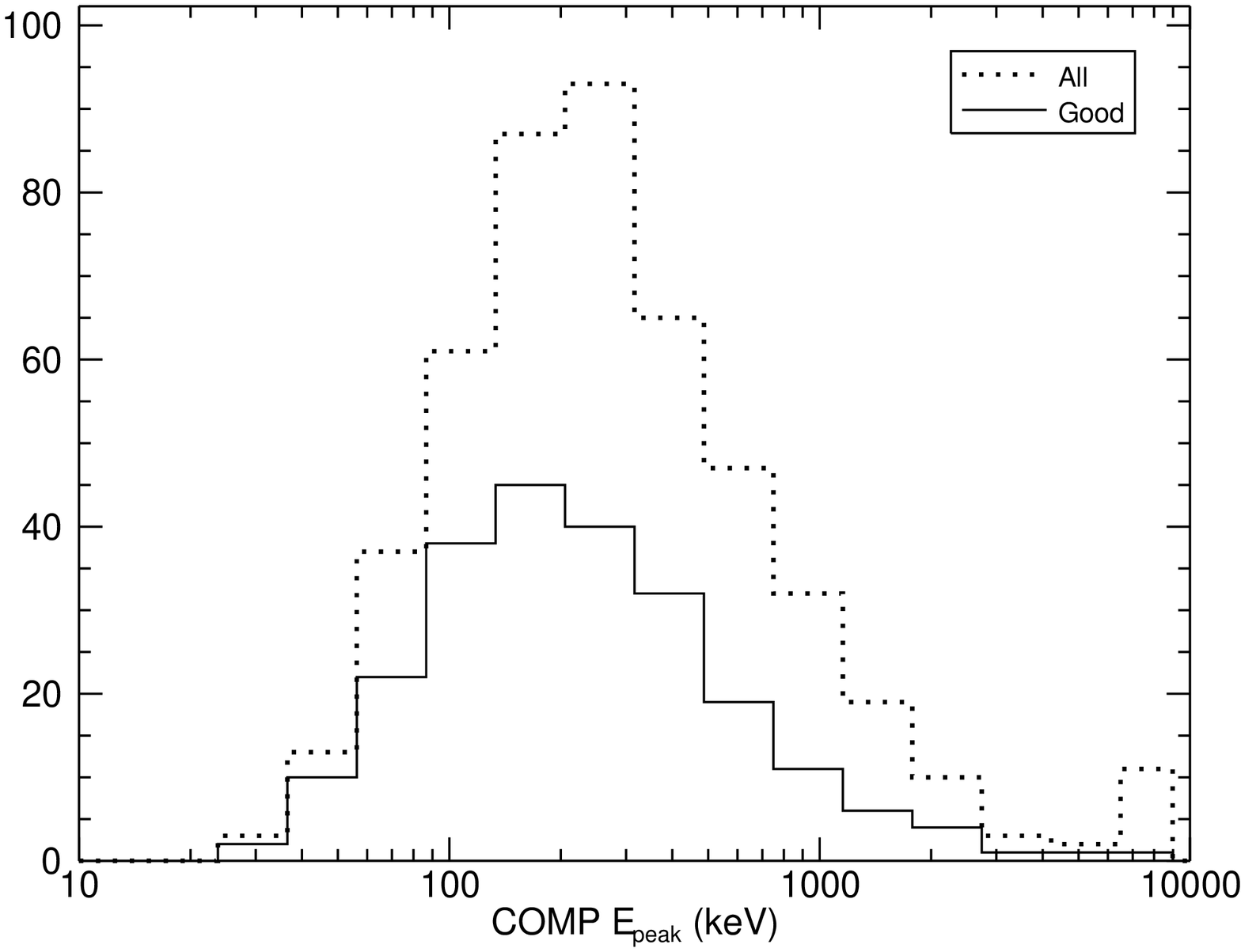}}
	\end{center}
\caption{Distributions of $E_{break}$ and $E_{peak}$ from fluence spectral fits.  \ref{ebreaksbplf} displays the comparison 
between the distribution of GOOD $E_{break}$ and $E_{break}$ with no data cuts.  \ref{epeakf} shows the distributions of 
GOOD $E_{peak}$ for BAND, SBPL, and COMP.  \ref{epeakbandf} and \ref{epeakcompf} display the comparison between the 
distribution of GOOD parameters and all parameters with no data cuts. \label{epeakebreakf}}
\end{figure}

\begin{figure}
	\begin{center}
		\includegraphics[scale=0.45]{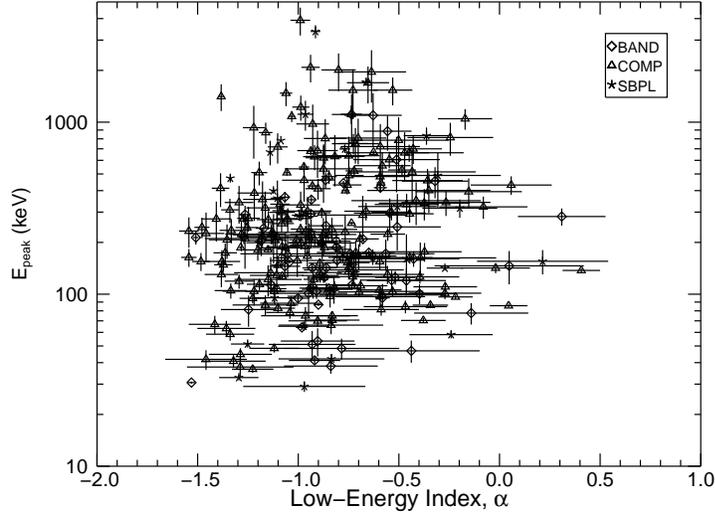}
	\end{center}
\caption{Comparison of the low-energy index and $E_{peak}$ for three models from the fluence spectral fits.  This comparison 
reveals a correlation between the $E_{peak}$ energy and the uncertainty on the low-energy index: generally a  lower energy 
$E_{peak}$ tends to result in a less constrained low-energy index. \label{alphaepeak}}
\end{figure}

In Figure \ref{epeakebreakf}, we show the distributions for the break energy, $E_{break}$ and the peak of the power density 
spectrum, $E_{peak}$.  $E_{break}$ is the energy at which the low- and high-energy power laws are joined, which is not 
necessarily representative of the $E_{peak}$.  The $E_{break}$ for the SBPL has a strong clustering about 100 keV, while 
the peak for the BAND distribution is closer to 200 keV.  The $E_{peak}$ distributions all generally peak around 200 keV  and 
cover just over two orders of magnitude, which is consistent with previous findings \citep{Mallozzi95, Lloyd00} from BATSE.  As 
discussed in \citet{Kaneko06}, although the SBPL is parametrized with $E_{break}$, the $E_{peak}$ can be derived from the 
functional form.  We have calculated the $E_{peak}$ for all bursts with low-energy index shallower than -2 and high-energy 
index steeper than -2, and we have used the covariance matrix to formally propagate and calculate the errors on the derived 
$E_{peak}$.  The peak and overall distribution of $E_{peak}$ is similar to that found by the BATSE Large Area Detectors, which 
had a much smaller bandwidth and larger collecting area.  This would seem to indicate that it is unlikely for there to be a hidden 
population undiscovered by either instrument within the ~20 keV - 2 MeV range.  Additionally, the value of $E_{peak}$ can 
strongly affect the measurement of the low-energy index of the spectrum, as shown in Figure \ref{alphaepeak}.  A general trend 
appears to show that lower $E_{peak}$ values tend to increase the uncertainty in the measurement of the low-energy index, 
mostly due to the fact that a spectrum with a low $E_{peak}$ will exhibit most of its curvature near the lower end of the instrument 
bandpass.  In many cases, if the low-energy index is found to be reasonably steep ($\lesssim -1$), the uncertainty of the index is 
minimized even if $E_{peak}$ is low.

\begin{figure}
	\begin{center}
		\includegraphics[scale=0.45]{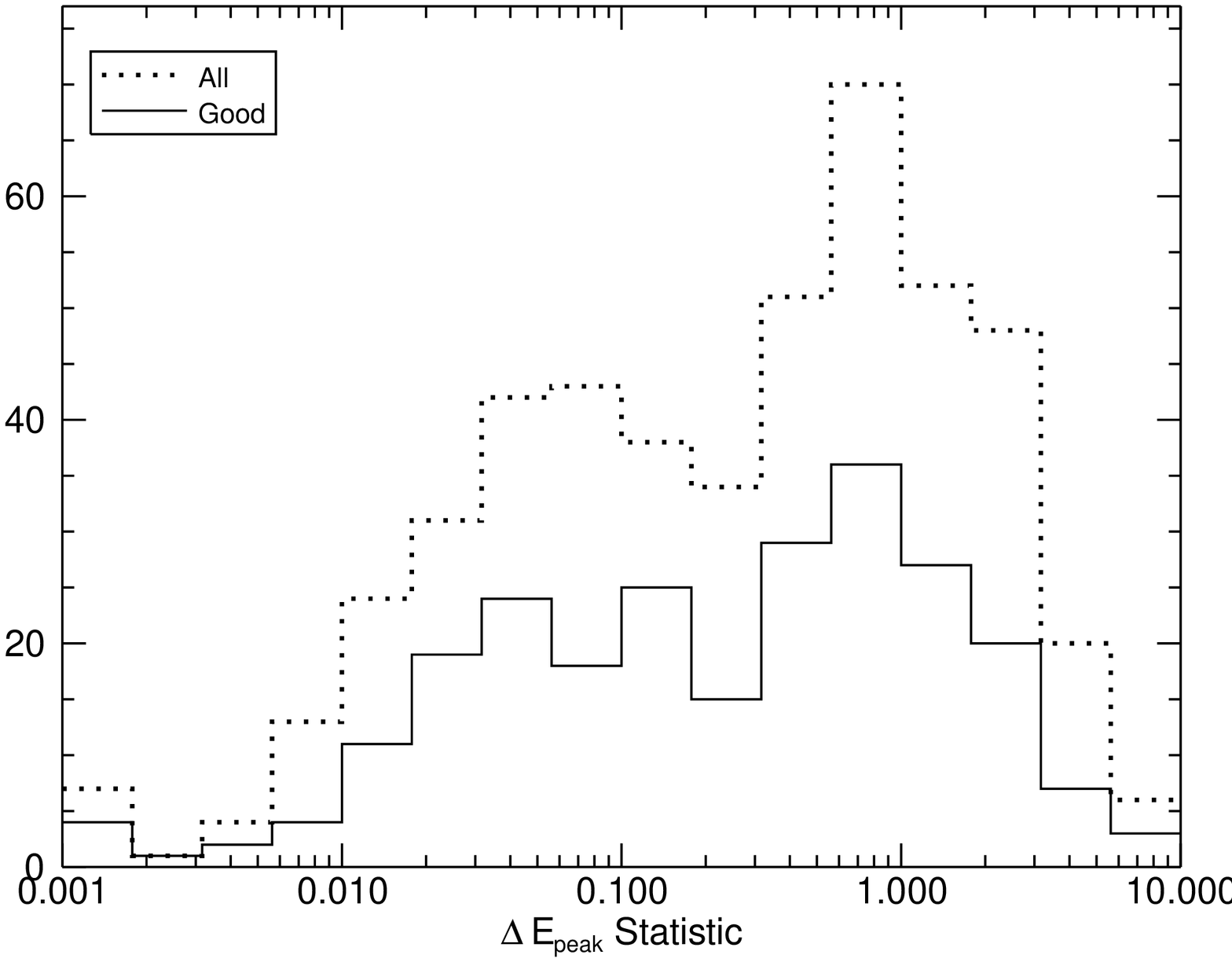}
	\end{center}
\caption{Distribution of the $\Delta E_{peak}$ statistic for the COMP and BAND models from fluence spectral fits.  A value less 
than 1 indicates the $E_{peak}$ values are within errors, while a value larger than 1 indicates the $E_{peak}$ values are not 
within errors. \label{deltaepeakf}}
\end{figure}

It is of interest to study the difference in the value of $E_{peak}$ between the BAND and COMP functions since they are the two 
main functions used to study GRB spectra, and COMP is a special case of BAND.  To study the relative deviation between the 
two values we calculate a statistic based on the difference between the values and taking into account their 1$\sigma$ errors.  
This statistic can be calculated by
\begin{equation}
	\Delta E_{peak}=\frac{|E_{peak}^{C}-E_{peak}^{B}|}{\sigma_{E_{peak}}^{C}+\sigma_{E_{peak}}^{B}}
\end{equation}
where C and B indicate the COMP and BAND values respectively.  This statistic has a value of unity when the deviation 
between the $E_{peak}$ values exactly matches the sum of the 1$\sigma$ errors.  A value less than one indicates the $E_
{peak}$ values are within errors, and a value greater than one indicates that the $E_{peak}$ values are not within errors of 
each other.  Figure \ref{deltaepeakf} depicts the distribution of the statistic and roughly 25\% of the BAND and COMP $E_{peak}
$ values are found to be outside the combined errors.  This indicates that, although COMP is a special case of BAND, a 
significant fraction of the $E_{peak}$ values can vary by more than 1$\sigma$ based on which model is chosen.

\begin{figure}
	\begin{center}
		\subfigure[]{\label{pflux1mevf}\includegraphics[scale=0.35]{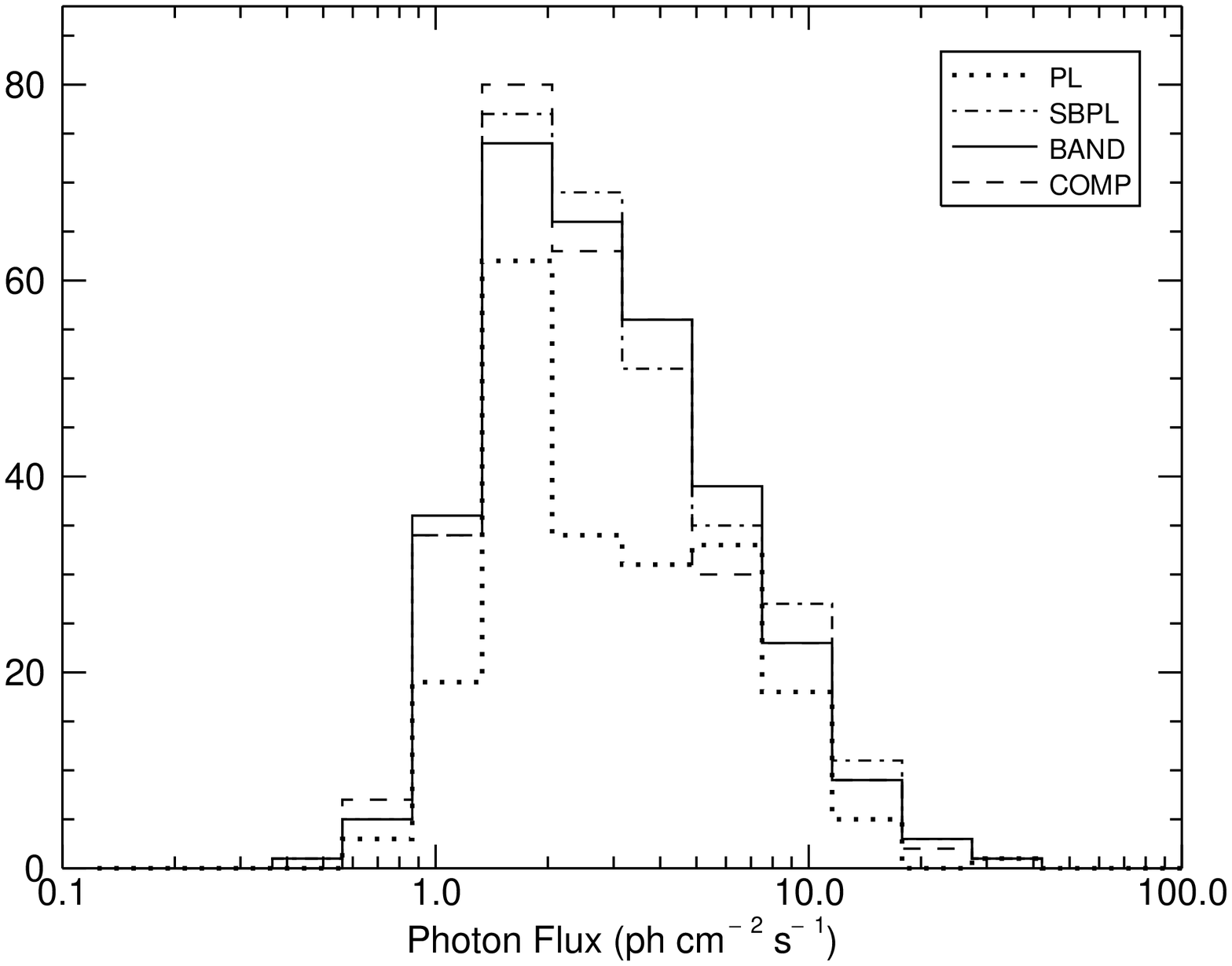}}
		\subfigure[]{\label{pflux40mevf}\includegraphics[scale=0.35]{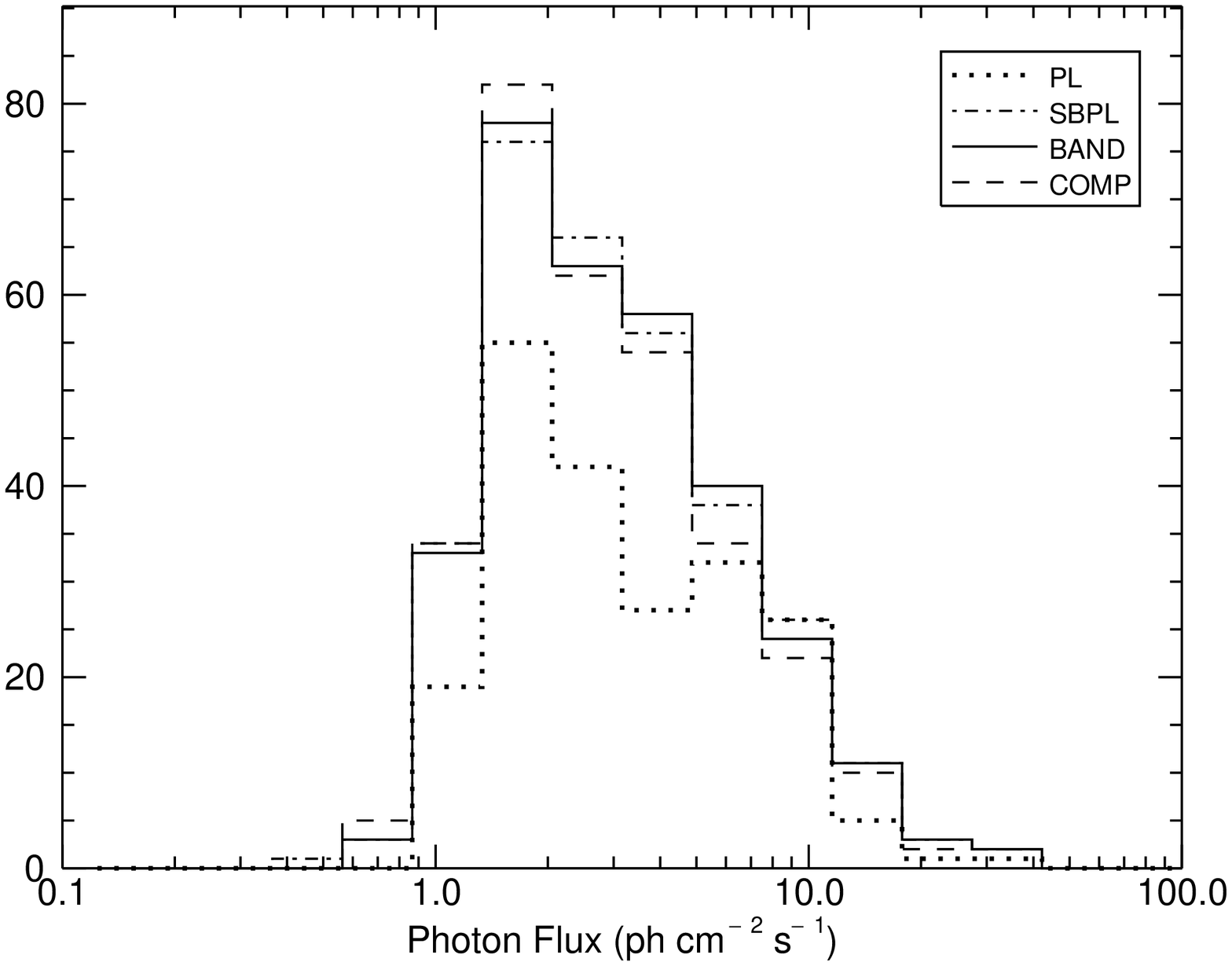}}\\
		\subfigure[]{\label{eflux1mevf}\includegraphics[scale=0.35]{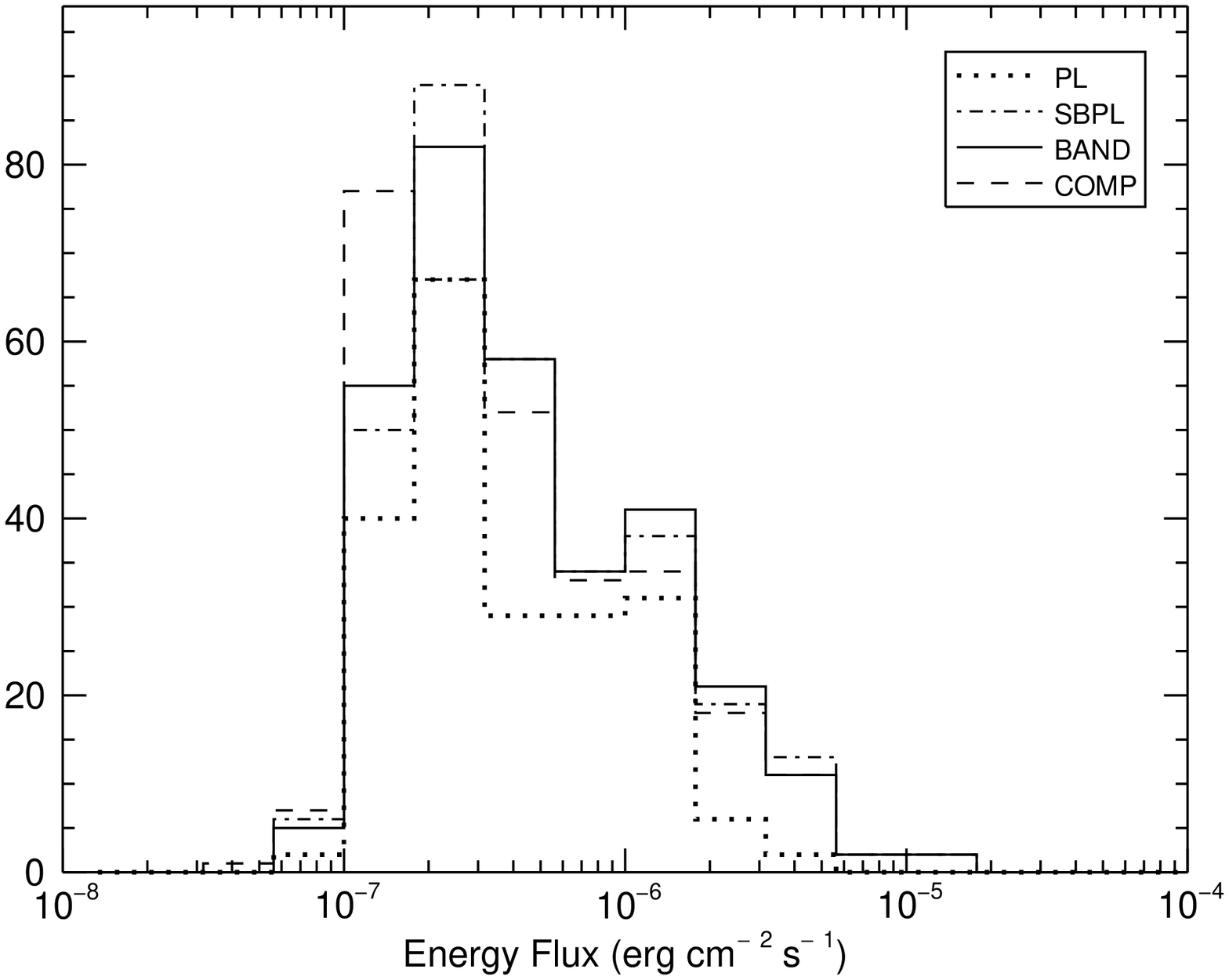}}
		\subfigure[]{\label{eflux40mevf}\includegraphics[scale=0.35]{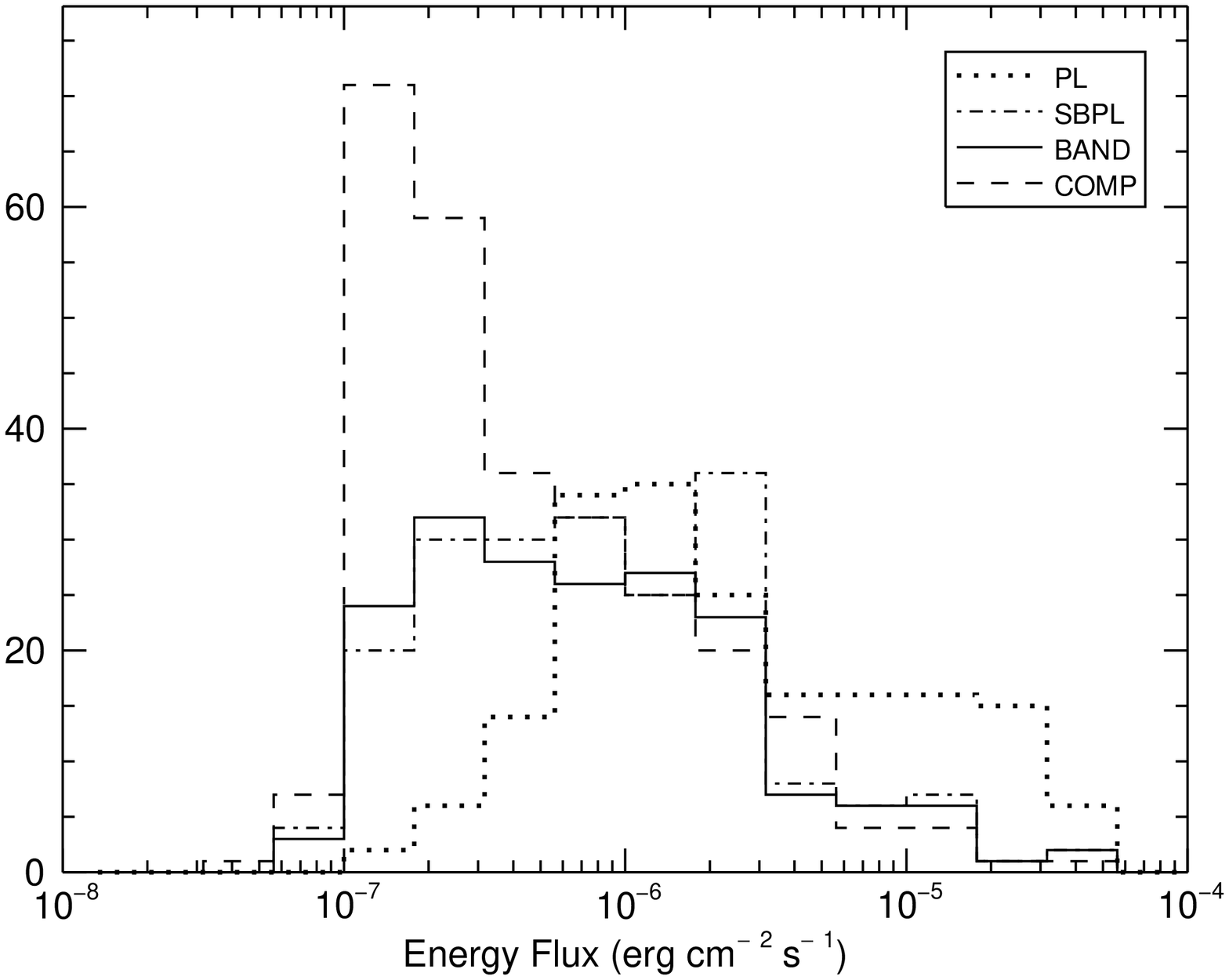}}
	\end{center}
\caption{Distributions of photon and energy flux from fluence spectral fits.  \ref{pflux1mevf} and \ref{eflux1mevf} display the flux 
distributions for the 8 keV--1 MeV band.  \ref{pflux40mevf} and \ref{eflux40mevf} display the flux distributions for the 8 keV--40 
MeV band.  Note that the plotted distributions contain the flux on two different timescales: 1024 ms and 64 ms. \label{fluxf}}
\end{figure}

\begin{figure}
	\begin{center}
		\subfigure[]{\label{pfluence1mev}\includegraphics[scale=0.35]{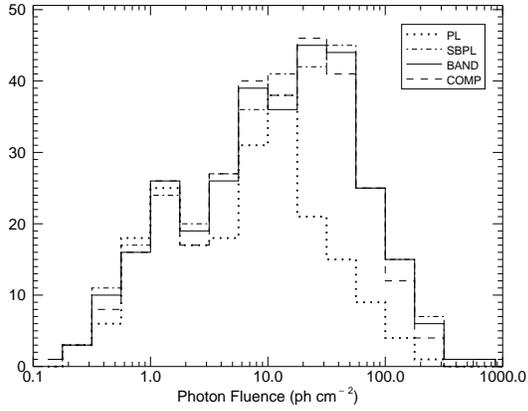}}
		\subfigure[]{\label{pfluence40mev}\includegraphics[scale=0.35]{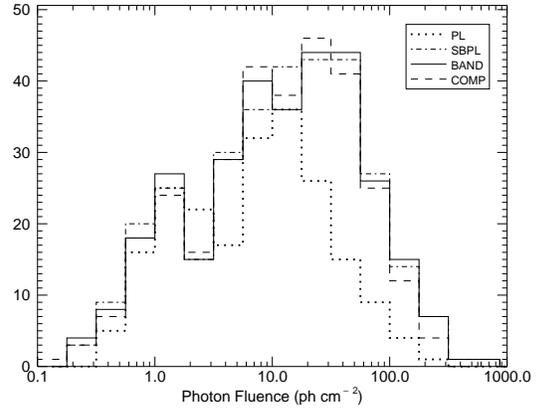}}\\
		\subfigure[]{\label{efluence1mev}\includegraphics[scale=0.35]{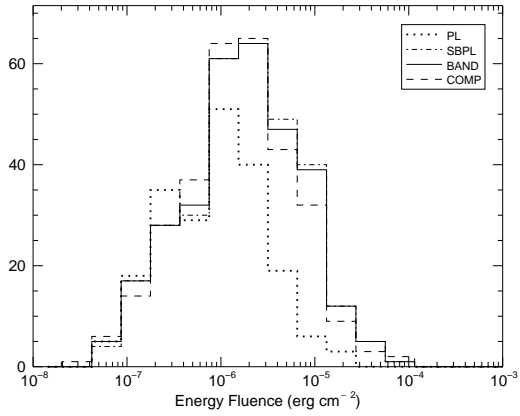}}
		\subfigure[]{\label{efluence40mev}\includegraphics[scale=0.35]{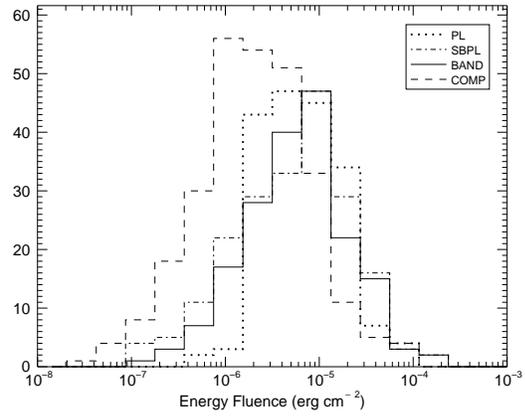}}
	\end{center}
\caption{Distributions of photon and energy fluence.  \ref{pfluence1mev} and \ref{efluence1mev} display the fluence 
distributions from the 8 keV--1 MeV band.  \ref{pfluence40mev} and \ref{efluence40mev} display the fluence distributions for the 
8 keV--40 MeV band.  \label{fluence}}
\end{figure}

The distributions for the time-averaged photon flux and energy flux are shown in Figure \ref{fluxf}.  The photon flux peaks 
around 2 photons cm$^{-2}$  s$^{-1}$, and the energy flux peaks at $2.5 \times 10^{-7}$ ergs cm$^{-2}$ s$^{-1}$ in the 8-1000 
keV band.  When integrating over the full GBM spectral band, 8 keV - 40 MeV, the energy flux broadens and for all but the 
simple power law model, approximates a top hat function with a small high-flux tail spanning about 2 orders of magnitude.  
Note that the low-flux cutoff is due to the sensitivity of the instrument, and therefore the peak of the distributions are directly 
affected by the instrument sensitivity.  A distribution of flux measurements with a more sensitive instrument will likely position the 
peak at a lower flux.  Similarly in Figure \ref{fluence}, the distributions for the photon fluence and energy fluence are depicted.  
The plots for the photon fluence appear to contain evidence of the duration bimodality of GRBs, and have discriminant peaks at 
about 1 and 20 photons cm$^{-2}$.  The energy fluence in the 8-1000 keV band peaks at about $1 \times 10^{-6}$ erg cm$^
{-2}$, while the peak in the 8 keV-40 MeV band shifts nearly an order of magnitude to about $1 \times 10^{-5}$ erg cm$^{-2}$ for 
all except the COMP model.  The COMP model is largely unaffected by the change in energy band due to the exponential cutoff.  
The brightest GRB contained in this catalog based on time-averaged photon flux is GRB 081009 with a flux of $\rm >30\ ph\ s^
{-1}\ cm^{-2}$ and the burst with the largest average energy flux is GRB 090227B with an energy flux of $\rm \sim1.3 \times 10^
{-5}\ ergs\ s^{-1} cm^{-2}$.  The most fluent burst (although not the longest duration) in the catalog is GRB 090618 with a photon 
fluence of $\rm >2600\ ph\ cm^{-2}$ and an energy fluence of $\rm >2.5 \times 10^{-4}\ ergs\ cm^{-2}$.

%% Peak Flux Spectra
\subsection{Peak Flux Spectra}
\begin{figure}
	\begin{center}
		\subfigure[]{\label{indexlop}\includegraphics[scale=0.35]{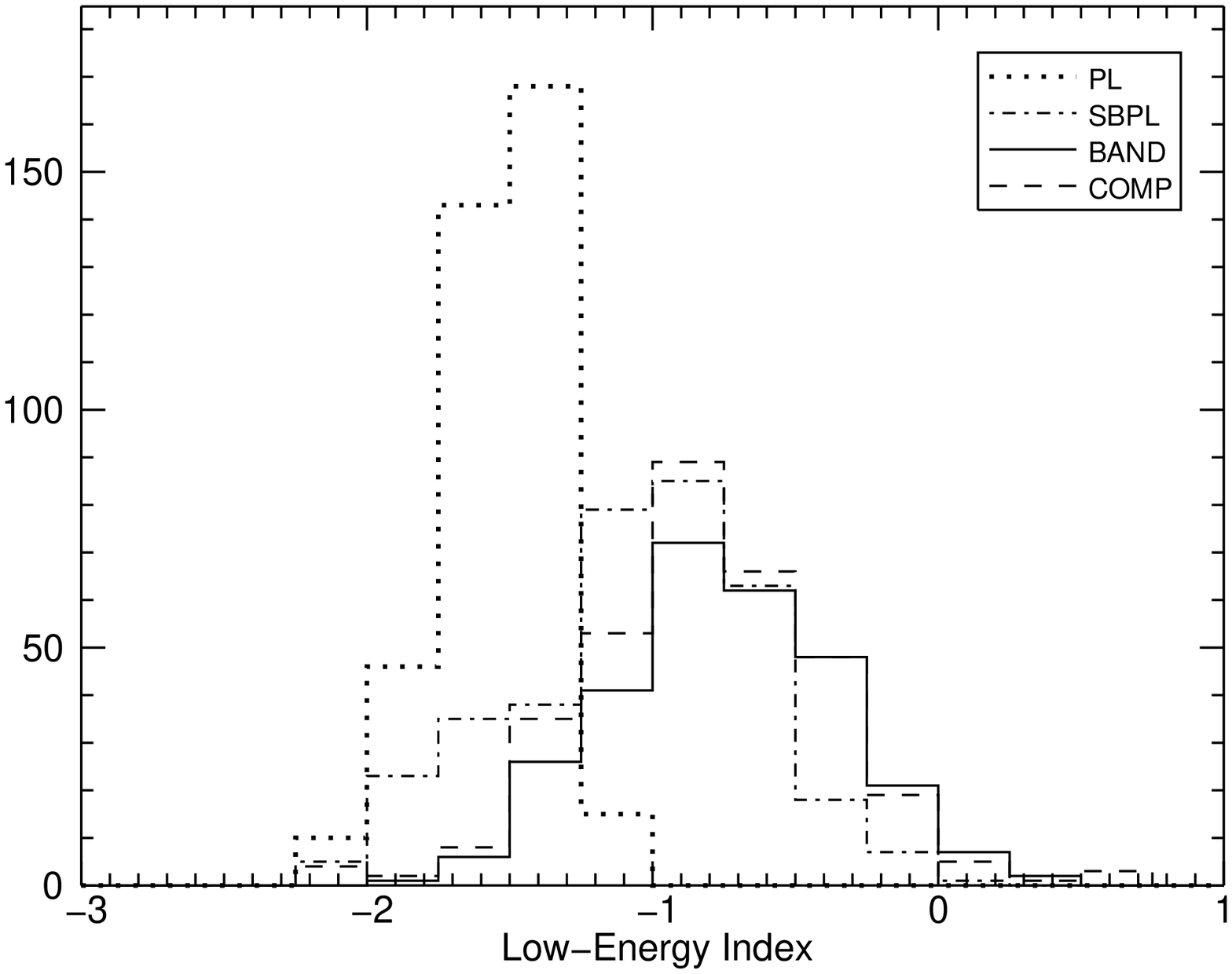}}
		\subfigure[]{\label{alphasbplp}\includegraphics[scale=0.35]{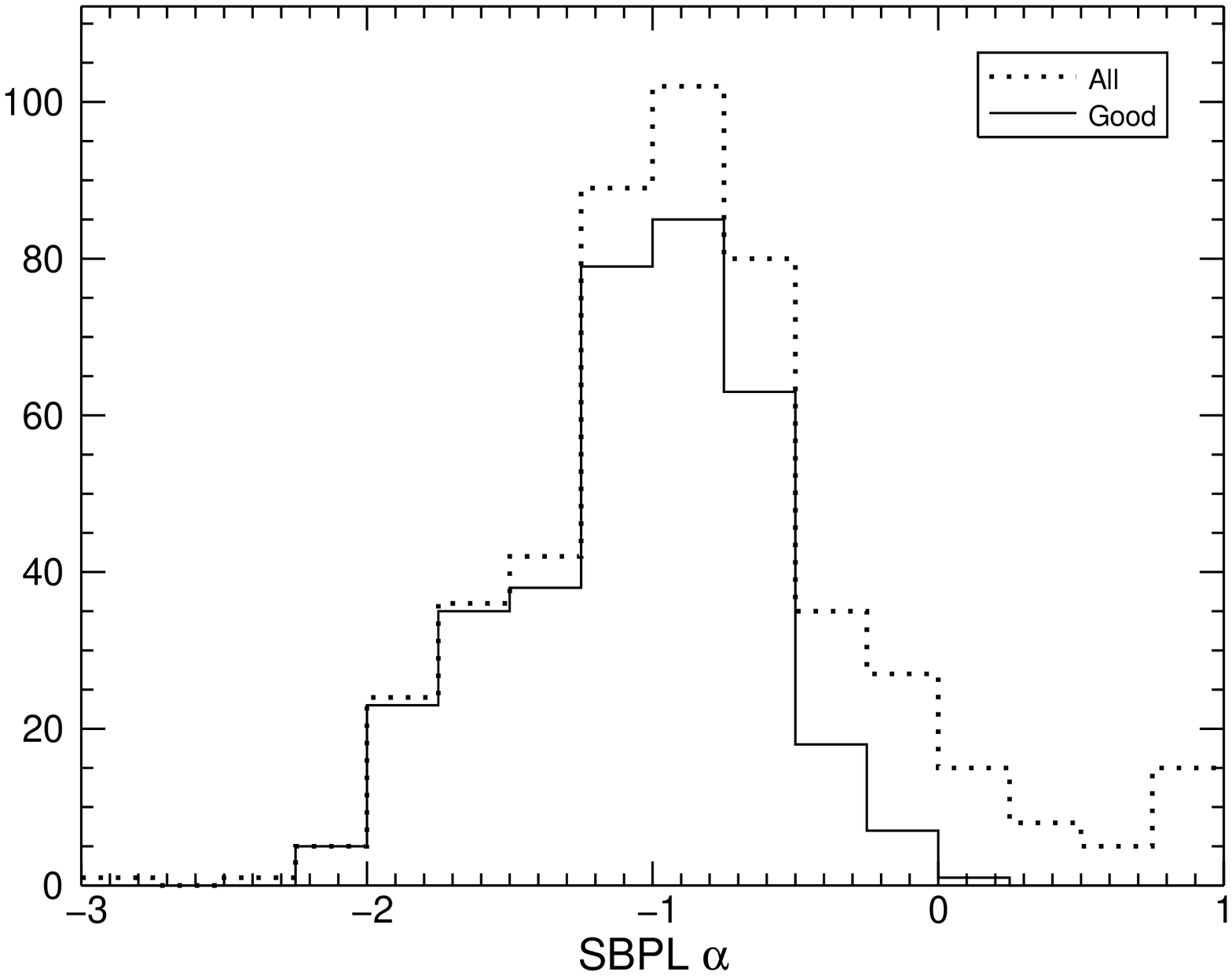}}\\
		\subfigure[]{\label{alphabandp}\includegraphics[scale=0.35]{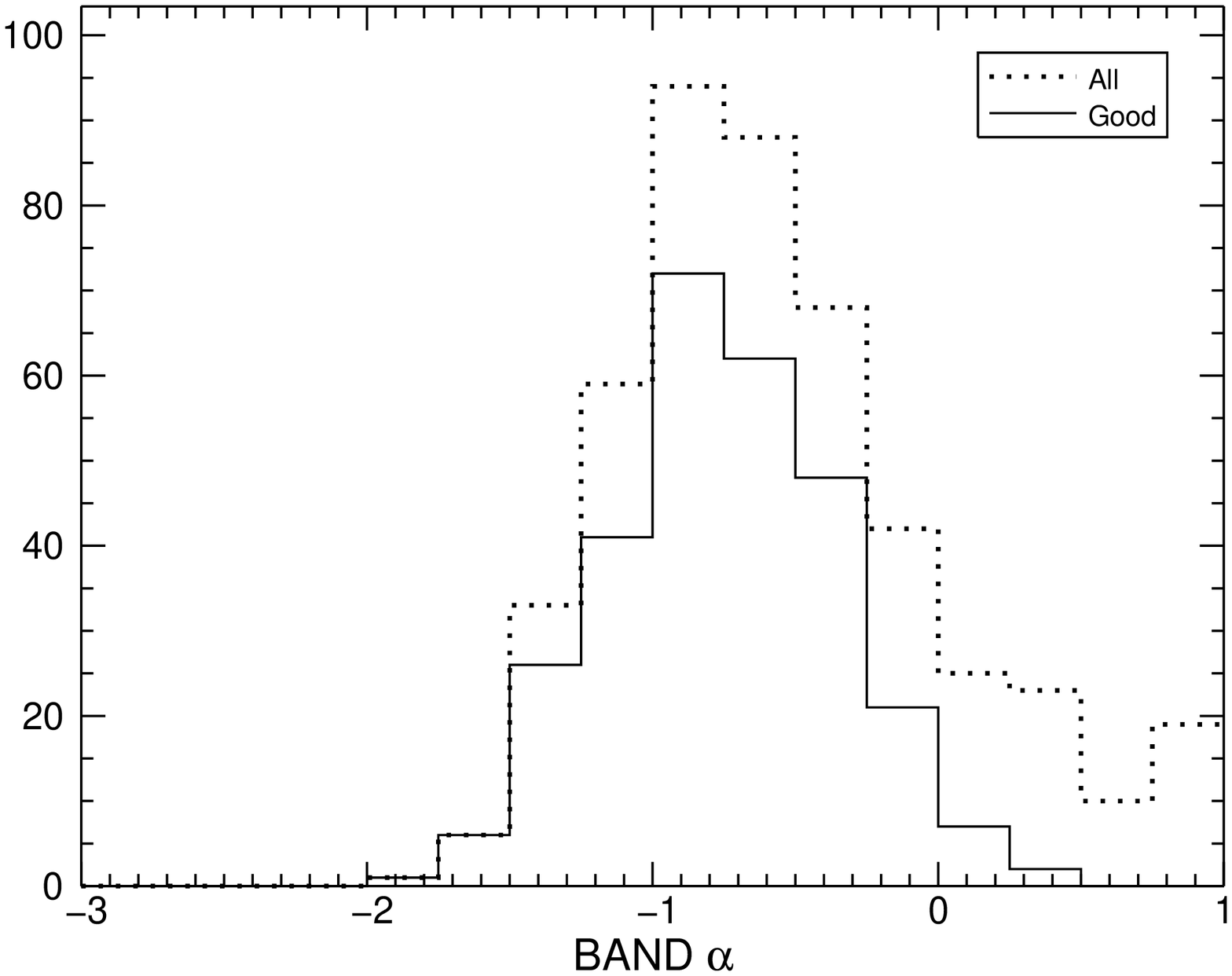}}
		\subfigure[]{\label{alphacompp}\includegraphics[scale=0.35]{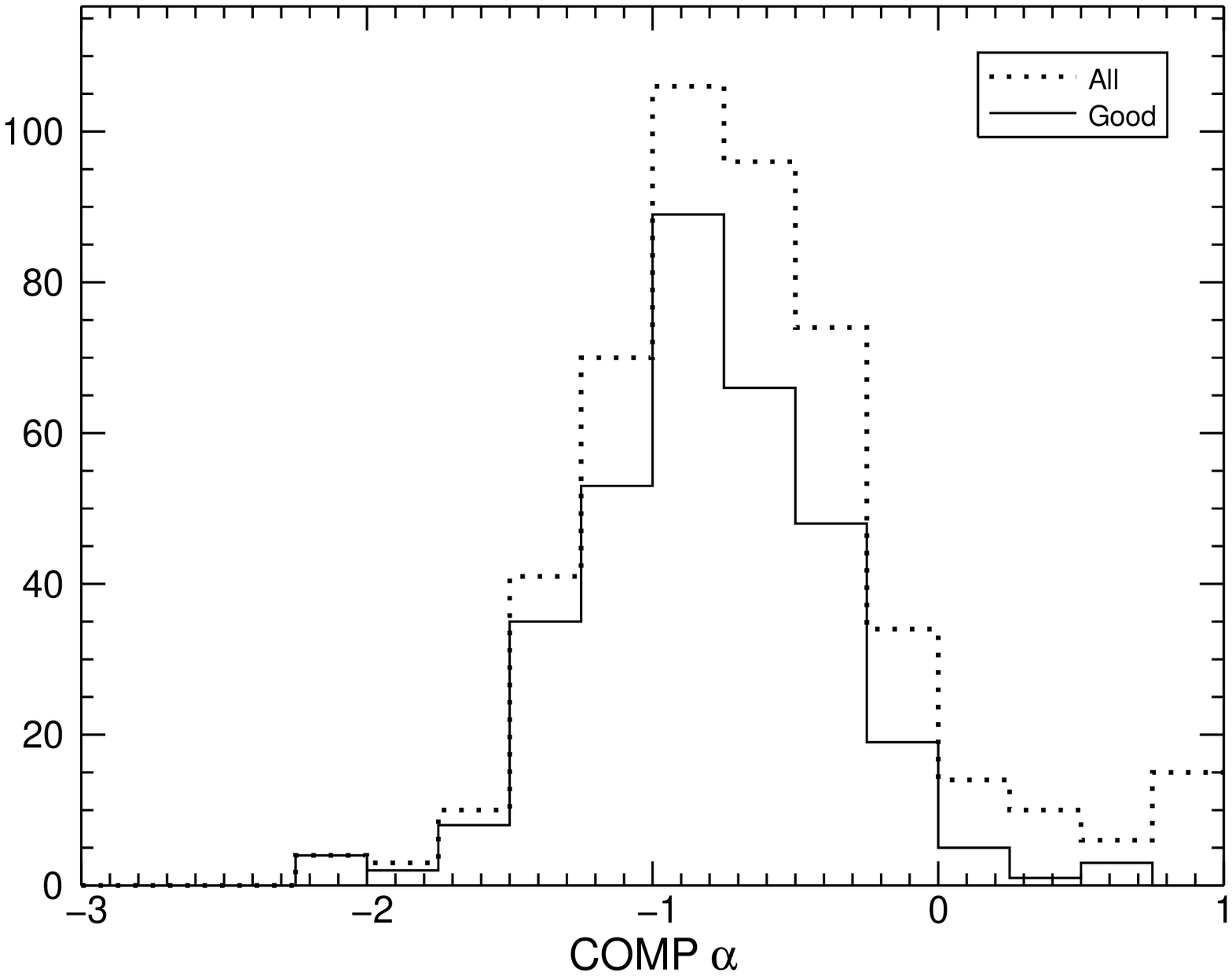}}
	\end{center}
\caption{Distributions of the low-energy spectral indices from peak flux spectral fits.  \ref{indexlop} shows the distributions of 
GOOD parameters and compares to the distribution of PL indices.  \ref{alphasbplp}--\ref{alphacompp} display the comparison 
between the distribution of GOOD parameters and all parameters with no data cuts.  The last bin includes values greater than 
1. \label{loindexp}}
\end{figure}

\begin{figure}
	\begin{center}
		\subfigure[]{\label{indexhip}\includegraphics[scale=0.35]{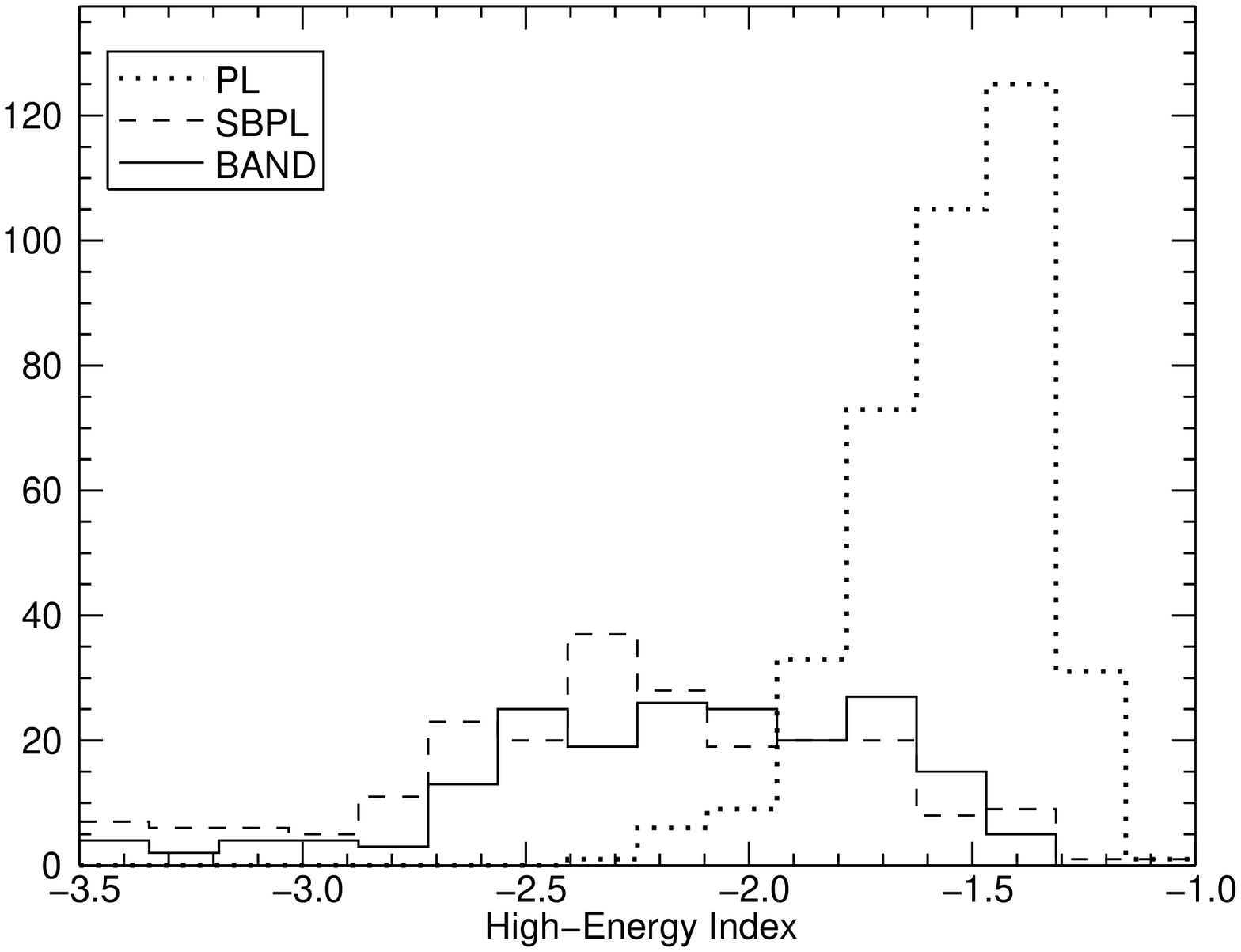}}
		\subfigure[]{\label{betasbplp}\includegraphics[scale=0.35]{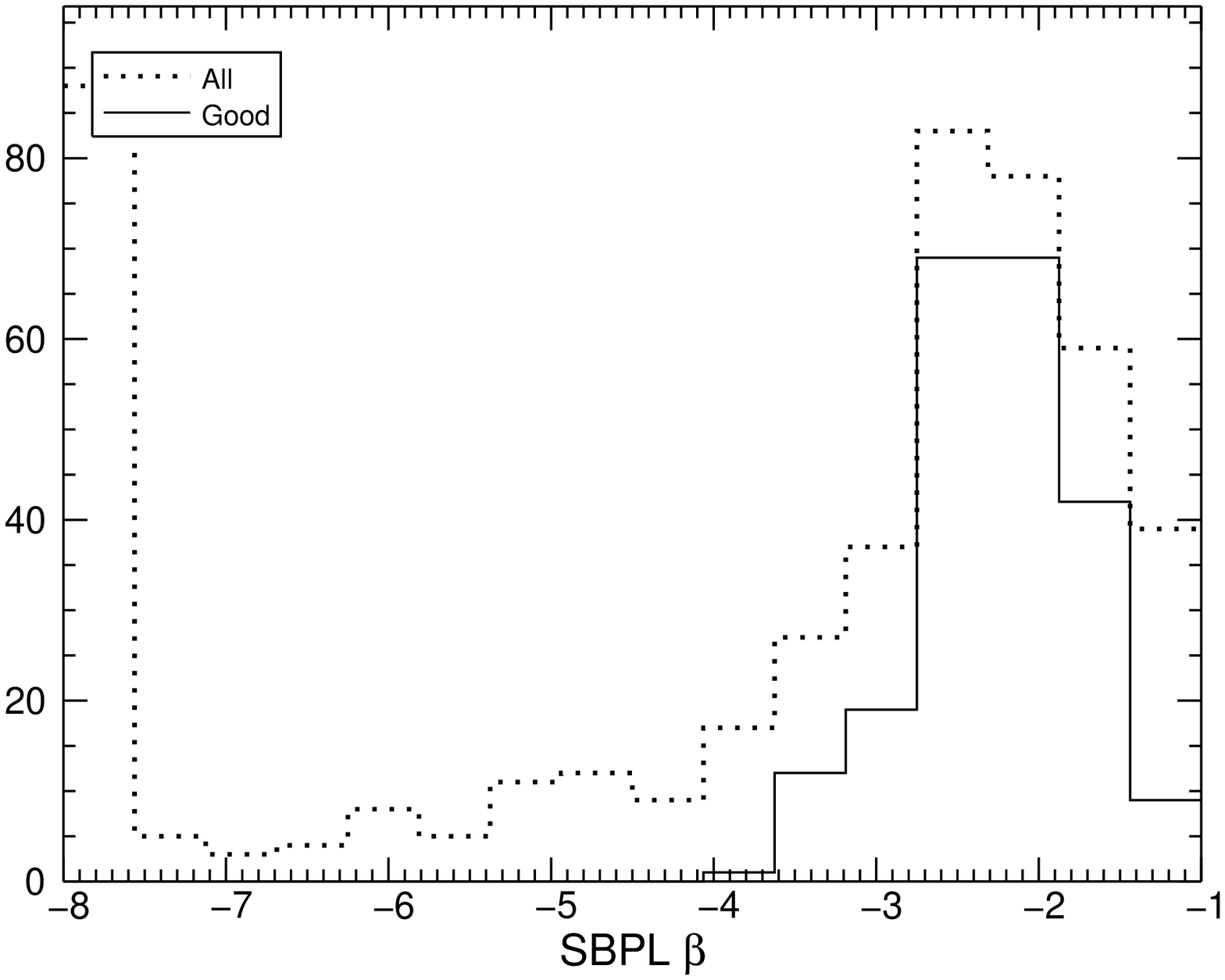}}\\
		\subfigure[]{\label{betabandp}\includegraphics[scale=0.35]{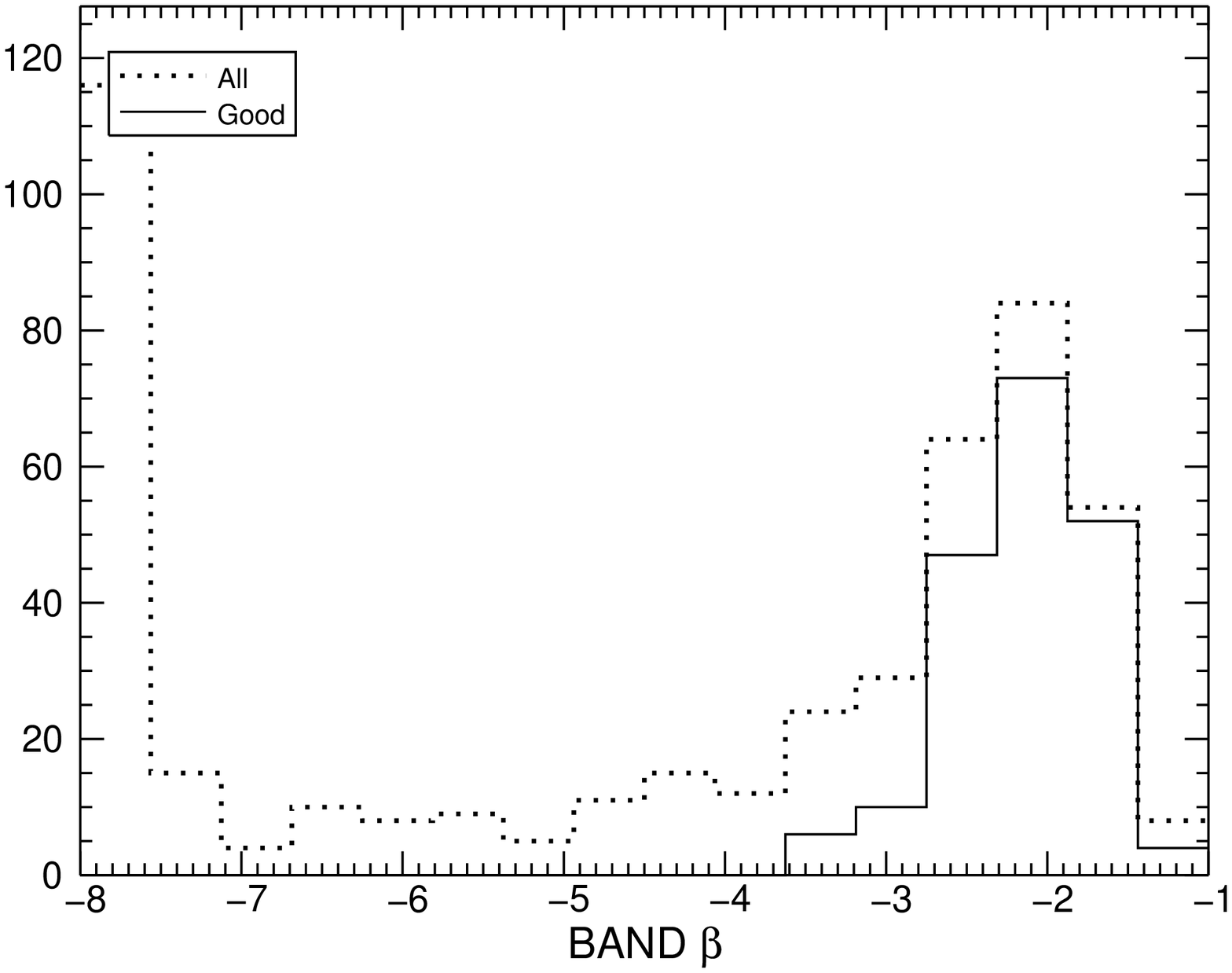}}
		\subfigure[]{\label{deltasp}\includegraphics[scale=0.35]{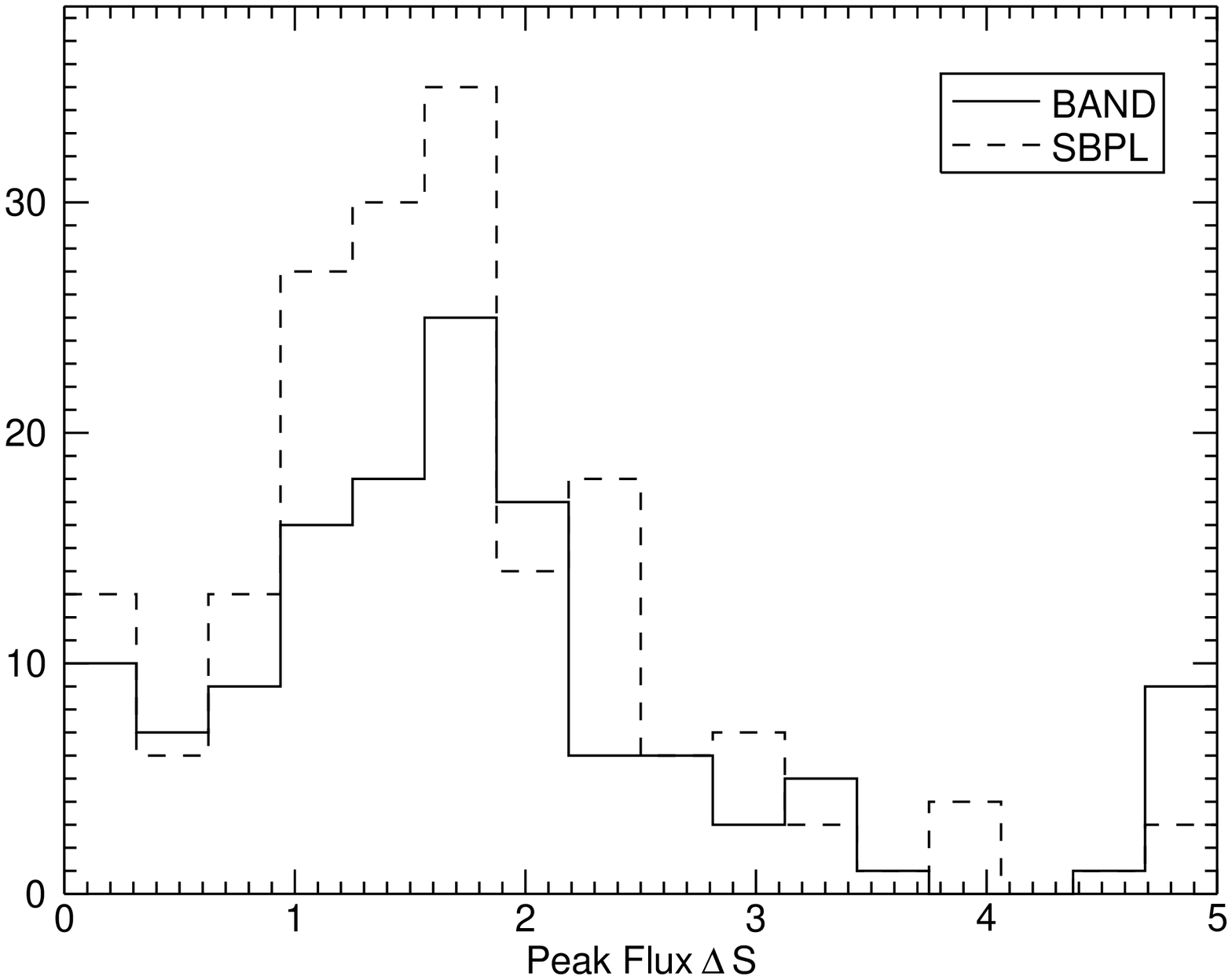}}
	\end{center}
\caption{\ref{indexhip} - \ref{betabandp} are distributions of the high-energy spectral indices from peak flux spectral fits.  \ref
{indexhip} shows the distributions of GOOD parameters and compares to the distribution of PL indices.  \ref{betasbplp} and \ref
{betabandp}  display the comparison between the distribution of GOOD parameters and all parameters with no data cuts.  The 
first bins include values less than -6 and the last bins includes values greater than -1. \ref{deltasp} shows the difference between 
the low- and high-energy indices.  The first bin contains values less than 0, indicating that the centroid value of alpha is 
steeper than the centroid value of beta.\label{hiindexp}}
\end{figure}

The following peak flux spectral distributions have been produced by fitting the GRB spectra over the 1024 ms and 64 ms peak 
flux duration of long and short bursts respectively.  Note that the results from both long and short bursts are included in the 
following figures.  The low-energy indices from the peak flux selections, as shown in Figure \ref{loindexp}, also distribute about 
the typical $-1$ power law.  42\% of the GOOD low-energy indices violate the $-2/3$ synchrotron ``line-of-death'', while an 
additional 40\% of the indices violate the $-3/2$ synchrotron cooling limit, both of which are significantly larger percentages 
than those from the fluence spectra.  The high-energy indices in Figure \ref{hiindexp} peak at a slope sightly steeper than $-2$ 
and  again have a long tail toward steeper indices.  The number of unconstrained high-energy indices increases when 
compared to the fluence spectra, likely due to the poorer statistics resulting from shorter integration times.  As shown with the 
fluence spectra, the PL index serves as an average between low- and high-energy indices for the BAND and SBPL functions. 
Shown in Figure \ref{deltasp} is the $\Delta S$ distribution for the peak flux spectra.  This distribution is roughly consistent with 
the BATSE results found previously \citep{Preece, Kaneko06}, but suffers from a deficit in values due in large part to the inability 
of the data to sufficiently constrain the high-energy power law index. 

\begin{figure}
	\begin{center}
		\subfigure[]{\label{ebreaksbplp}\includegraphics[scale=0.35]{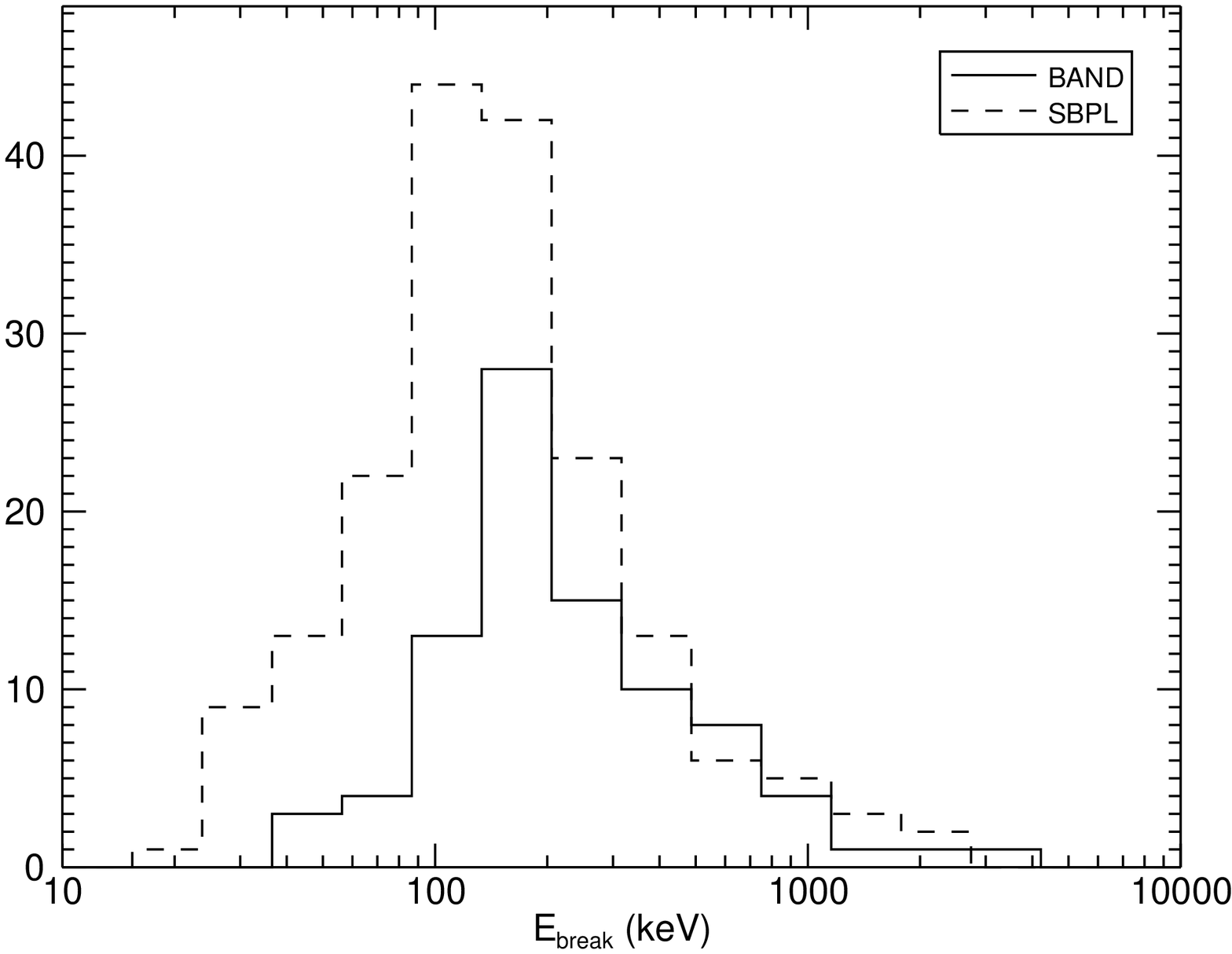}}
		\subfigure[]{\label{epeakp}\includegraphics[scale=0.35]{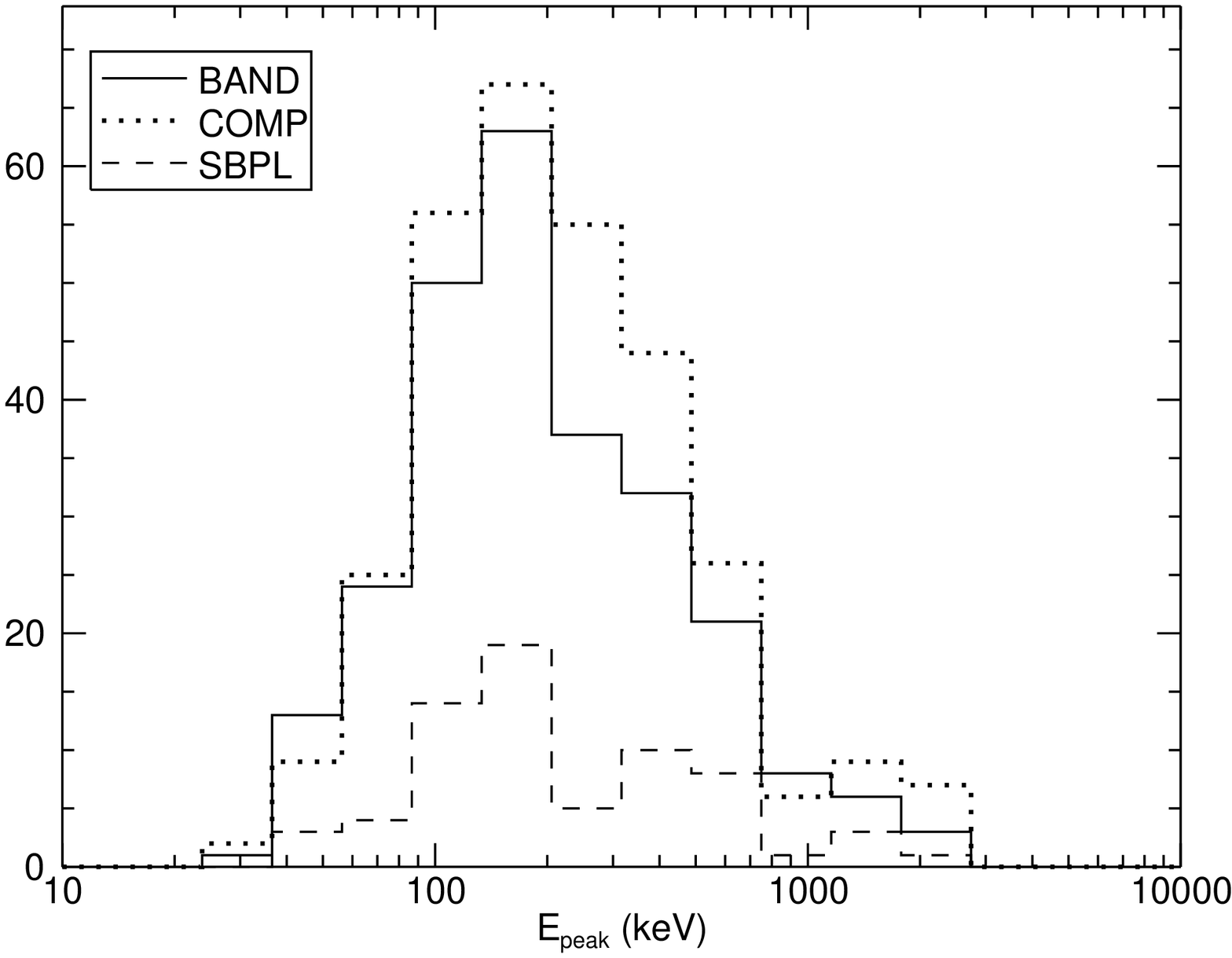}}\\
		\subfigure[]{\label{epeakbandp}\includegraphics[scale=0.35]{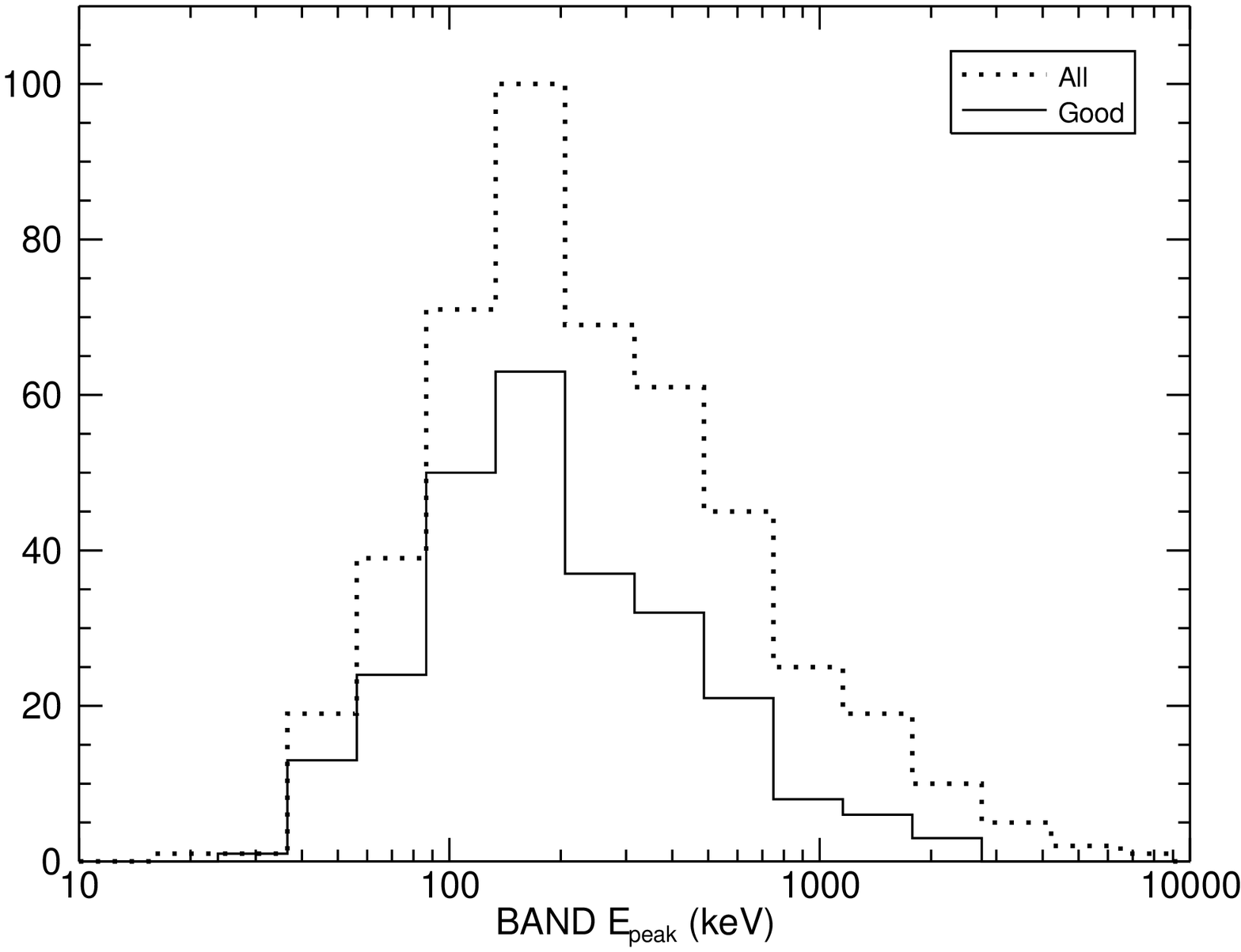}}
		\subfigure[]{\label{epeakcompp}\includegraphics[scale=0.35]{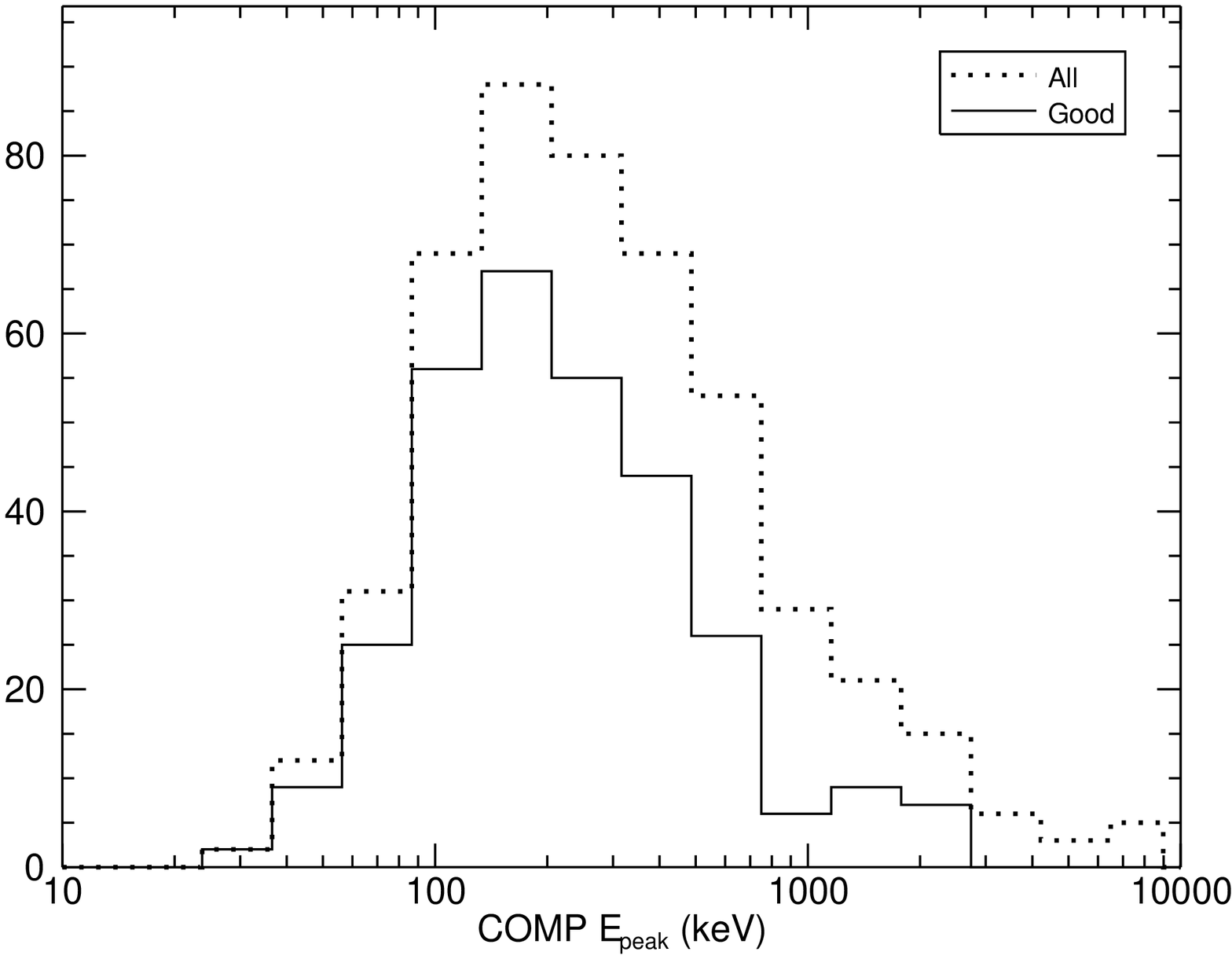}}
	\end{center}
\caption{Distributions of $E_{break}$ and $E_{peak}$ from peak flux spectral fits.  \ref{ebreaksbplp} displays the comparison 
between the distribution of GOOD $E_{break}$ and $E_{break}$ with no data cuts.  \ref{epeakp} shows the distributions of 
GOOD $E_{peak}$ for BAND, SBPL,  and COMP.  \ref{epeakbandp} and \ref{epeakcompp} display the comparison between the 
distribution of GOOD parameters and all parameters with no data cuts. \label{epeakebreakp}}
\end{figure}

\begin{figure}
	\begin{center}
		\includegraphics[scale=0.7]{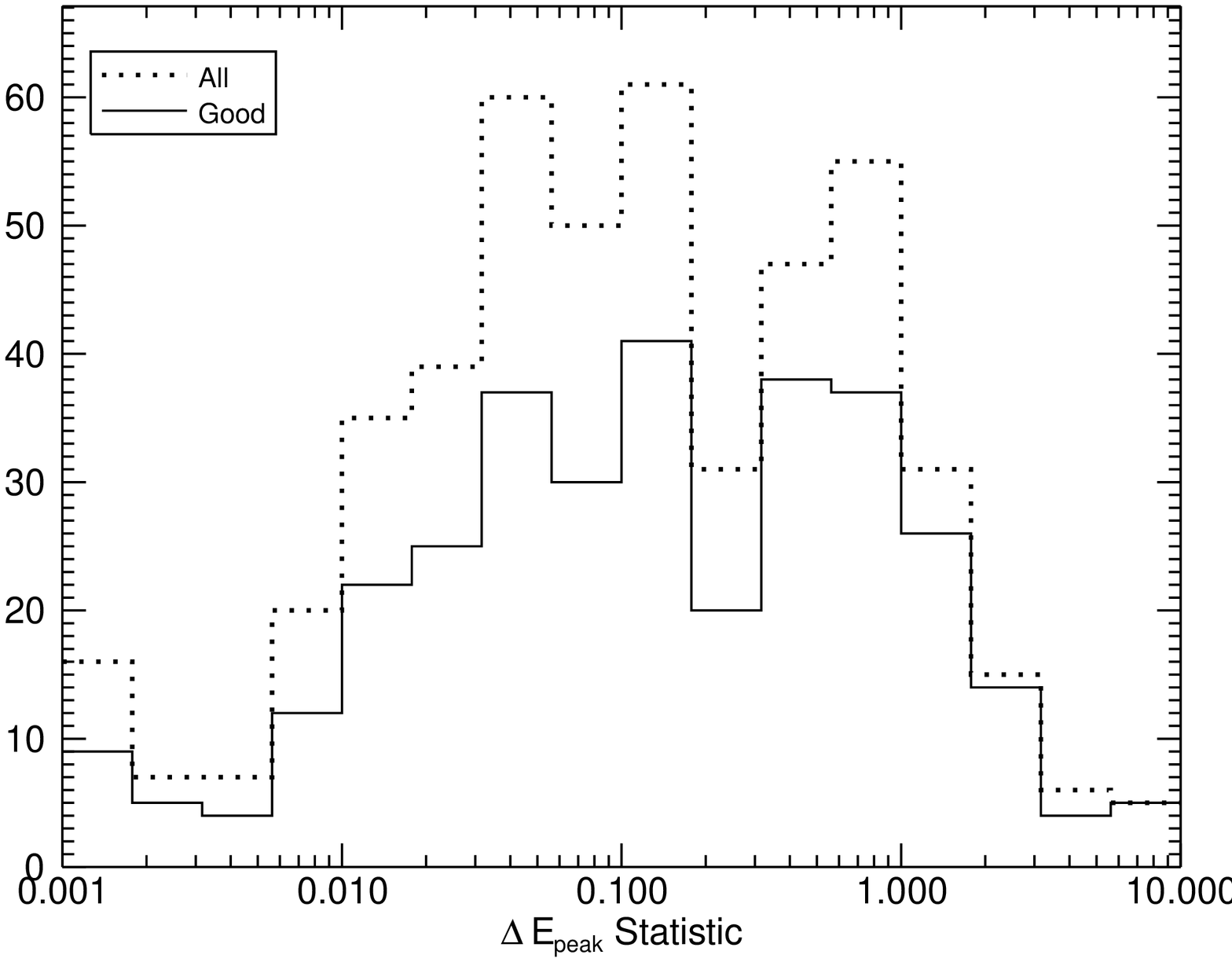}
	\end{center}
\caption{Distribution of the $\Delta E_{peak}$ statistic for the COMP and BAND models from peak flux spectral fits.  A value less 
than 1 indicates the $E_{peak}$ values are within errors, while a value larger than 1 indicates the $E_{peak}$ values are not 
within errors. \label{deltaepeakp}}
\end{figure}

In Figure \ref{epeakebreakp}, we show the distributions for $E_{break}$ and  $E_{peak}$.  As was evident from the fluence 
spectra, the $E_{break}$ from the SBPL fits appears to peak at 100 keV, meanwhile the $E_{break}$ from BAND is harder and 
peaks at about 200 keV.  The $E_{peak}$ distributions for all models peak around 150 keV and cover just over two orders of 
magnitude, which is consistent with previous findings \citep{Preece98, Kaneko06}.  It should be noted that the data over the 
short timescales in the peak flux spectra do not often favor the SBPL model, resulting in large parameter errors.  Using the 
propagation of errors scheme to calculate the $E_{peak}$ for SBPL therefore causes many of the values to be largely 
unconstrained.  This results in the much smaller $E_{peak}$ distribution for SBPL seen in Figure \ref{epeakp}.  In addition, we 
calculate and show the $\Delta E_{peak}$ statistic in Figure \ref{deltaepeakp} and, as was the case with the fluence spectra, 
$\sim$25\% of the BAND and COMP $E_{peak}$ values are found to be outside the combined errors.

\begin{figure}
	\begin{center}
		\subfigure[]{\label{pflux1mevp}\includegraphics[scale=0.35]{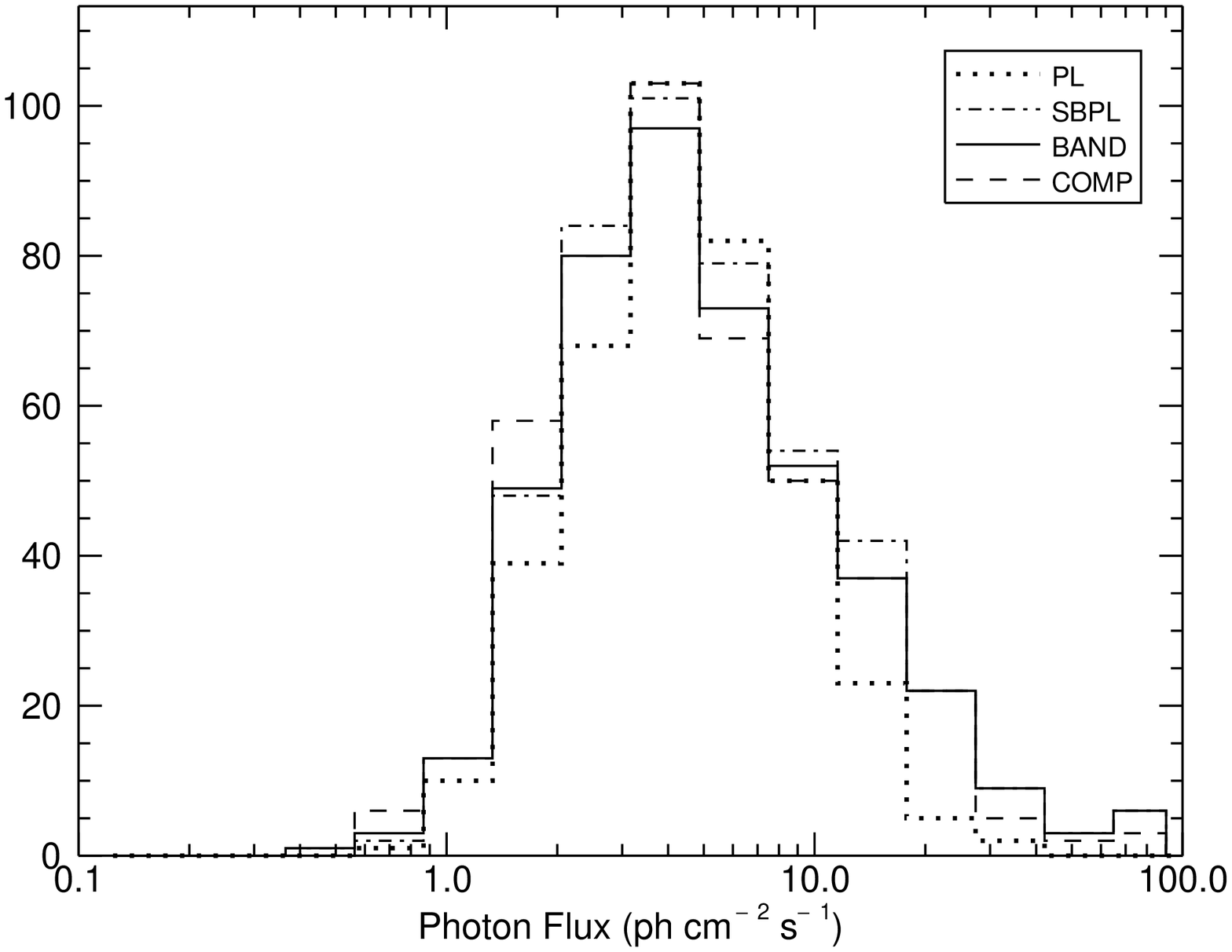}}
		\subfigure[]{\label{pflux40mevp}\includegraphics[scale=0.35]{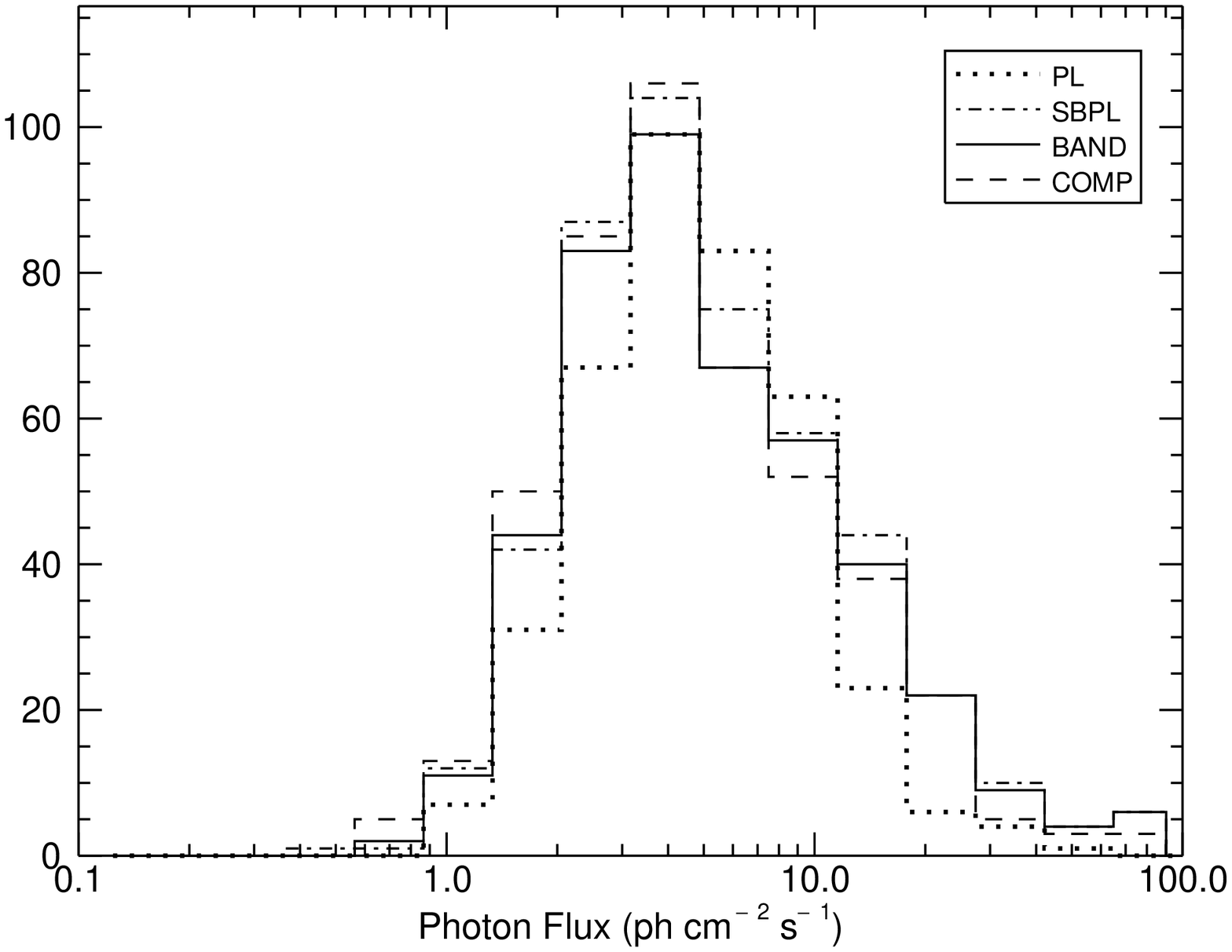}}\\
		\subfigure[]{\label{eflux1mevp}\includegraphics[scale=0.35]{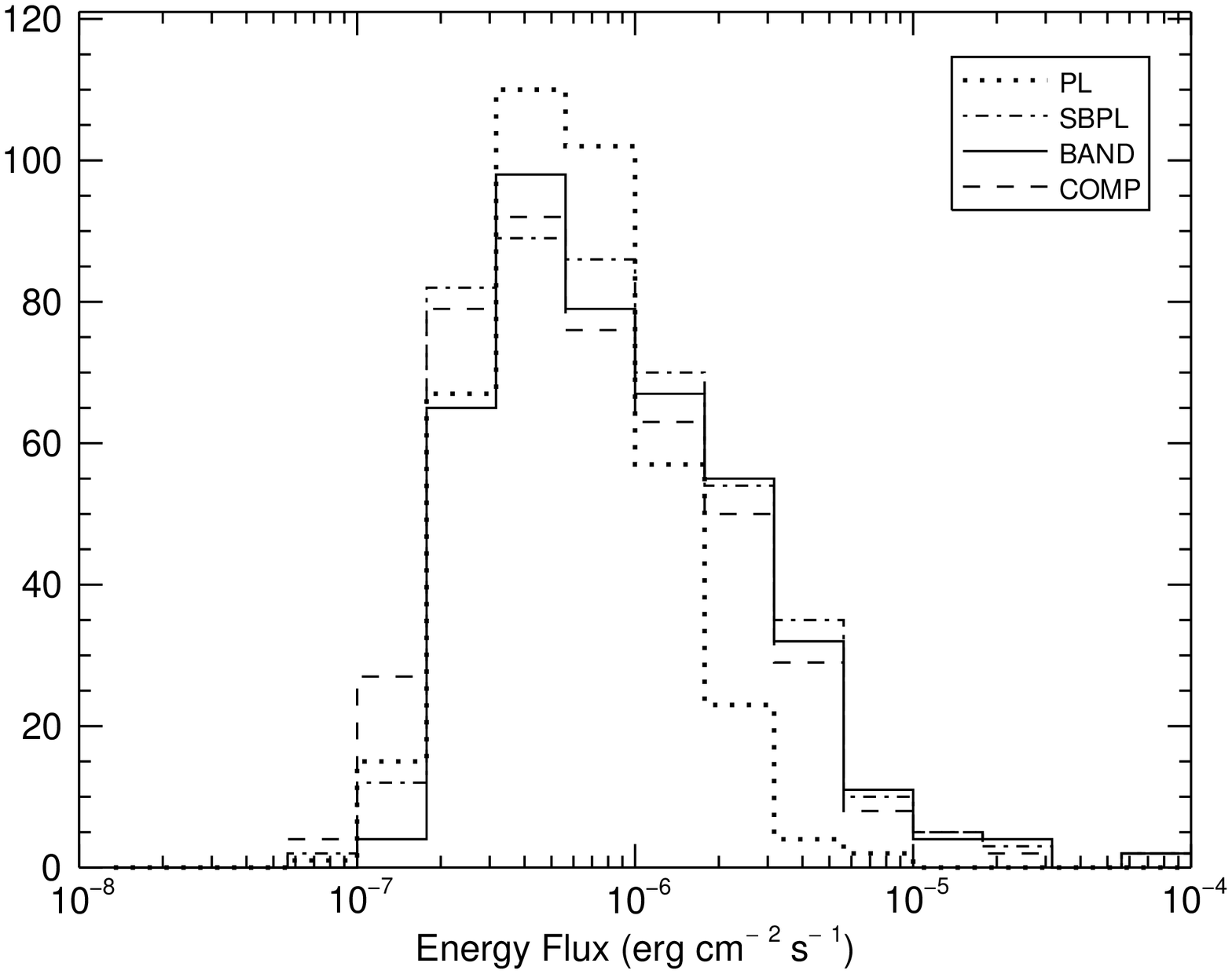}}
		\subfigure[]{\label{eflux40mevp}\includegraphics[scale=0.35]{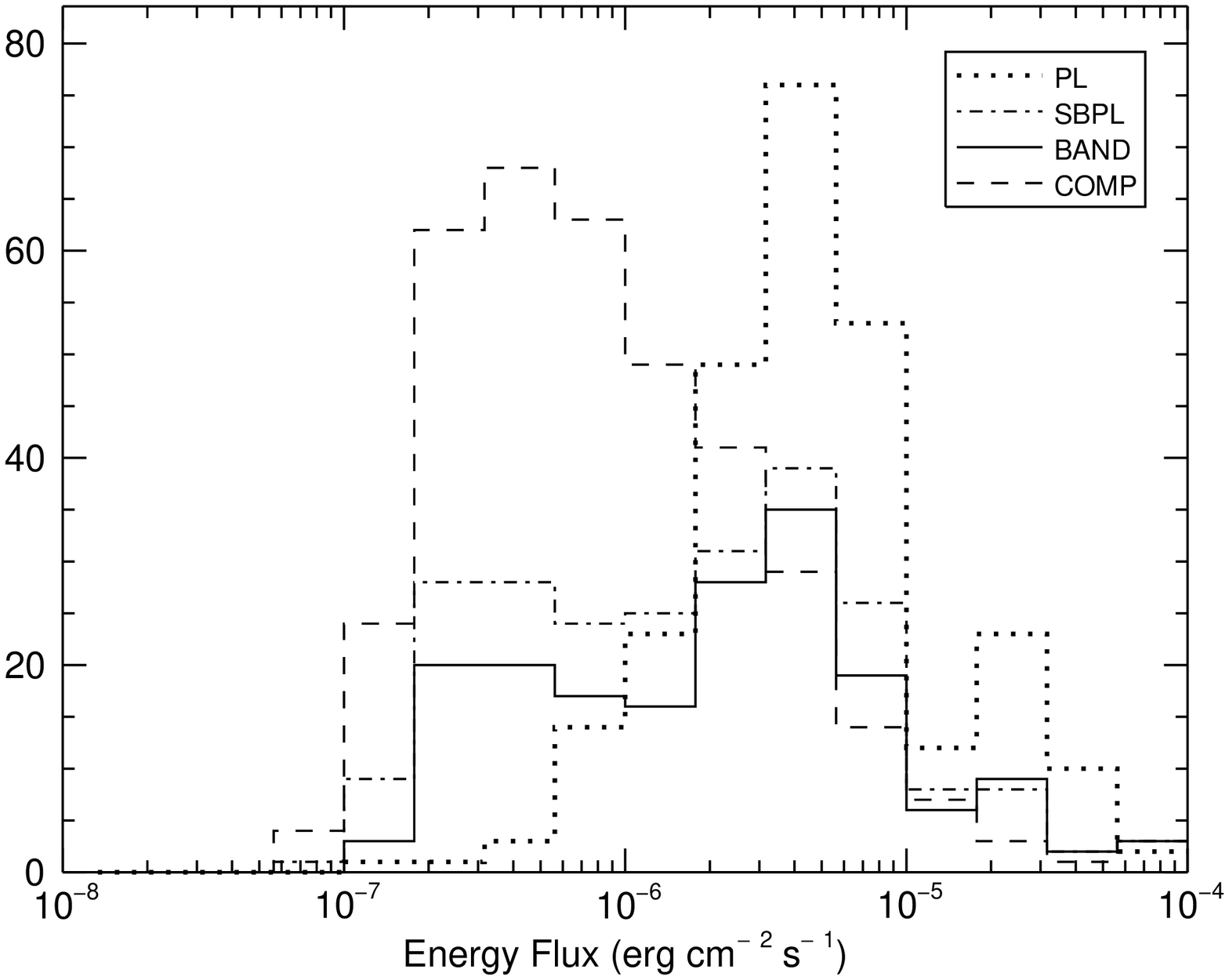}}
	\end{center}
\caption{Distributions of photon and energy flux from peak flux spectral fits.  \ref{pflux1mevp} and \ref{eflux1mevp} display the 
flux distributions for the 8 keV--1 MeV band.  \ref{pflux40mevp} and \ref{eflux40mevp} display the flux distributions for the 8 
keV--40 MeV band.  Note that the plotted distributions contain the flux on two different timescales: 1024 ms and 64 ms. \label
{fluxp}}
\end{figure}

The distributions for the peak photon flux and energy flux are shown in Figure \ref{fluxp}.  The photon flux peaks around 4 
photons cm$^{-2}$  s$^{-1}$, and the energy flux peaks at $7.4 \times 10^{-7}$ ergs cm$^{-2}$ s$^{-1}$ in the 8-1000 keV 
band.  When integrating over the full GBM spectral band, 8 kev-40 MeV, the dispersion in the energy flux increases for all 
but the simple power law model and approximates a top hat function with a small high-flux tail spanning about 2 orders of 
magnitude.  The GRB with the brightest peak photon flux is GRB 081009 at $\rm >130\ ph\ s^{-1}\ cm^{-2}$.  This GRB is also the 
brightest GRB in terms of time-averaged photon flux which makes it an excellent burst for calibration studies. Similar to the 
time-averaged energy flux, the burst with highest peak energy flux is GRB 090227B at $\rm >8.0 \times 10^{-5}\ ergs\ s^{-1}\ cm^
{-2}$.

\begin{figure}
	\begin{center}
		\subfigure[]{\label{pffalpha}\includegraphics[scale=0.35]{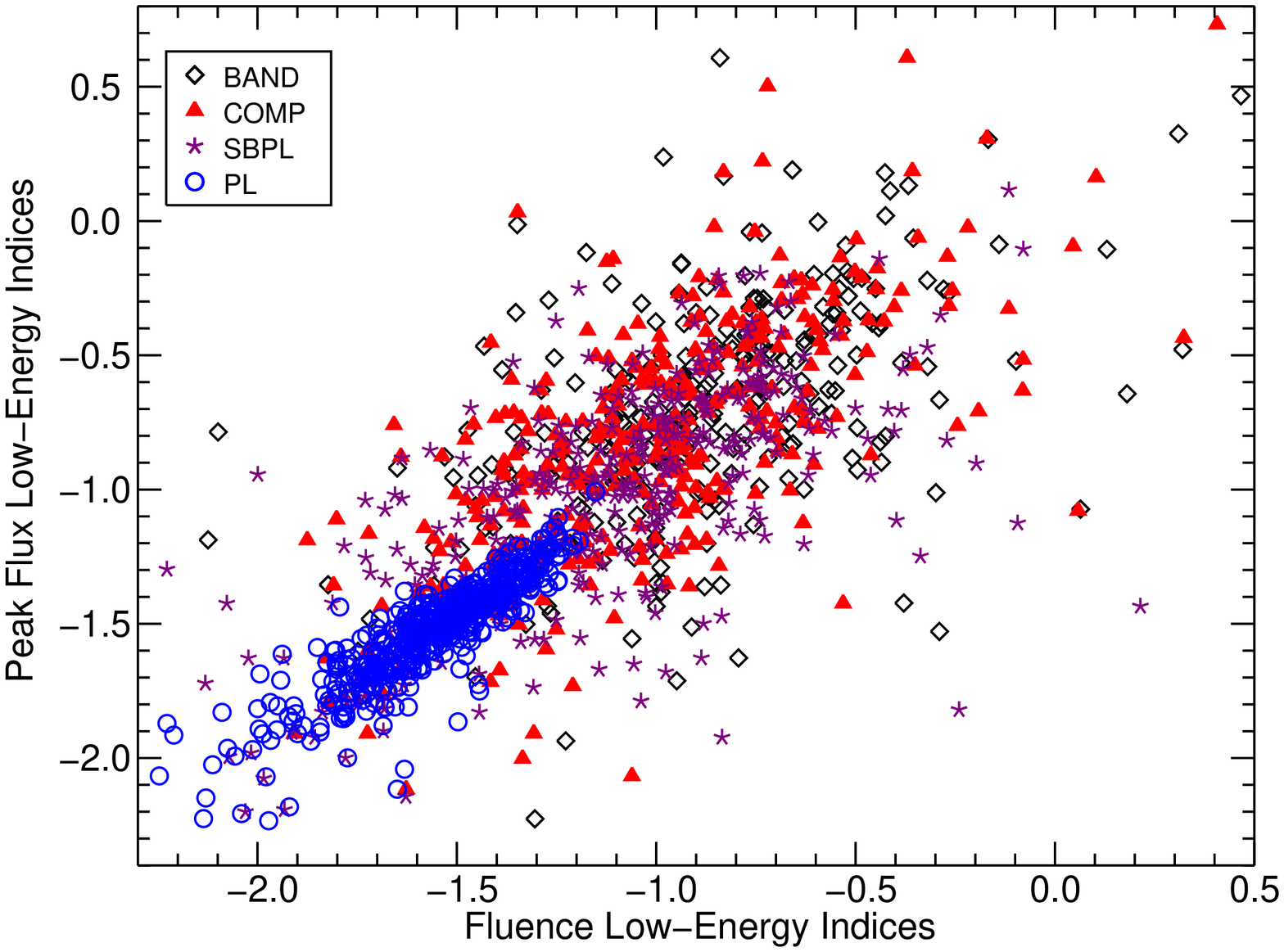}}
		\subfigure[]{\label{pffbeta}\includegraphics[scale=0.35]{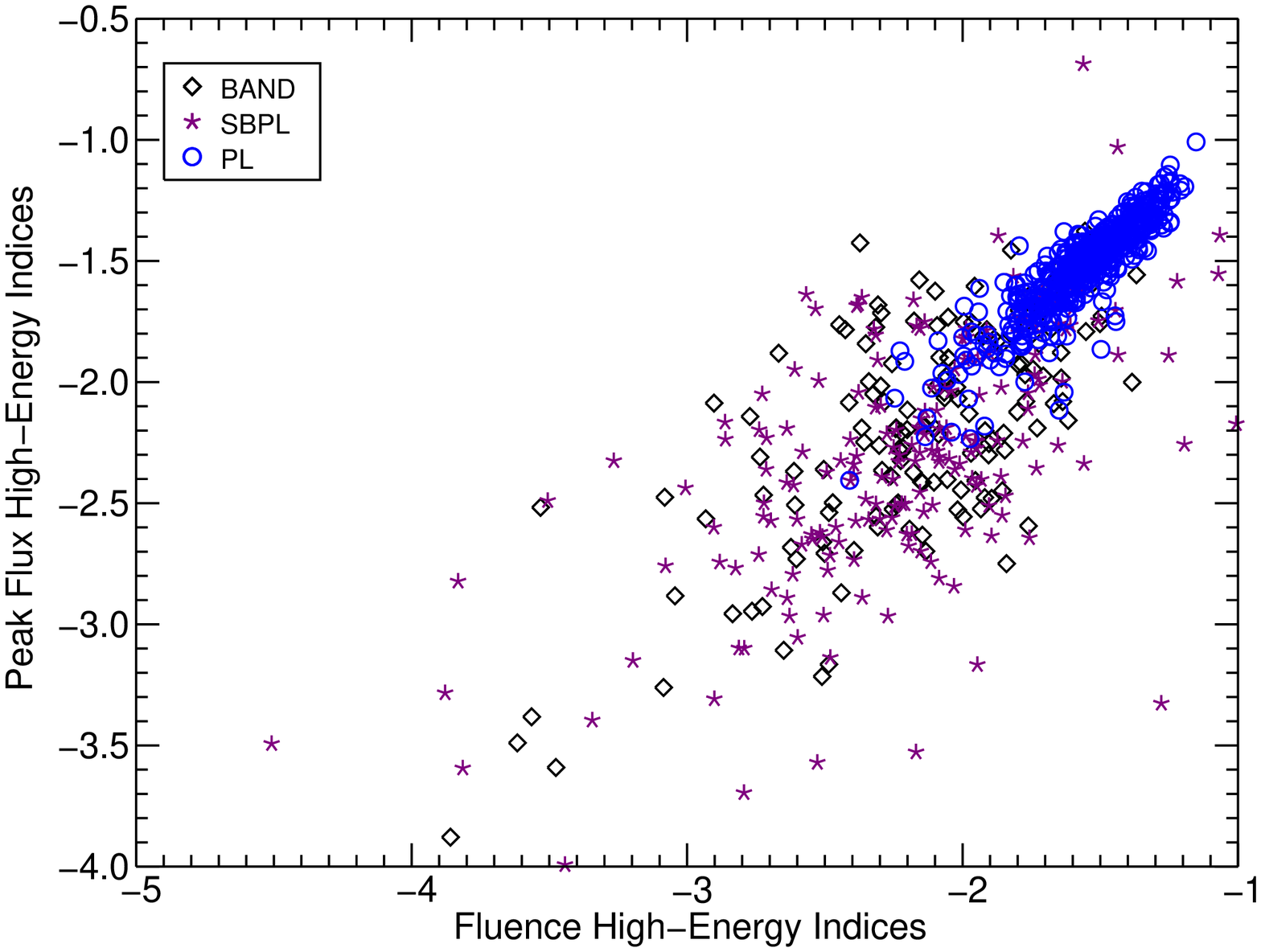}}\\
		\subfigure[]{\label{pffepeak}\includegraphics[scale=0.35]{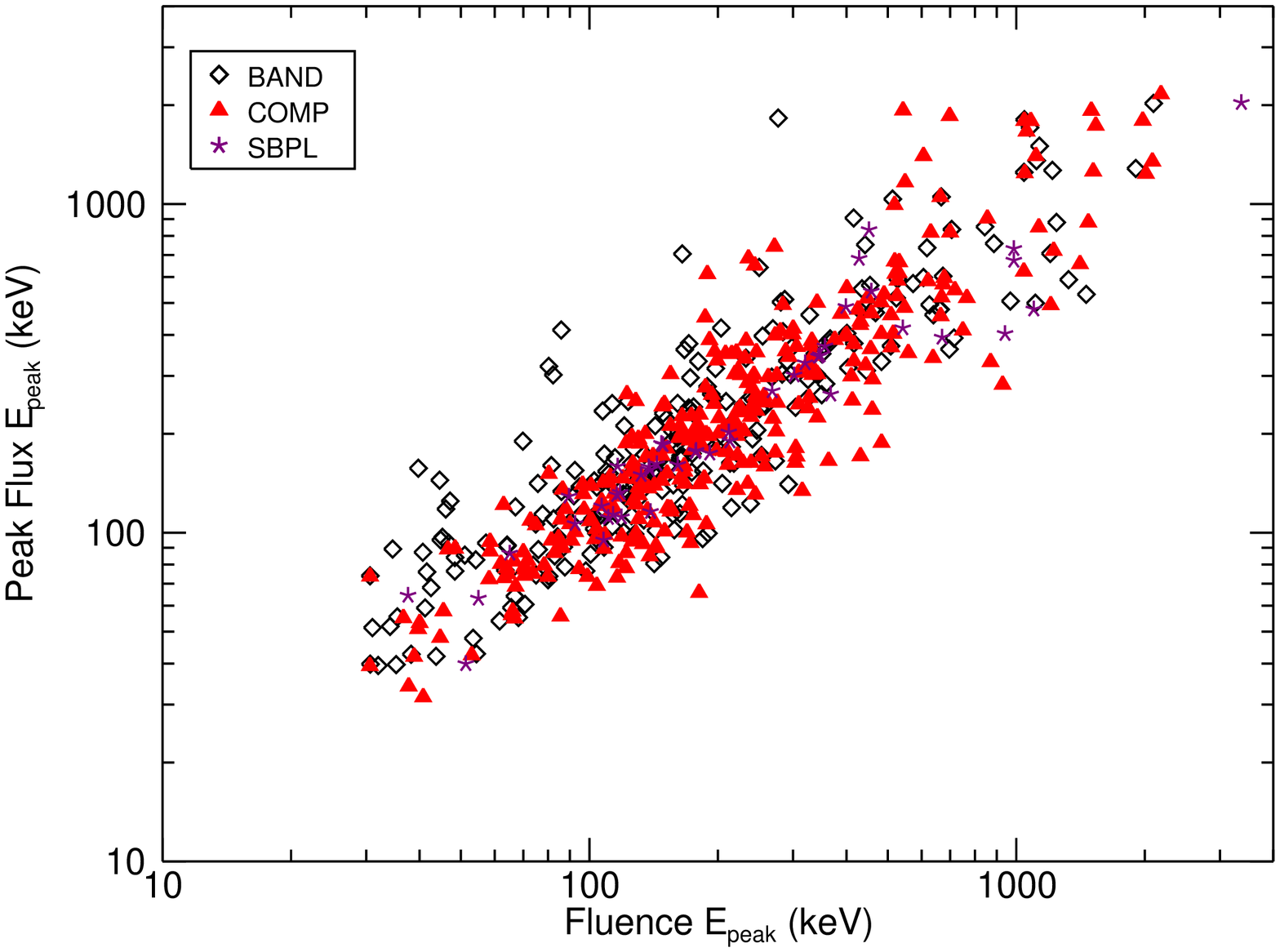}}
	\end{center}
\caption{Peak flux spectral parameters as a function of the fluence spectral parameters. For all three parameters there is 
evidence for a strong correlation between the parameters found for the fluence spectra and those for the peak flux spectra.  Note that the PL index is shown in both \ref{pffalpha} and \ref{pffbeta} for comparison. \label
{pff}}
\end{figure}

When studying the two types of spectra in this catalog, it is instructive to study the similarities and differences between the 
resulting parameters.  Plotted in Figure \ref{pff} are the low-energy indices, high-energy indices, and $E_{peak}$ energies of the 
peak flux spectra as a function of the corresponding parameters from the fluence spectra.  Most of the peak flux spectral 
parameters correlate with the fluence spectral parameters on the order of unity.  There are particular regions in each plot where 
outliers exist, and these areas are indications that either the GRB spectrum is poorly sampled or there exists significant spectral 
evolution in the fluence measurement of the spectrum that skews the fluence spectral values.  Examples of the former case are 
when the low-energy index is atypically shallow ($\gtrsim-0.5$) or the high-energy index is steeper than average ($\lesssim-3$).  
An example where spectral evolution may skew the correlation between the two types of spectra is apparent in the comparison 
of $E_{peak}$.  Here, it is likely that a fluence spectrum covering significant spectral evolution will produce a lower energy $E_
{peak}$ than is measured when inspecting the peak flux spectrum.   

%% BEST Sample
\subsection{The BEST Sample}
The BEST parameter sample produces the best estimate of the observed properties of GRBs.  By using model comparison, the 
preferred model is selected, and the parameters are reviewed for that model.  The models contained herein and in most 
GRB spectral analyses are empirical models, based only on the data received; therefore the data from different GRBs tend to 
support different models.  Perhaps it will be possible to determine the physics of the emission process by investigating the 
tendencies of the data to support a particular model over others.  It is this motivation, as well as the motivation to provide a 
sample that contains the best picture of the global properties of the data, that prompts the investigation of the BEST sample.

\begin{figure}
	\begin{center}
		\subfigure[]{\label{indexbestf}\includegraphics[scale=0.35]{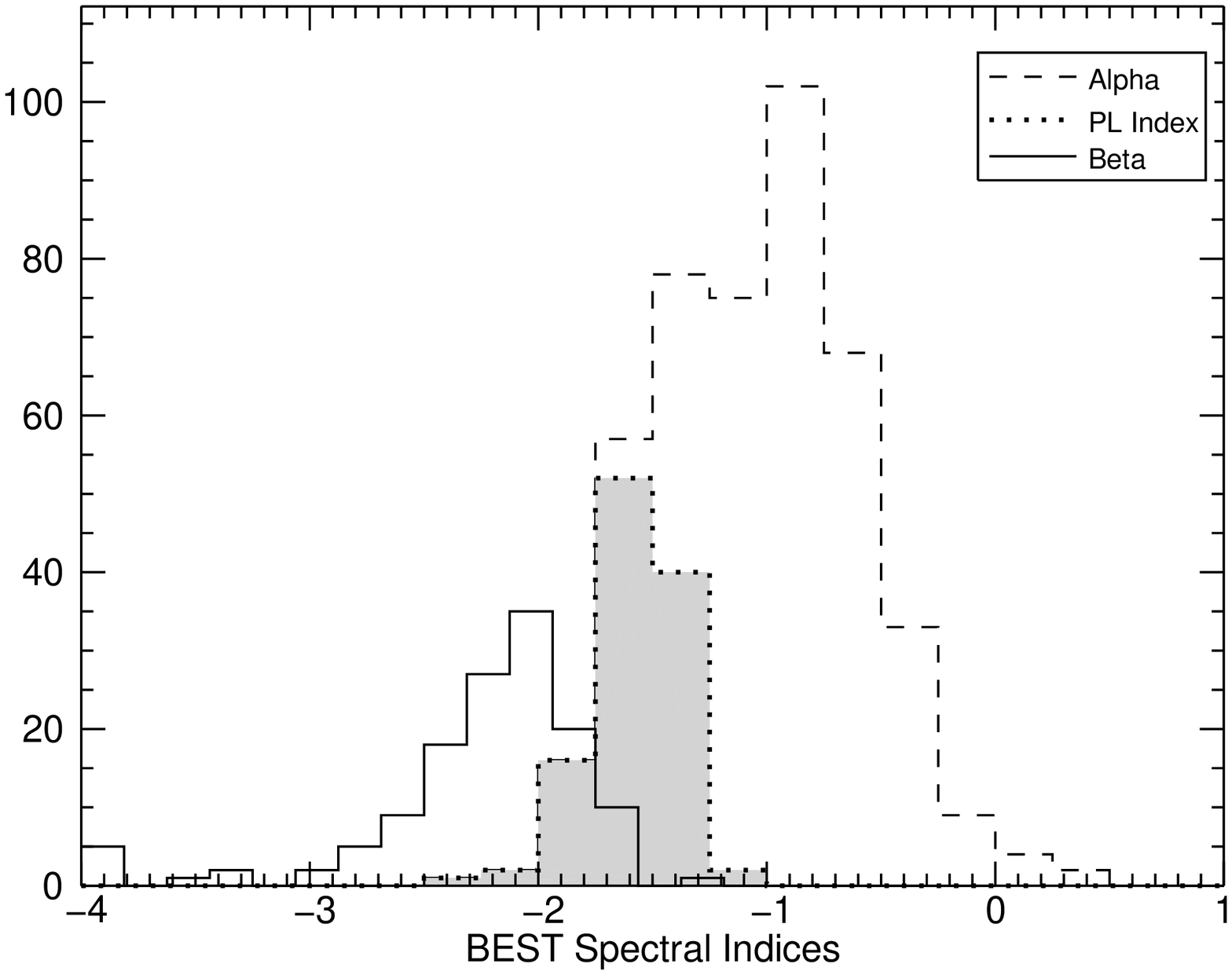}}
		\subfigure[]{\label{epeakbestf}\includegraphics[scale=0.35]{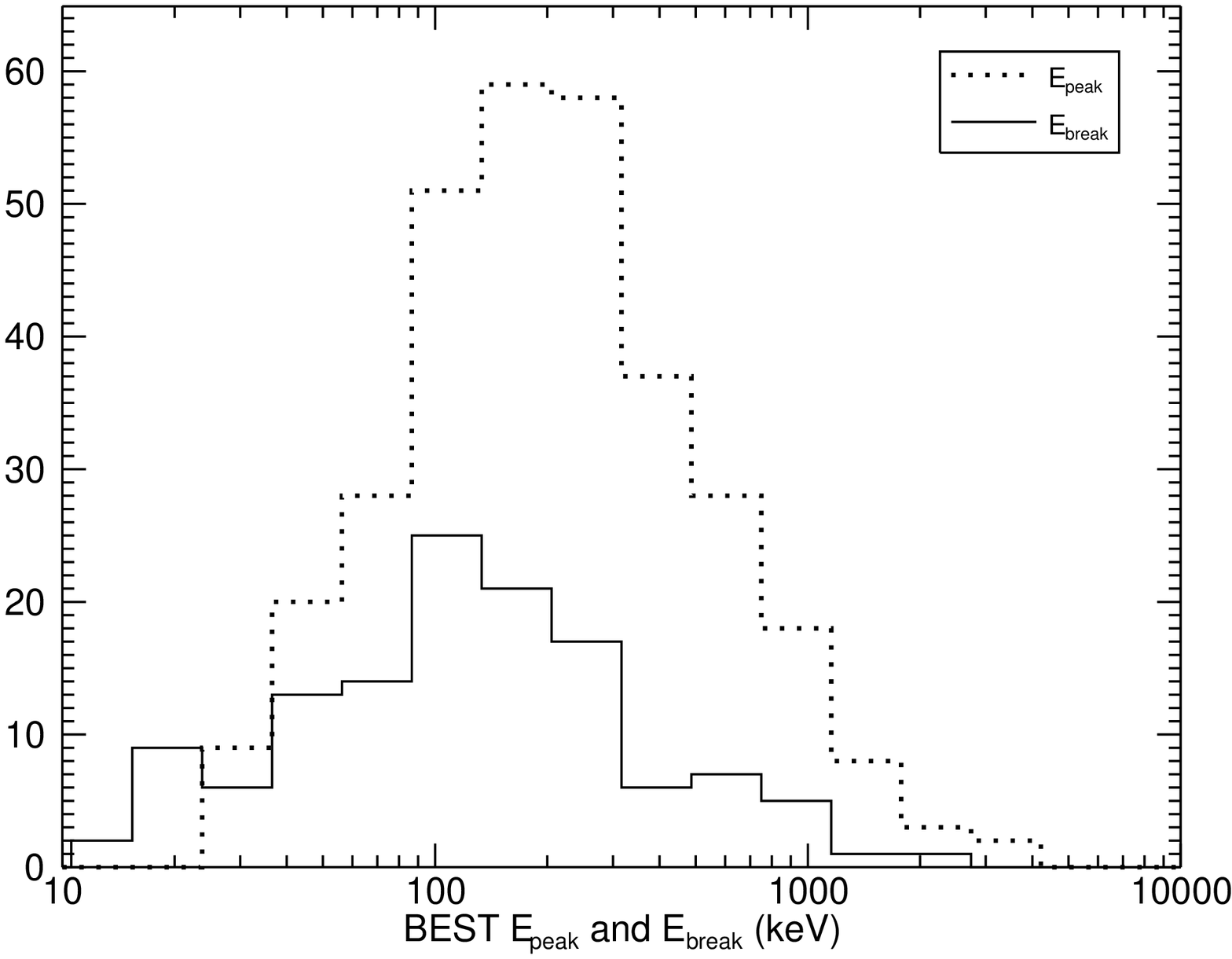}}\\
		\subfigure[]{\label{pfluxbestf}\includegraphics[scale=0.35]{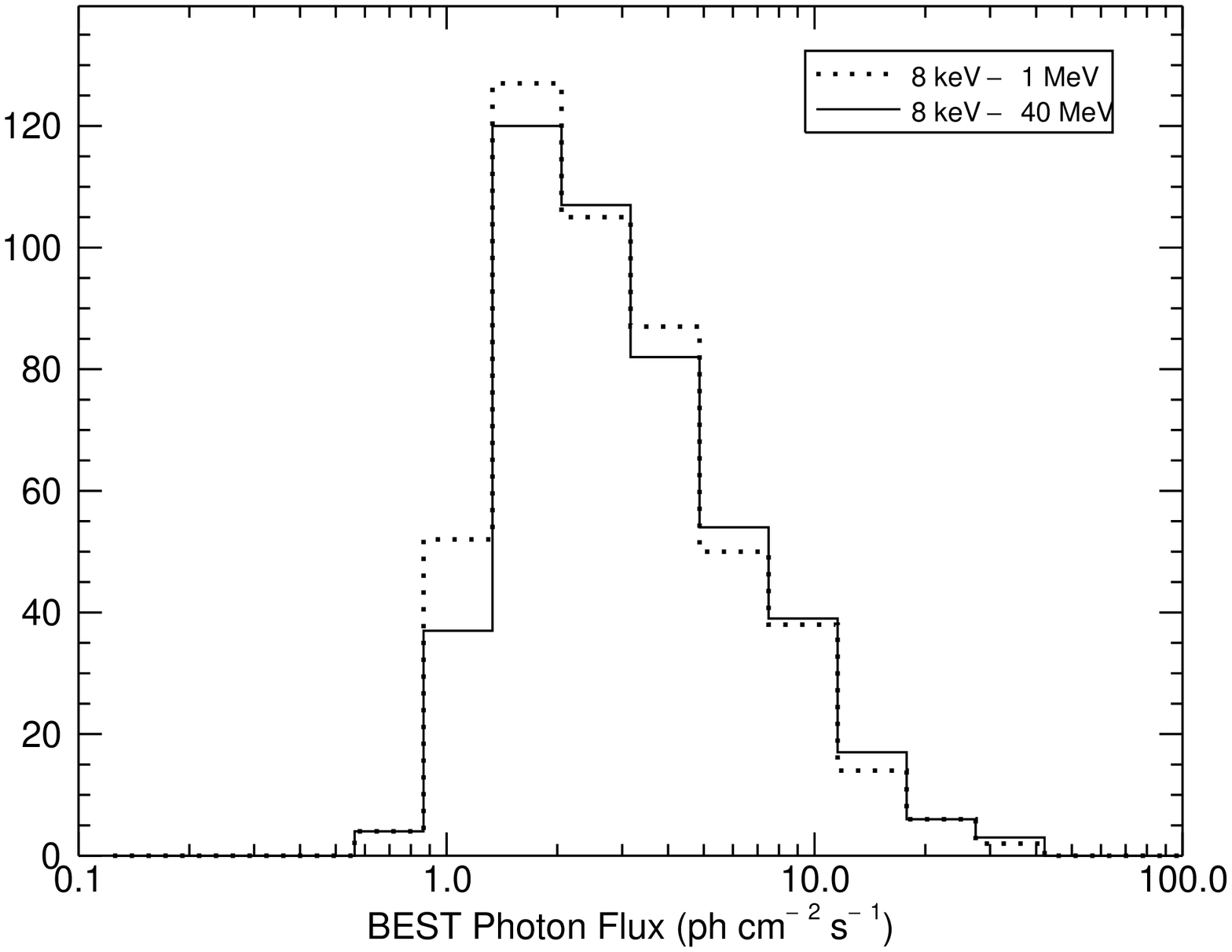}}
		\subfigure[]{\label{efluxbestf}\includegraphics[scale=0.35]{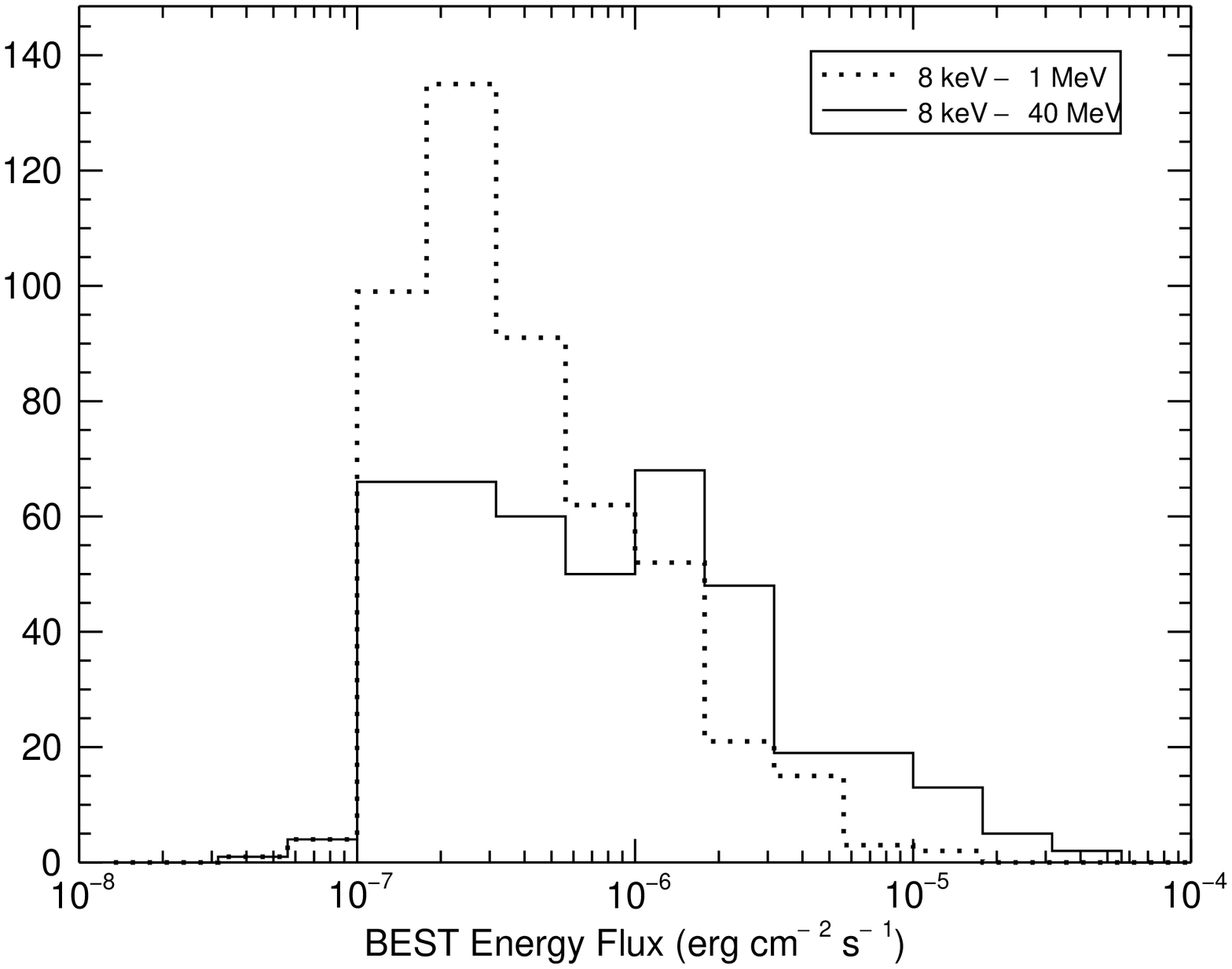}}
	\end{center}
\caption{Distributions of the BEST spectral parameters for the fluence spectra.  \ref{indexbestf} displays the selection of best 
low-energy and high-energy spectral indices.  The shaded distribution depicts the location of the distribution of the PL index.  
\ref{epeakbestf} shows the selection of the best $E_{peak}$ and $E_{break}$.  \ref{pfluxbestf} and \ref{efluxbestf} show the 
selection of the best photon flux and energy flux respectively.  \label{bestf}}
\end{figure}

\begin{figure}
	\begin{center}
		\subfigure[]{\label{indexbestp}\includegraphics[scale=0.35]{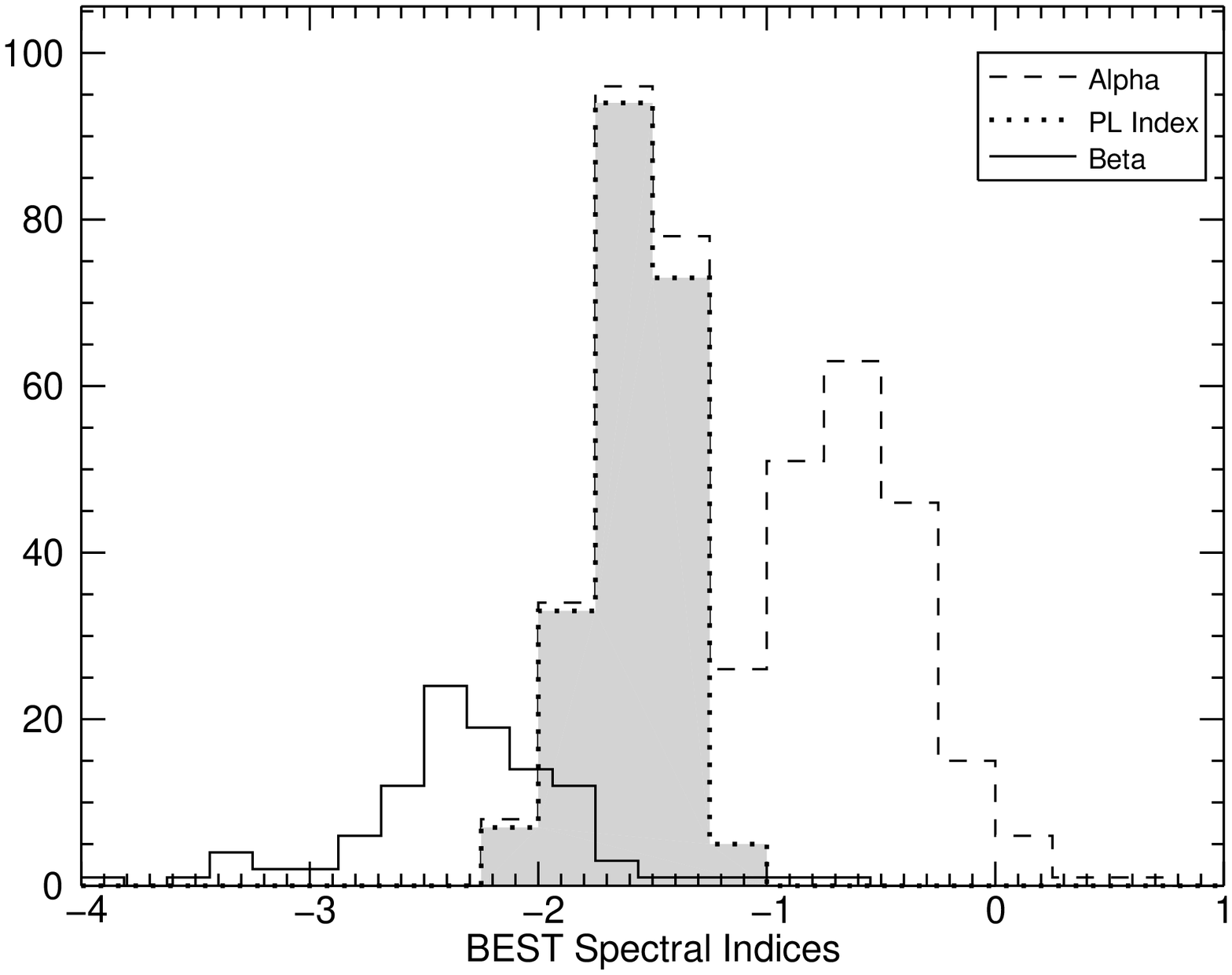}}
		\subfigure[]{\label{epeakbestp}\includegraphics[scale=0.35]{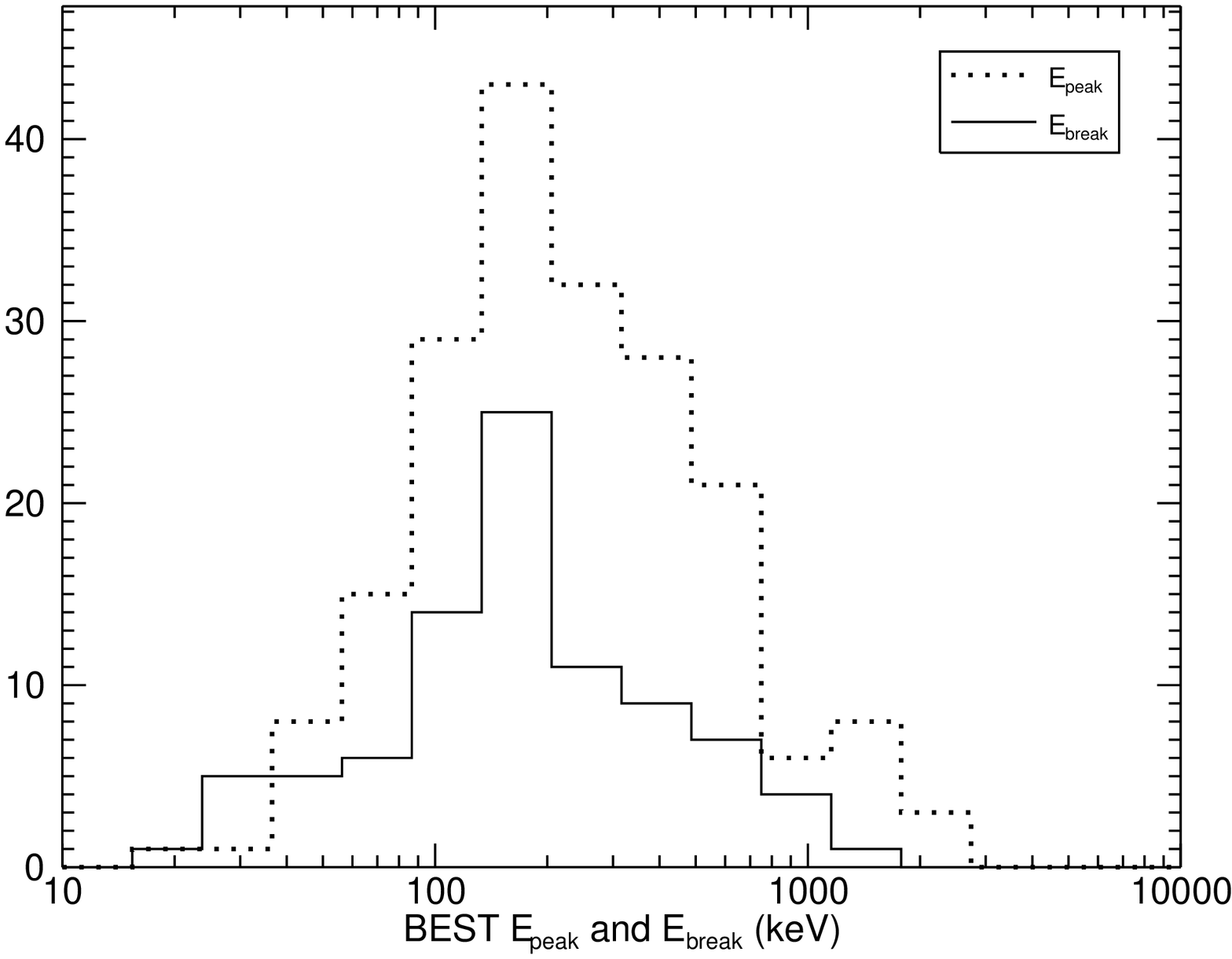}}\\
		\subfigure[]{\label{pfluxbestp}\includegraphics[scale=0.35]{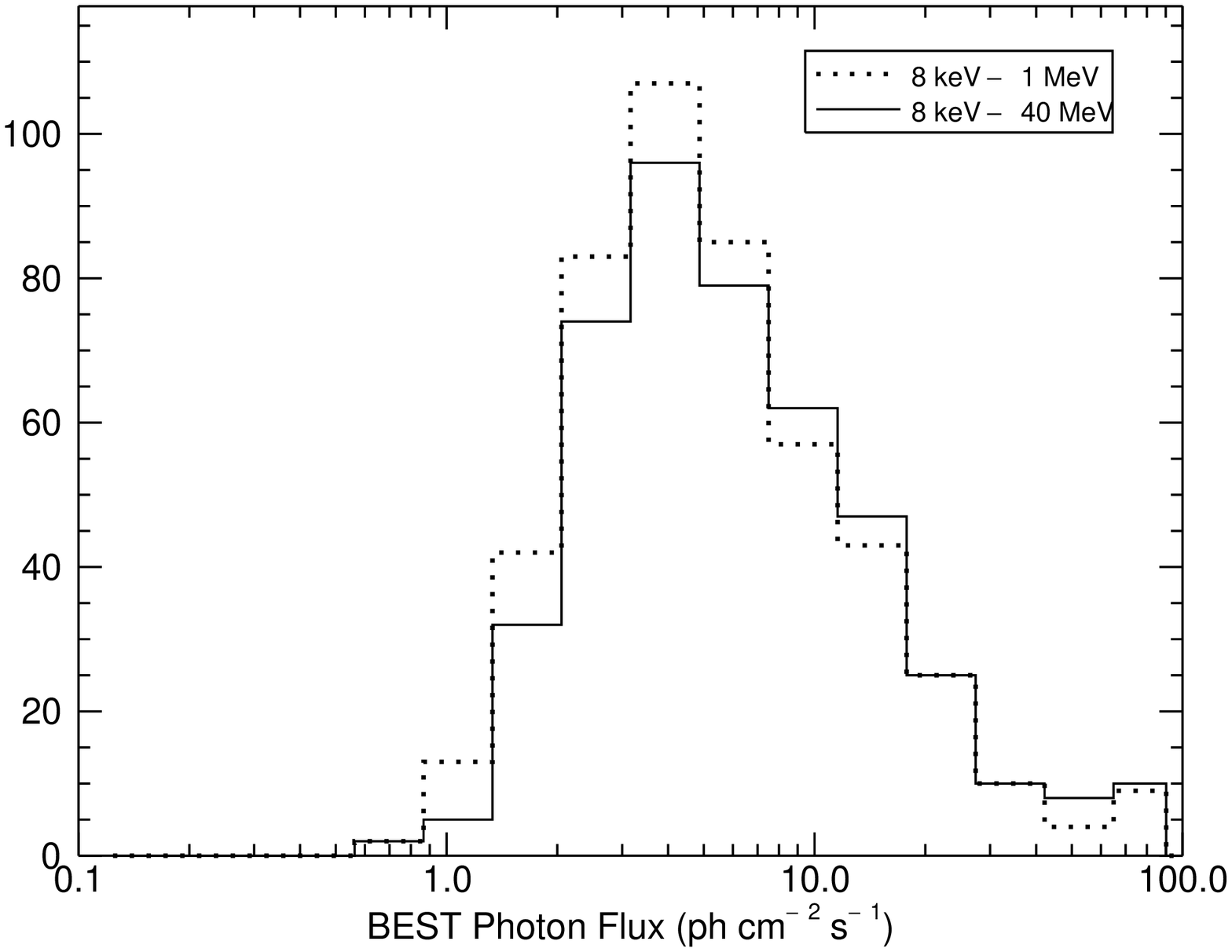}}
		\subfigure[]{\label{efluxbestp}\includegraphics[scale=0.35]{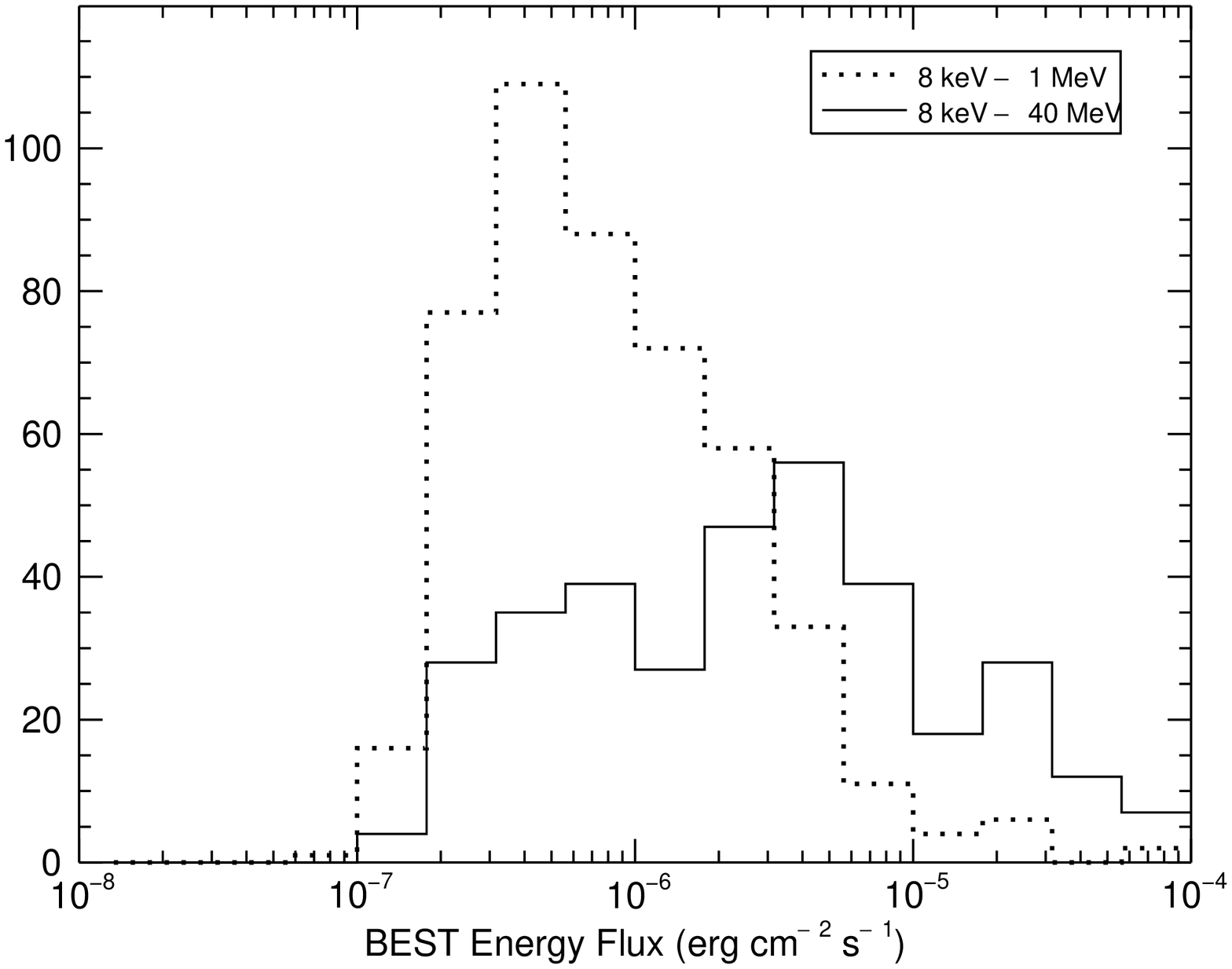}}
	\end{center}
\caption{Distributions of the BEST spectral parameters for the peak flux spectra.  \ref{indexbestp} displays the selection of best 
low-energy and high-energy spectral indices.  The shaded distribution depicts the location of the distribution of the PL index.  
\ref{epeakbestp} shows the selection of the best $E_{peak}$ and $E_{break}$.  \ref{pfluxbestp} and \ref{efluxbestp} show the 
selection of the best photon flux and energy flux respectively.  \label{bestp}}
\end{figure}

\begin{figure}
	\begin{center}
		\subfigure[]{\label{deltasbestf}\includegraphics[scale=0.35]{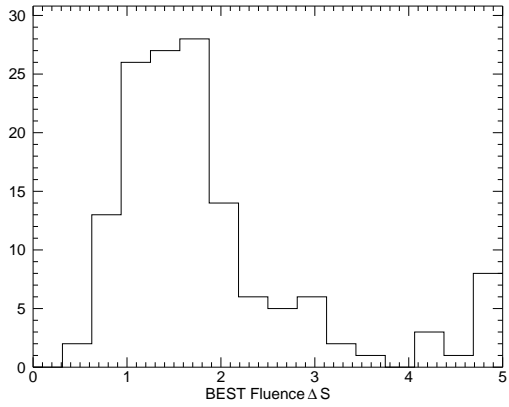}}
		\subfigure[]{\label{deltasbestp}\includegraphics[scale=0.35]{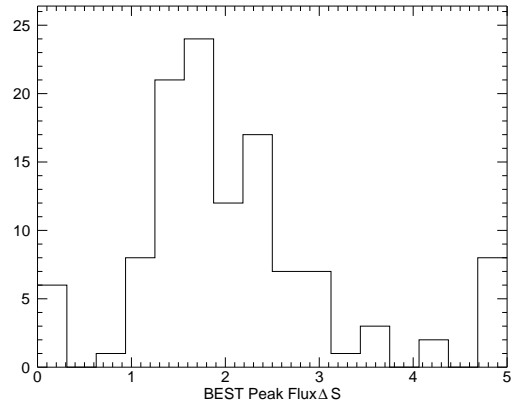}}
	\end{center}
\caption{Distributions of the difference between the BEST low- and high-energy spectral indices.  The first bin contains values 
less than 0, indicating that the centroid value of alpha is steeper than the centroid value of beta.  \label{deltasbest}}
\end{figure}

In Table \ref{BestTable} we present the composition of models for the BEST samples.  From this table, it is apparent that the 
fluence spectral data from GBM strongly favors the COMP model over the others in nearly half of all GRBs.  The BAND and SBPL 
are favored by relatively few GRBs in the catalog.  It should be noted that the number of GRBs best fit by PL increases in the 
peak flux spectra mainly due to the fact that the smaller statistics from the short integration time are unable to support a model 
more complex than the PL.  In Figures \ref{bestf} and \ref{bestp} the same error cuts used in the GOOD samples were also used 
for the BEST parameters.  Note that the PL index is statistically an averaging of the low- and high-energy power laws, and that, 
due to the fact that GBM spectral responses have a peak effective area at lower energies (below 100 keV), we have included 
PL indices in the BEST low-energy index distribution.  Although in a number of cases the PL model is statistically preferred 
over the other models in this catalog, the spectral shape represented by the PL is inherently different from the shape of the 
other models.  Therefore, the PL index is not necessarily representative of either the low- or high-energy indices from the other 
models.  In Figures \ref{indexbestf} and \ref{indexbestp} we show where the distribution of PL indices exists relative to the alpha 
and beta distributions.  The PL index represents 25\% of the BEST fluence alpha distribution and 50\% of the BEST peak flux 
alpha distribution.  The fluence spectra, on the whole, have a steeper measured alpha and shallower beta than the peak flux 
spectra.  The alpha distribution for the fluence spectra peaks at about $-1$, while the peak flux low-energy spectral index peaks 
at about $-0.6$.  Conversely, the beta distribution for the fluence spectra peaks at $-2.1$ and the peak flux high-energy spectral 
index peaks at $-2.4$.  As shown before for the GOOD spectra, Figure \ref{deltasbest} shows the $\Delta S$ distributions for the 
BEST fluence and peak flux spectra.  Both distributions are found to be consistent with previous investigations into BATSE time-
integrated and time-resolved spectra \citep{Preece, Kaneko06}

Additionally, the $E_{peak}$ (Figure \ref{epeakbestf}) and $E_{break}$ (\ref{epeakbestp}) distributions differ between the two 
spectra.  The fluence spectra $E_{peak}$ peaks near 200 keV and the $E_{break}$ peaks at about 150 keV, although the 
kurtosis of the distribution is much lower than that of the $E_{peak}$ distribution.  Meanwhile, the $E_{peak}$ and $E_{break}$ 
for the peak flux spectra both peak at about 150 keV.  As is shown in Figures \ref{pfluxbestf} and \ref{pfluxbestp} the photon flux 
does not significantly change when widening the energy band from 8 keV--1 MeV to 8 keV--40 MeV.  The fluence spectra 
photon flux peaks around 1.5 photons $\rm cm^{-2} \ s^{-1}$ and the peak flux photon flux peaks around 4 photons $\rm cm^{-2} 
\ s^{-1}$.  Alternatively, the energy flux is greatly affected by expanding the energy band, as is shown in Figures \ref{efluxbestf} 
and \ref{efluxbestp}.  Specifically, the distribution in the 8 keV--1 MeV band peak around a few$\rm \times 10^{-7} \ ergs \ cm^
{-2} \ s^{-1}$, but in the 8 keV--40 MeV band the distribution broadens significantly and becomes much less peaked, 
approaching what appears to be a top hat distribution.

\begin{figure}
	\begin{center}
		\subfigure[]{\label{fluencealpha}\includegraphics[scale=0.35]{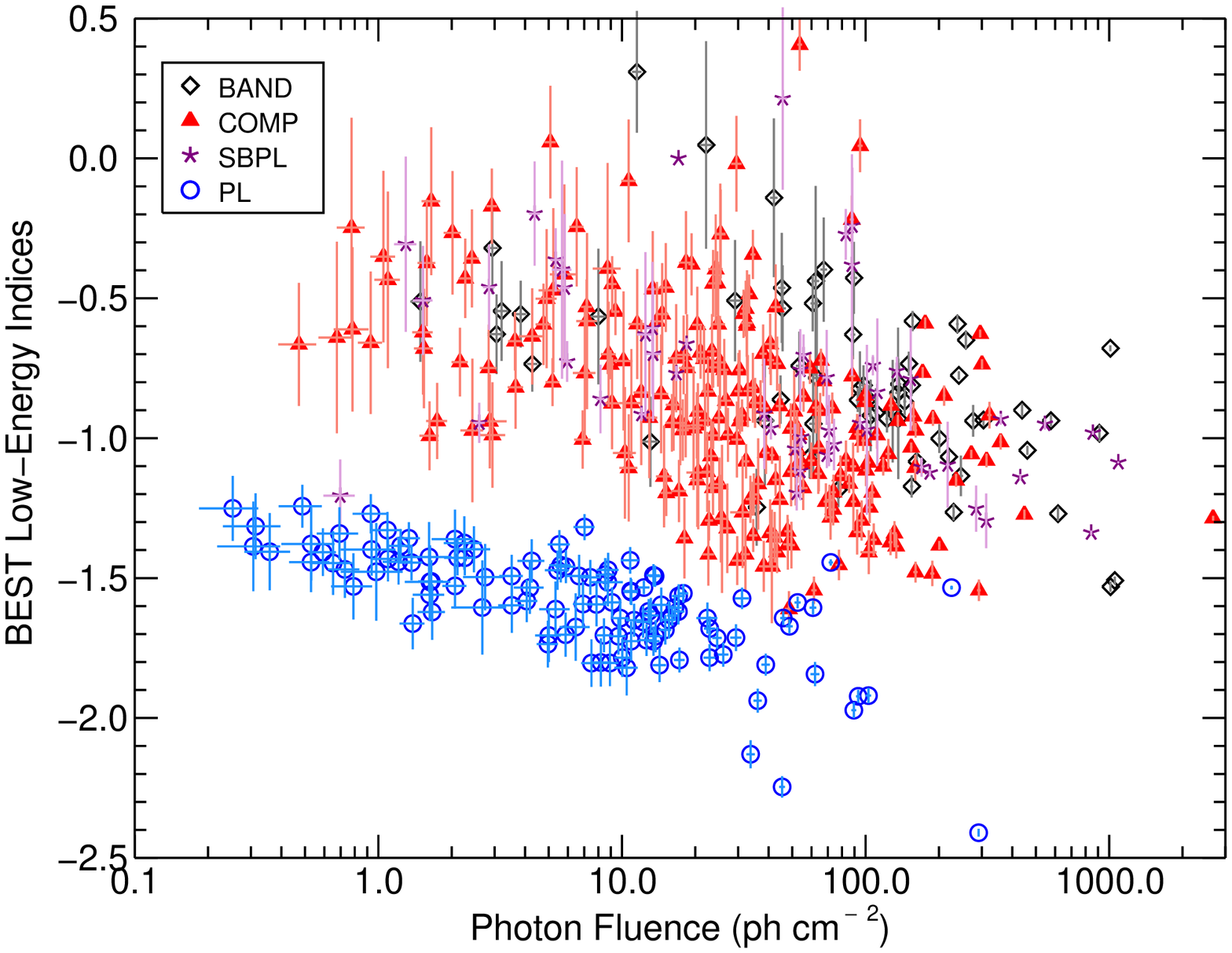}}
		\subfigure[]{\label{fluencebeta}\includegraphics[scale=0.35]{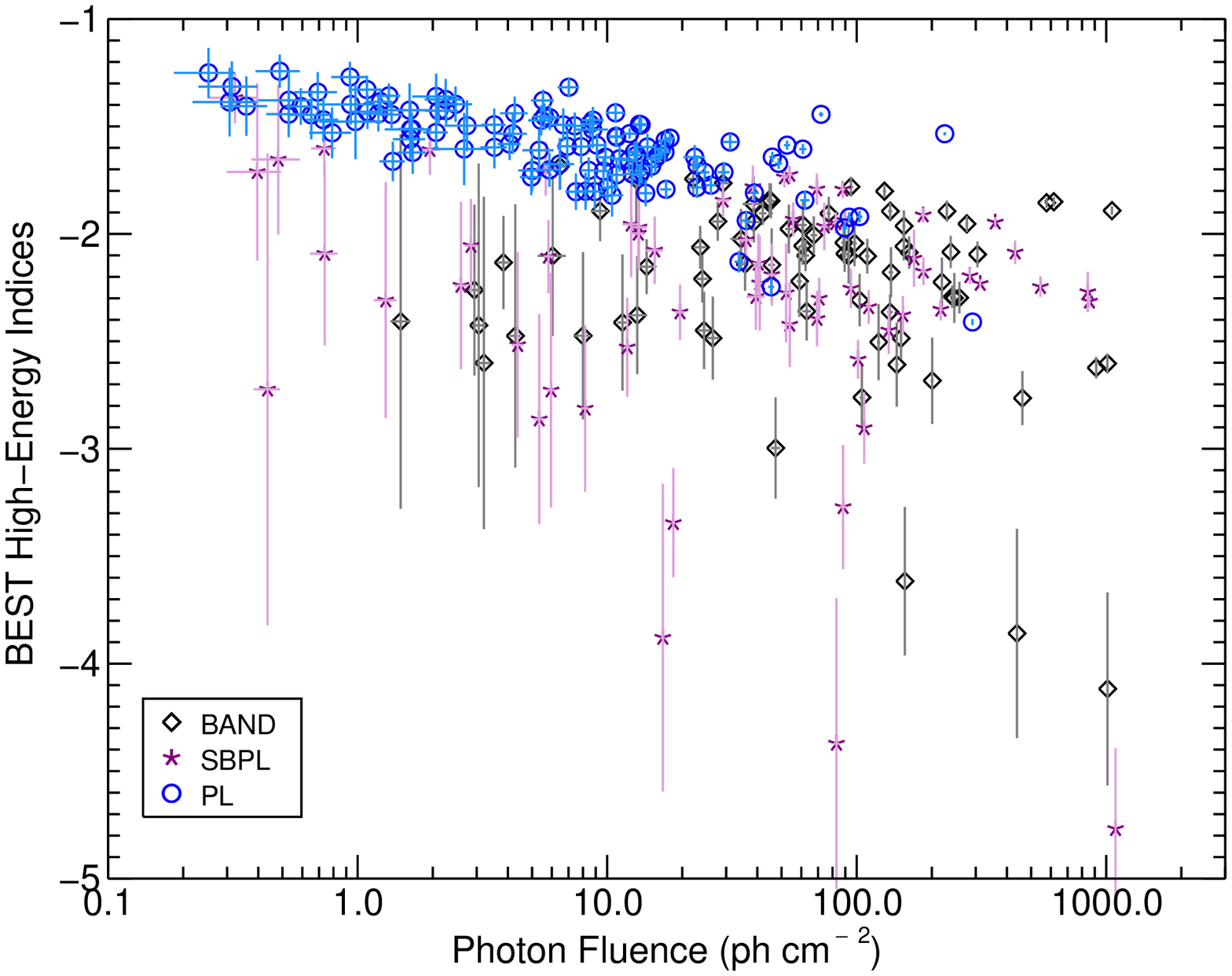}}\\
		\subfigure[]{\label{fluenceepeak}\includegraphics[scale=0.35]{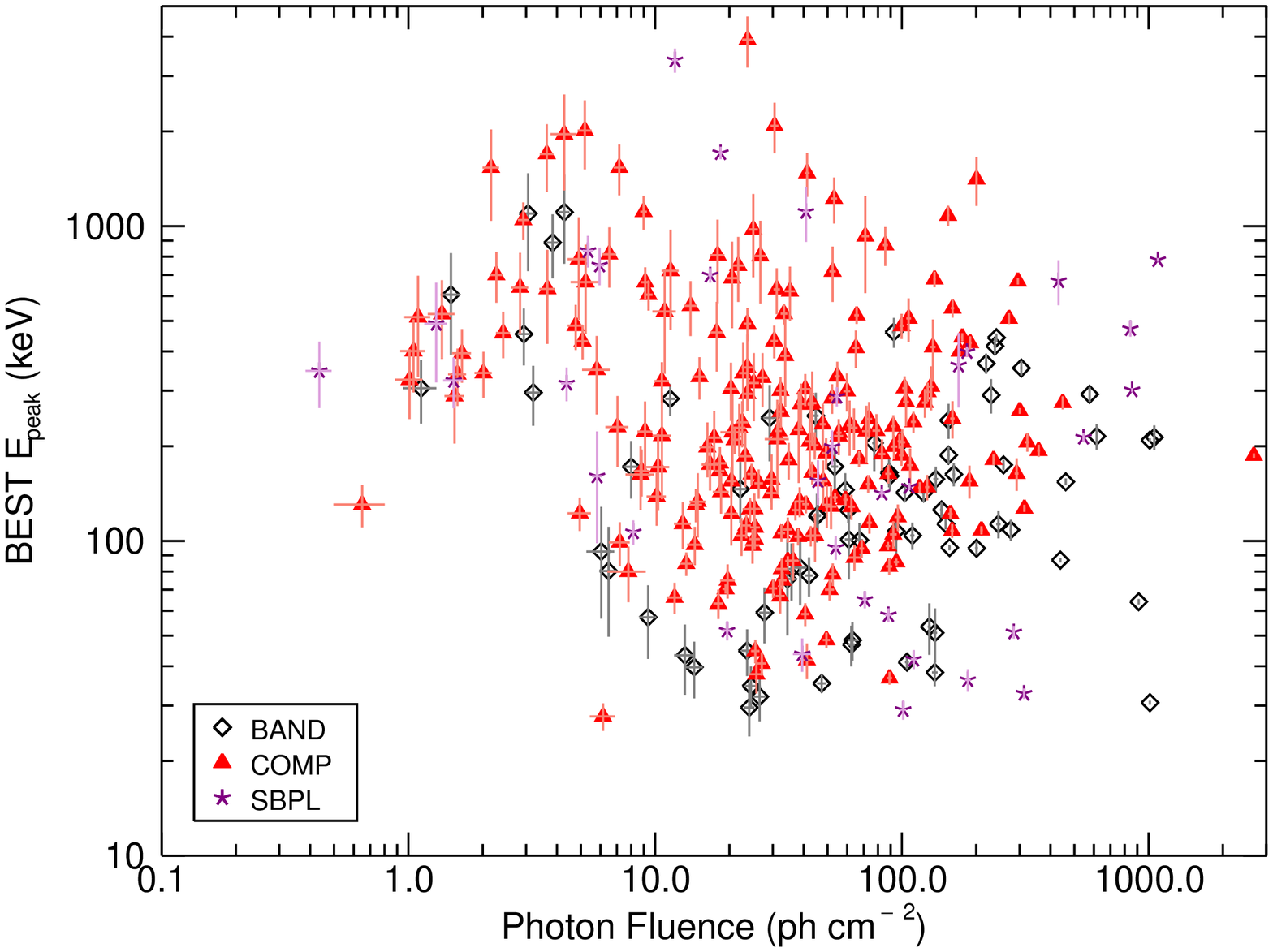}}
	\end{center}
\caption{BEST fluence spectral parameters as a function of the model photon fluence.  Note that the PL index is shown in both 
\ref{fluencealpha} and \ref{fluencebeta} for comparison. \label{fluenceparms}}
\end{figure}

\begin{figure}
	\begin{center}
		\subfigure[]{\label{fluxalpha}\includegraphics[scale=0.35]{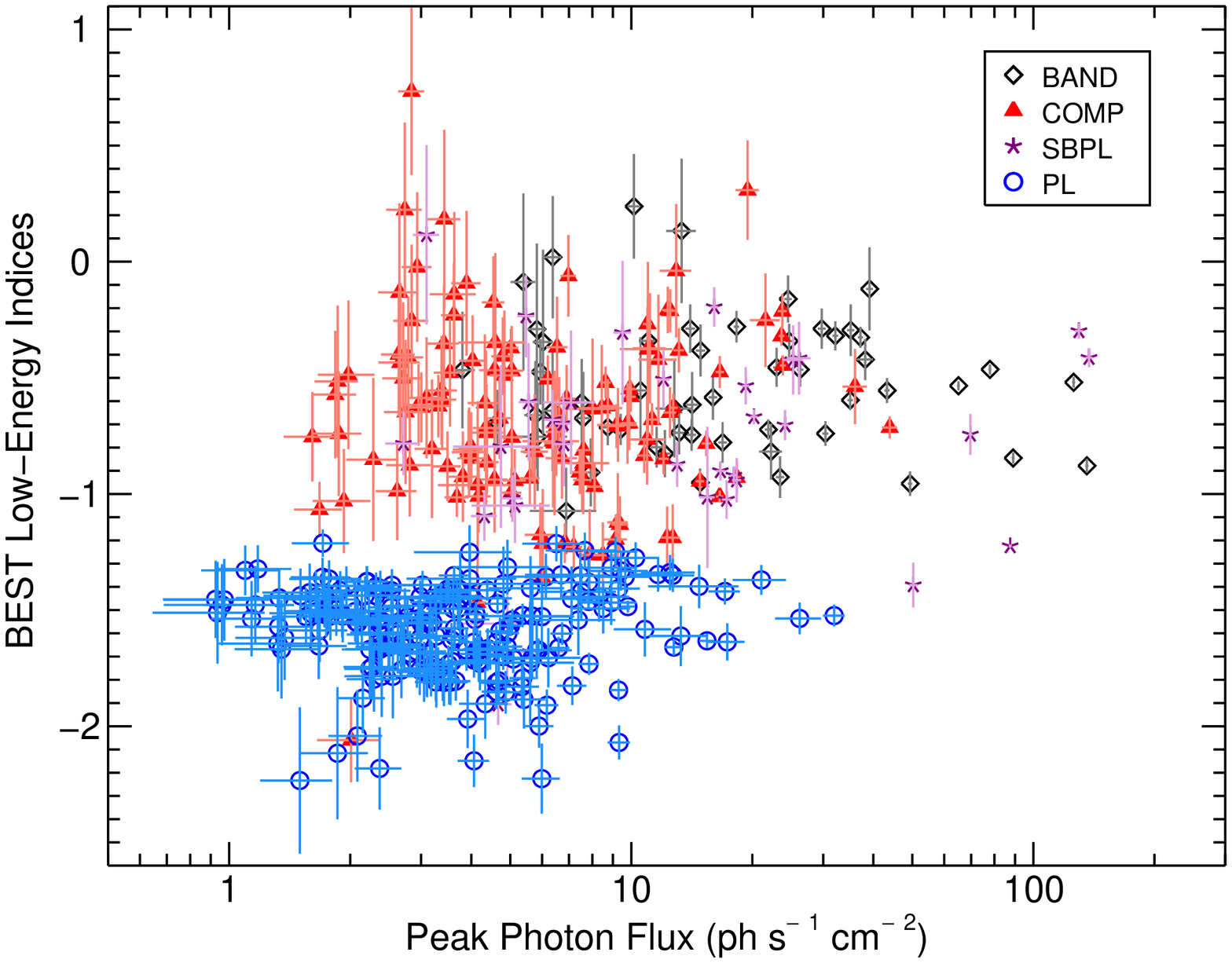}}
		\subfigure[]{\label{fluxbeta}\includegraphics[scale=0.35]{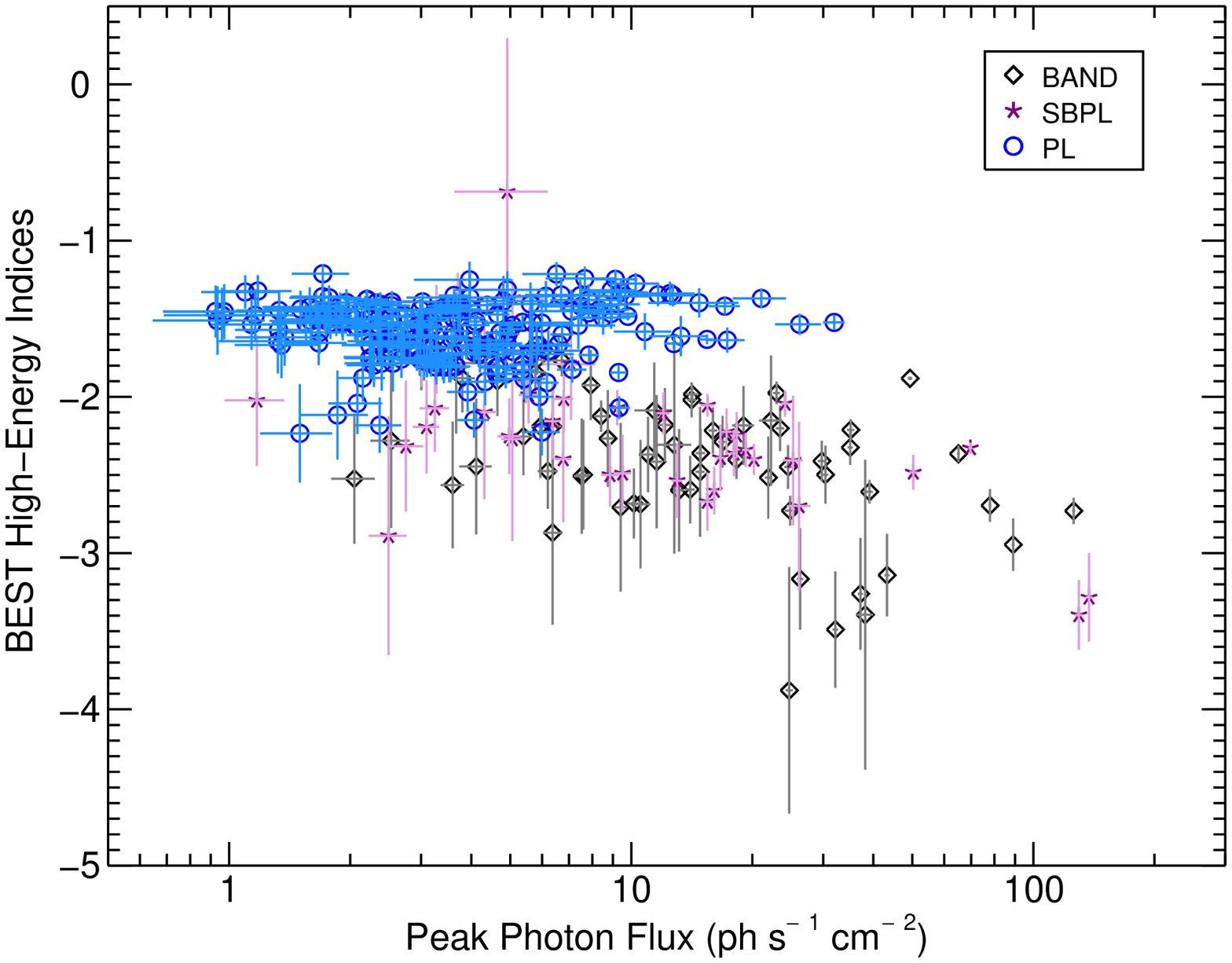}}\\
		\subfigure[]{\label{fluxepeak}\includegraphics[scale=0.35]{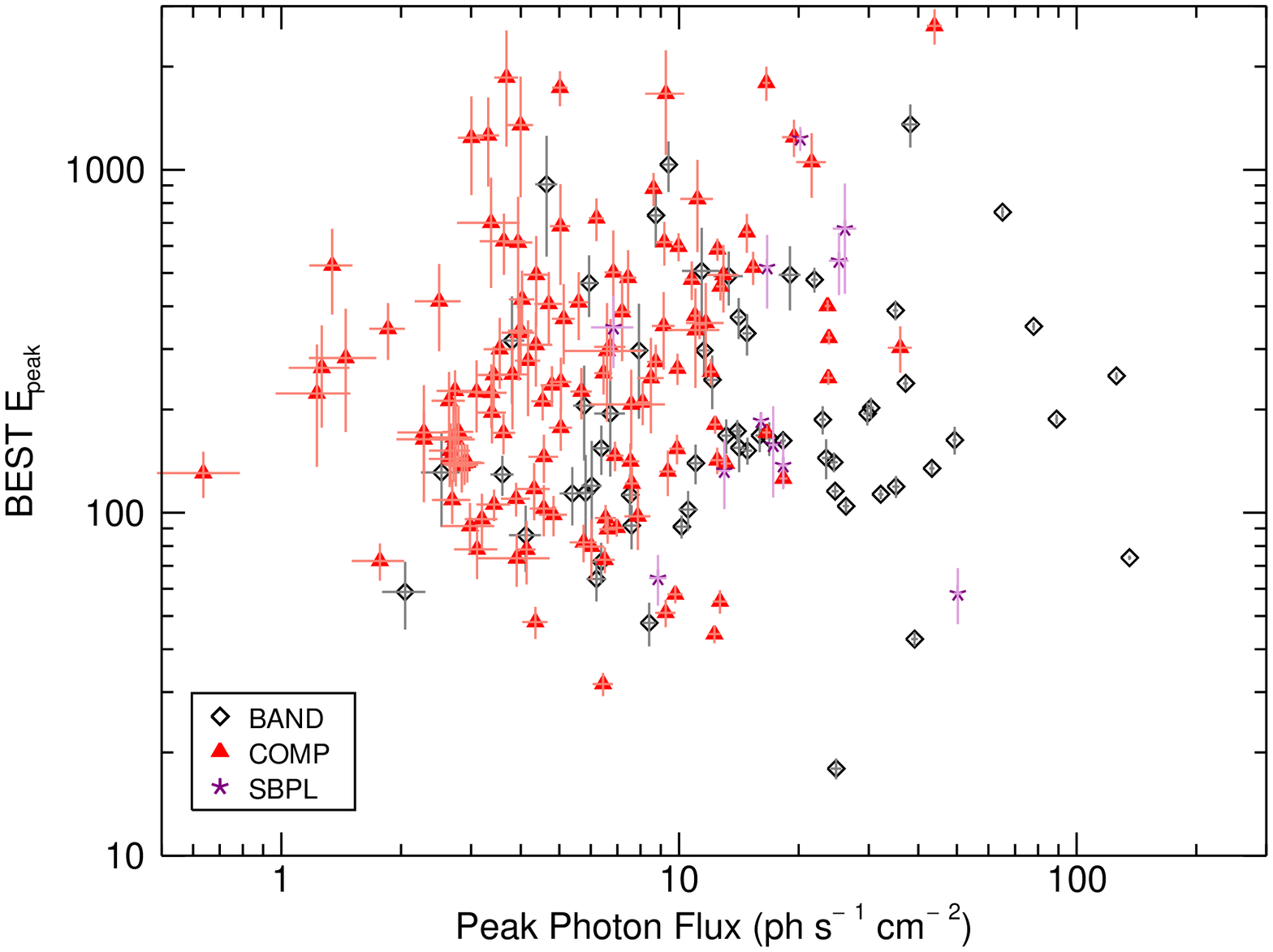}}
	\end{center}
\caption{BEST flux spectral parameters as a function of the model peak photon flux.  Note that the PL index is shown in both \ref
{fluxalpha} and \ref{fluxbeta} for comparison. \label{fluxparms}}
\end{figure}

To aid in the study of the systematics of the parameter estimation, as well as the garner the effect statistics has on the fitting 
process, we investigate the behavior parameter values as a function of the photon fluence and peak photon flux for the fluence 
and peak flux BEST spectra respectively.  These distributions are shown in Figures \ref{fluenceparms} and \ref{fluxparms}.  When  
fitting the time-integrated spectrum of a burst, we find the low- and high-energy indices trend toward steeper values for 
exceedingly more fluent spectra.  The simple PL index trends from shallow value of $\sim-1.3$ to a steeper value of $\sim-2$.  
The low-energy index for a spectrum with curvature tends to exhibit an unusually shallow value of $\sim-0.4$ for 
extremely low fluence spectra, and steepens to $\sim-1.5$.  Similarly, the high-energy index trends from $\sim-1.6$ at low 
fluence to $\sim-2.7$ at high fluence, although this is complicated by unusually steep and poorly constrained indices that 
indicate that an exponential cutoff may result in a more reliable spectral fit.  When inspecting the $E_{peak}$ as a function of 
photon fluence, a trend is much less apparent.  If a burst is assumed to have significant spectral evolution, then obviously the 
$E_{peak}$ will change values through the time history of the burst, typically following the traditional hard-to-soft energy 
evolution.  For this reason, spectra that integrate over increasingly more time will tend to suppress the highest energy of $E_
{peak}$ within the burst, so a general decrease in $E_{peak}$ is expected with longer integration times.  However, the photon 
fluence convolves the integration time with the photon flux so that an intense burst with a short duration may have on the order 
the same fluence as a much longer but less intense burst but results in a higher $E_{peak}$.  This causes significant broadening 
to the decreasing trend as shown in Figure \ref{fluenceepeak}.  The distribution of parameters as a function of the peak photon 
flux, however, is much less clear.  The distributions shown in Figure \ref{fluxparms} are more susceptible to uncertainty because 
of the generally smaller statistics involved in study the peak flux section of the GRB, except in some cases where the peak 
photon flux is on the order of the photon fluence.  Ignoring the regions where the parameters are poorly constrained, another 
trend emerges from the low-energy indices; they appear to become slightly more shallow as the photon flux increases.  The 
high-energy indices, however, appear to be unaffected by the photon flux, except those that are unusually steep and indicate 
that an exponential cutoff may be preferred.  Finally, no obvious trend is visible for $E_{peak}$ as a function of the peak photon 
flux.  This likely results from the fact that neither spectral index in the peak flux spectrum is affected much by the photon flux.

%% Chi^2 Distributions
\subsection{$\bf \chi^2$ Distributions}
\begin{figure}
	\begin{center}
		\subfigure[]{\label{redchisqf}\includegraphics[scale=0.35]{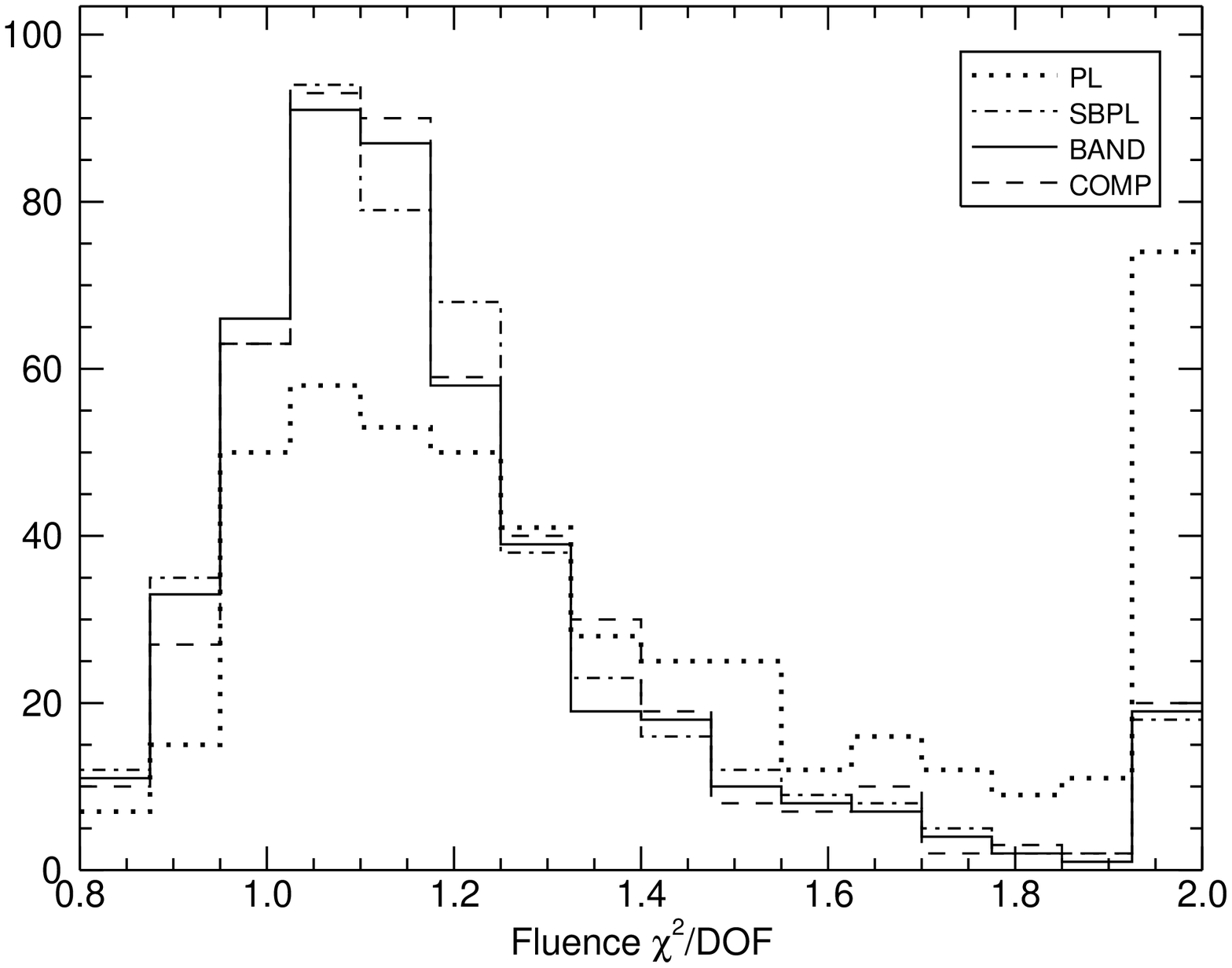}}
		\subfigure[]{\label{redchisqp}\includegraphics[scale=0.35]{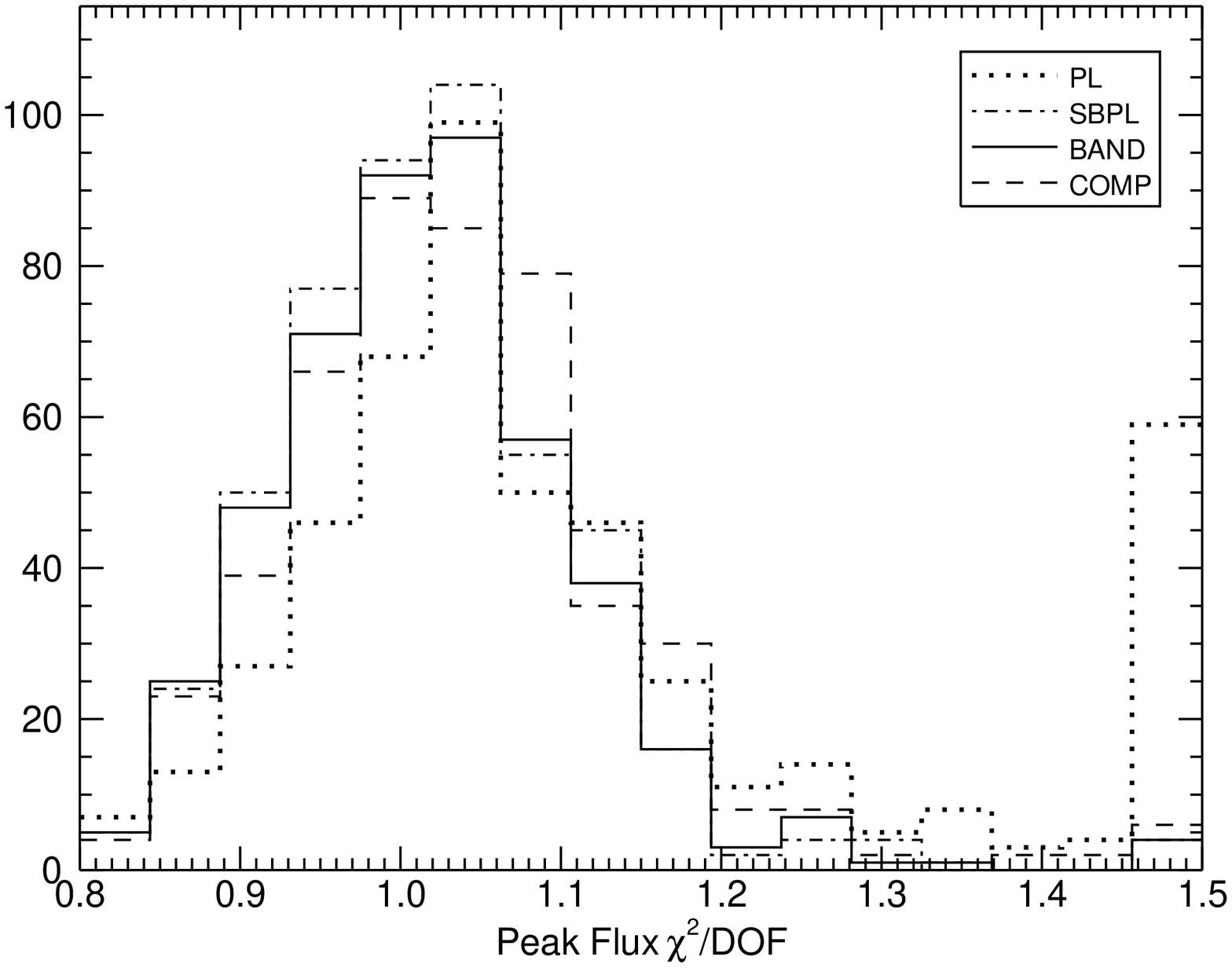}}\\
		\subfigure[]{\label{redchisqbestf}\includegraphics[scale=0.35]{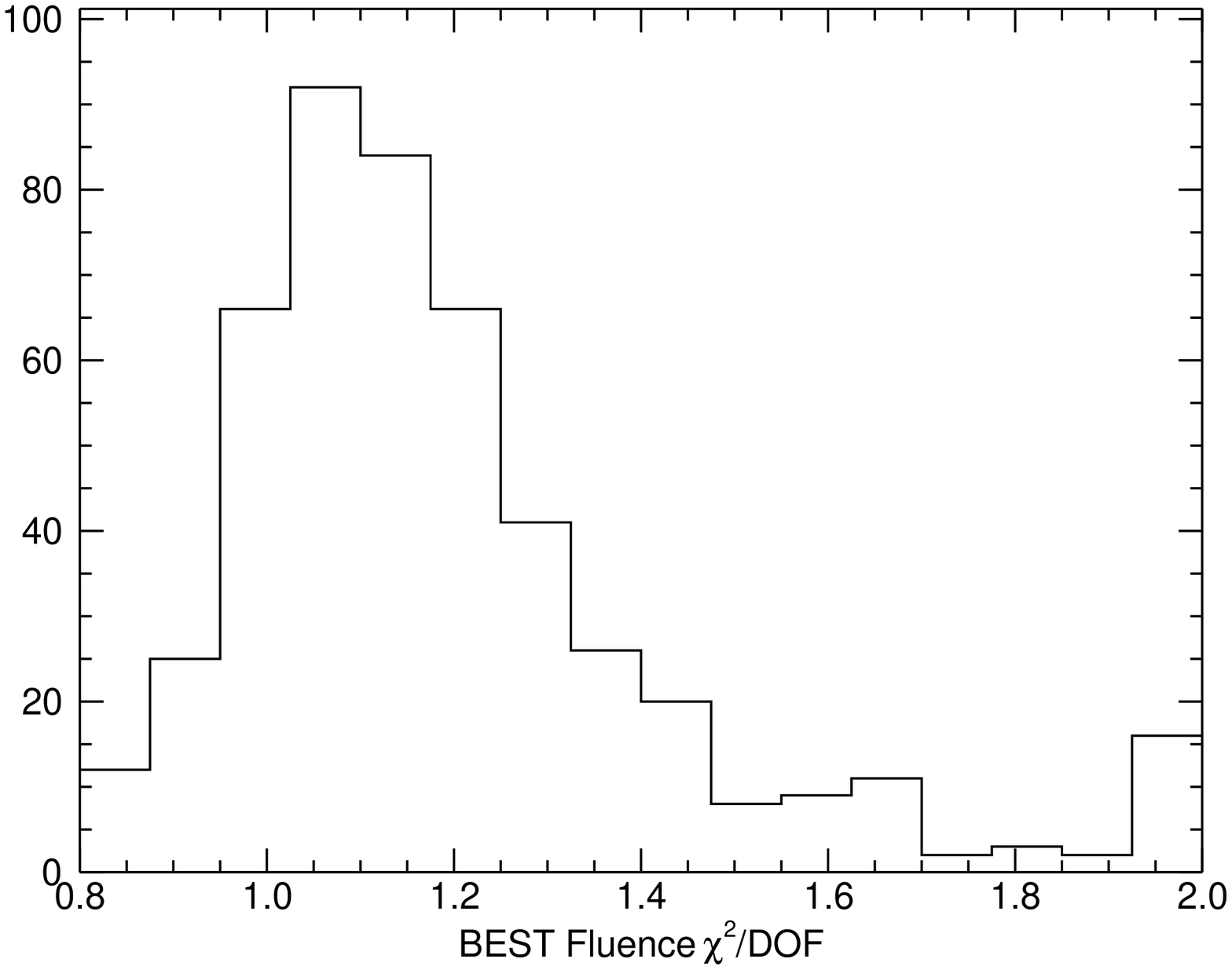}}
		\subfigure[]{\label{redchisqbestp}\includegraphics[scale=0.35]{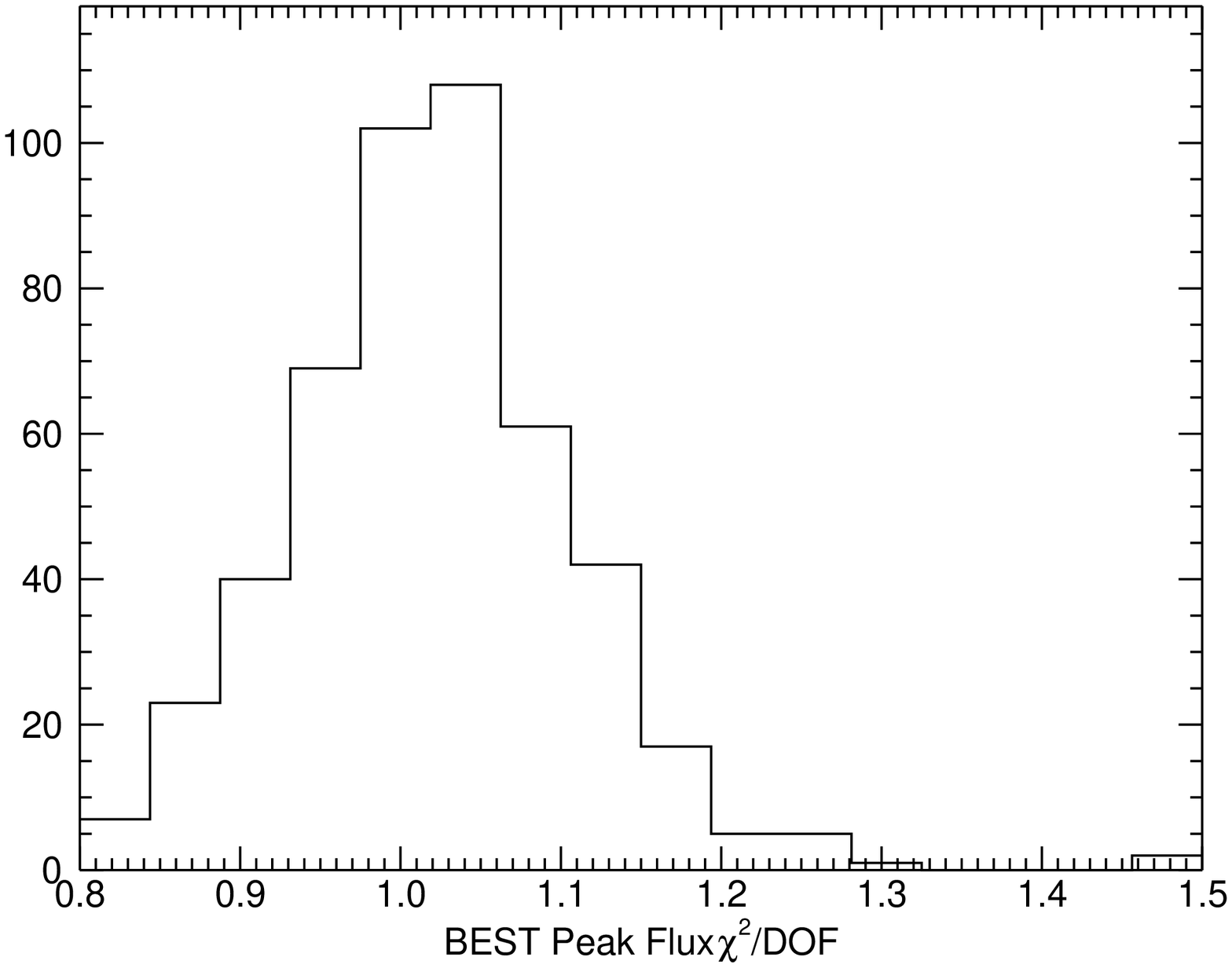}}
	\end{center}
\caption{Distributions of the reduced $\chi^2$ for each model.  \ref{redchisqf} and \ref{redchisqp} are distributions of the 
reduced $\chi^2$ for each model for each burst.  Note that these distributions include all fits, including unconstrained fits.  \ref
{redchisqbestf} and \ref{redchisqbestp} show the distribution of BEST reduced $\chi^2$ values.  The first and last bins in all 
distributions contain overflow values.  Note that the peak of the distributions are slightly larger than 1.  \label{redchisq}}
\end{figure}

Although the least-squares fitting process did not minimize $\chi^2$ as a figure of merit, we can calculate the $\chi^2$ 
goodness-of-fit statistic comparing the model to the data.  To do this, we difference the background-subtracted count rates from 
the model rates, summing first over all energy channels in each detector, and then over all detectors.  This is shown by
\begin{equation}
\chi^2= \sum_i \sum_j \Biggl[ \frac{O_{ij} - B_{ij} - M_{ij}}{\sqrt{\sigma_{M_{ij}}^2}} \Biggr]^2,
\end{equation}
where the $O_{ij}$ are the observed count rates, $B_{ij}$ are the background rates, $M_{ij}$ are the model rates, and $
\sigma_{M_{ij}}^2$ are the derived model variances.  In the ideal situation and assuming acceptable spectral fits (i.e. when 
performing spectral analysis of simulated data), the reduced $\chi^2$ value ($\chi^2$/d.o.f.) will tend to distribute around a 
value of 1.  In this way, $\chi^2$ gives an estimate on the acceptability of the fit.  Figure \ref{redchisq} shows the reduced $
\chi^2$ distributions for both the fluence and peak flux spectra, as well as the corresponding BEST distributions.  It is 
important to note that the distributions all peak at slightly larger values than 1, which is acceptable since the fits did not minimize 
$\chi^2$ and even the BEST spectral fits represent an approximation to the actual spectra of GRBs due to their empirical 
nature.  The fluence $\chi^2$ distributions appear to be much broader and shifted farther from the nominal reduced $\chi^2$ 
value than the peak flux distributions.  This is likely due to the longer time integration intervals in the fluence spectra and the fact 
that many GRBs experience spectral evolution.  The peak flux sample captures the spectra of all bursts in a small slice of time at 
the same stage of the lightcurve, meanwhile the fluence sample integrates over the duration of the emission, in many cases over 
several pulses.  The proximity of most of the reduced $\chi^2$ values to the nominal value is indicative of acceptable spectral fits.

\begin{figure}
	\begin{center}
		\subfigure[]{\label{fluencechisq}\includegraphics[scale=0.35]{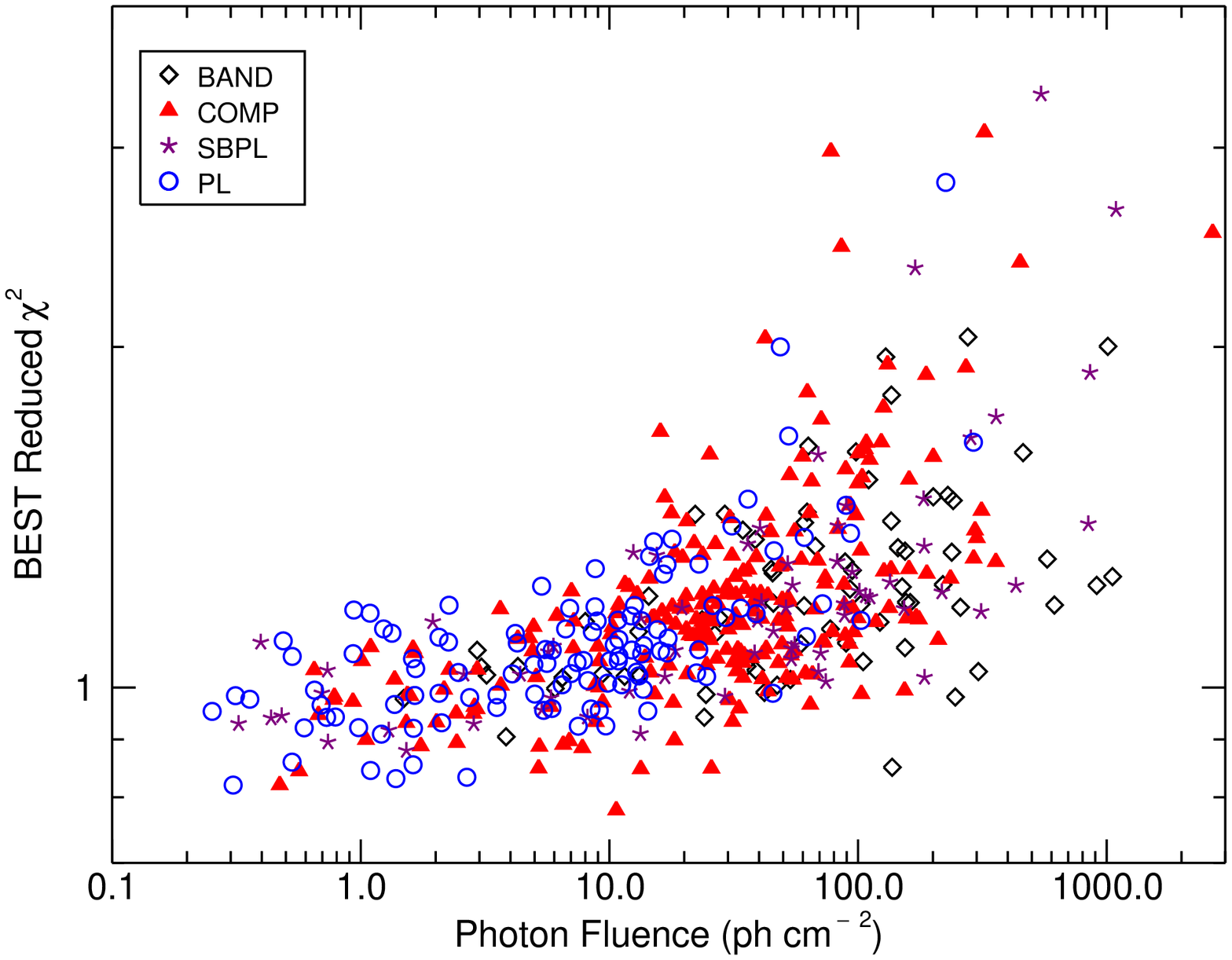}}
		\subfigure[]{\label{fluxchisq}\includegraphics[scale=0.35]{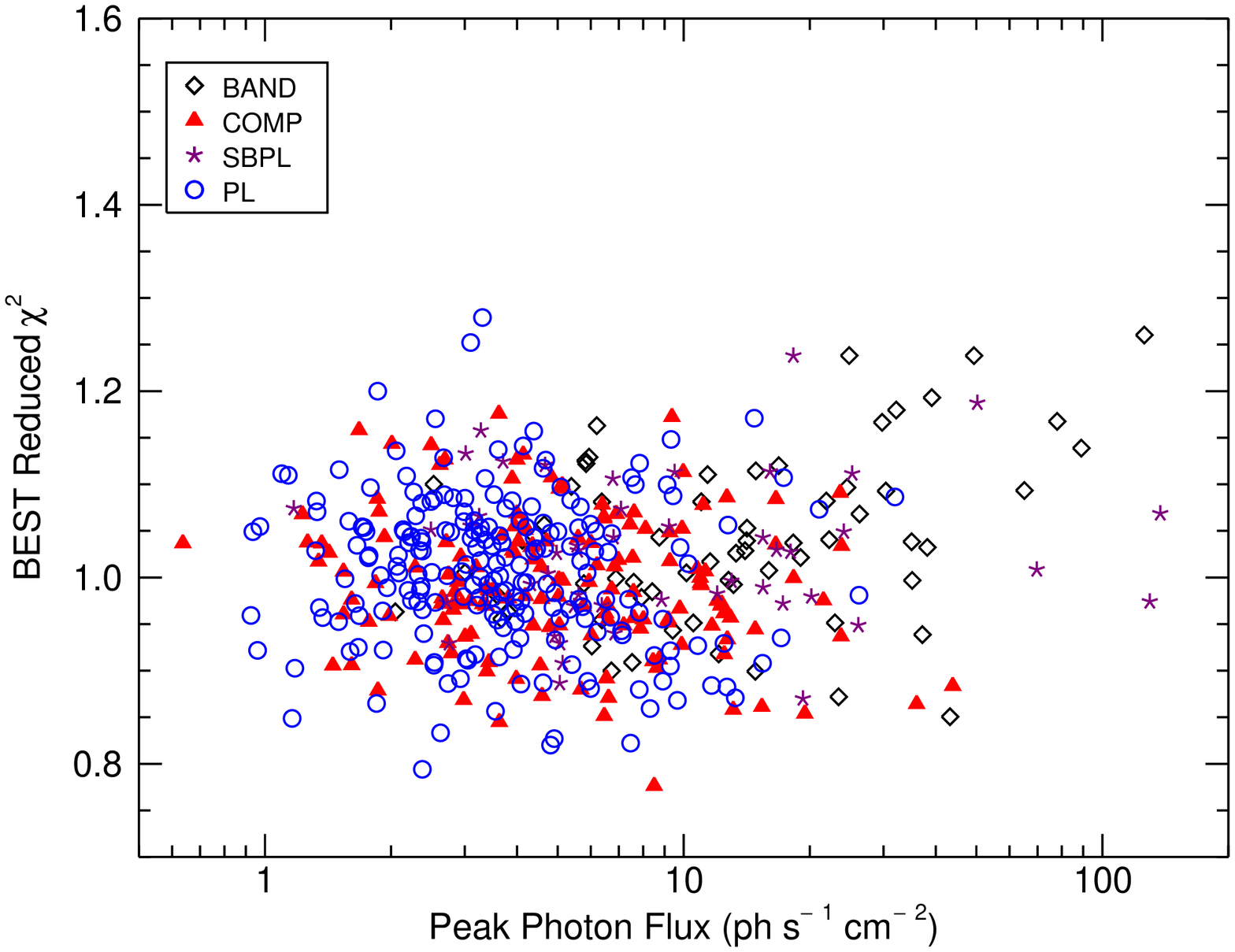}}
	\end{center}
\caption{\ref{fluencechisq} Reduced $\chi^2$ as a function of the model photon fluence.  \ref{fluxchisq} Reduced $\chi^2$ as a 
function of the model peak photon flux.   \label{ffredchisq}}
\end{figure}

In addition, Figure \ref{ffredchisq} plots the BEST reduced $\chi^2$ as a function of photon fluence and peak photon flux for the 
fluence and peak flux spectra respectively.  The reduced $\chi^2$ for the fluence fits shows a marked upward trend as the 
photon fluence increases.  The average reduced $\chi^2$ starts at approximately a value of unity at a low fluence of $\sim$0.2 $
\rm photons \ cm^{-2}$ and increases to a value of $\sim$2 at $\sim$1000 $\rm photons \ cm^{-2}$.  Additionally there are several 
outliers to the trend that exist at high fluence and exhibit even larger reduced $\chi^2$ values.  This indicates that the goodness-
of-fit is increasingly worse the more fluent the burst.  This follows from the fact that in most cases extremely fluent bursts are long 
and may exhibit significant spectral evolution, therefore the time-integrated spectral fit will average over the evolution and will 
produce a significantly worse fit.  When inspecting the reduced $\chi^2$ as a function of the photon flux, we find that the trend is 
not as dramatic, with low flux spectra producing a marginally better goodness-of-fit than the high flux spectra.  This indicates that 
the peak flux spectra do not suffer from the same systematic effects as the fluence spectra.  Most of the large reduced $\chi^2$ 
values from the peak flux spectra result from high flux bursts where the problem around the K--edge dominates the statistics of 
the goodness-of-fit.

%%Summary
\section{Summary \& Discussion}
GBM has allowed the observation of a large number of GRB spectra over a broad energy range that has not been fully covered 
previously.  The broad energy range of GBM translates to parameter distributions that are variants of distributions found by 
previous instruments \citep[e.g.][Goldstein et al. in prep. on the final BATSE spectral catalog]{Kaneko06}.  The distributions 
contained here are similar to those shown by previous studies, yet contain differences that display the usefulness of studying 
GRBs over different energy bands and sensitivities.  Although the typical average GRB spectrum has largely remained 
unchanged within the GBM energy band, the distributions of spectral parameters are no longer as narrow as previously thought.  
We have shown in many cases that the fitted spectrum of a GRB depends in large part on its intensity as well as the detector 
sensitivity.  This observation implies that weak GRBs may have the same inherent spectrum as their more intense counterparts, 
yet we are unable to accurately determine their spectrum, reinforcing the importance of comparing the spectral parameter 
distributions from different acceptable models.  Additionally, the high-energy coverage provided by the BGO detectors allows the 
extension of fine spectral analysis into a new regime.  For example, the inclusion of high-energy data in many cases causes the 
high-energy power law to soften, and in some cases provides evidence for spectral cutoffs at even higher energies that is being 
investigated by the {\it Fermi}/LAT (Omodei et al. in prep.).  Because of the high-energy coverage, we have uncovered several 
GRBs with peak energies greater than 1 MeV, extending the $E_{peak}$ distribution to higher energies.  Observations of high-
$E_{peak}$ GRBs have been observed previously \citep[e.g. GRB 990123;][]{Briggs}, but were typically not from time-integrated 
spectra.  According to our BEST samples, 18 GRBs have a time-integrated $E_{peak}$ larger than 1 MeV, which is $\sim$6\% of 
the GRB fluence spectra in the catalog that can be fitted with a curved function.  Similarly, 8 GRBs have a peak flux $E_{peak}$ 
greater than 1 MeV, which is $\sim$5\% of the peak flux spectra that can be fit by a curved function \citep[for details on three short 
hard bursts see][]{Guiriec}.  

Another interesting result of the parameter distributions is the $\Delta$S parameter, the difference between the low- and 
high-energy spectral indices.  This can be an important quantity because current models for the GRB prompt emission 
mechanism can be broken into two categories: magnetic \citep[e.g.][]{Lee} or internal/external shock \citep[e.g.][]{Rees} driven.  
Implications about the shock driven case can be drawn directly from $\Delta$S.  \citet{Sari} presented a model for the prompt 
emission spectrum where the emission of synchrotron photons by shock accelerated electrons evolves due to radiative and 
adiabatic losses. A key prediction of this model is a series of relationships between the photon spectral indices above and 
below the synchrotron peak. \citet{Preece} found that this relationship can be used to constrain the synchrotron cooling regime 
of the electrons by looking at the difference of the Band function spectral indices across the $\nu F_{\nu}$ peak.  Therefore, 
investigating $\Delta$S potentially provides insight into the emission mechanism of GRBs by comparing $\Delta$S to the 
prediction offered by the synchrotron shock model.

A comparison between different catalogs can be instructive, especially when noting the difference in methodologies and 
analysis.  For example,  the methodology and fit results for GBM bursts presented in \citet{Nava} are somewhat different than the 
analysis and results given in this GBM catalog.  Simple differences in data selection and methodology, such as analysis on 
detectors with spacecraft blockage and usage of detectors with detector-to-source angles larger than $60^{\circ}$, can 
significantly affect the results of the spectral analysis.  Additionally, in the \citet{Nava} sample, signal selections were not 
performed in a systematic way, but rather subjectively selected, based on a subjective estimate of where the signal fell below the 
background.  These differences as well as many others can lead to large differences in results.  We have compared the overall 
average values and distributions between this catalog and the analysis of \citet{Nava}.  \citet{Nava} report which model is most 
preferred for each burst in their sample; however they do this by comparing C-Stat, which does not allow for model comparison 
without Monte Carlo simulation for each burst.  Unfortunately, a $\chi^2$ goodness-of-fit is also not reported so there is no clue 
to the accuracy of their model selection method, nor how well their models actually fit the data.  This should also be noted as a 
caveat when comparing the two catalogs.  Table \ref{BestComparison} shows the sample mean and standard deviation of \citet
{Nava} and a comparison to the results of this catalog, of which there are many notable differences.  Although their average best 
low-energy index value for the time-integrated spectra is relatively similar, their average high-energy index is found to be steeper 
and represents a narrower distribution, although this may be due to their stated inability to constrain the high-energy index.  
Their average $E_{peak}$ value is slightly softer and does not capture as many of the extended high-energy $E_{peak}$ values 
that are found in this catalog.  The mean time-averaged energy flux is also softer by about 25\% compared to our derived values.  
The differences in the peak flux spectra are even more divergent.  The average low-energy index found by \citet{Nava} is 
surprisingly shallow and the high-energy index is somewhat steeper and represents a much broader distribution than what is 
presented here.  They found the average $E_{peak}$ of the peak flux spectra to be somewhat harder, but their average peak 
energy flux is nearly an order of magnitude larger than reported in this catalog.  These differences are representative of 
divergent methodologies and samples; therefore the reader should take care to understand how the values in the catalogs are 
derived.

Additionally, when comparing our results to the BATSE catalog of bright GRBs \citep{Kaneko06}, there are some marked 
differences as well, although many of these arise because the \citet{Kaneko06} sample studied only the 350 brightest BATSE 
GRBs.  We investigated the sample mean of the time-integrated spectra for comparison (see Table \ref{BestComparison}) and 
found that the average low-energy index was relatively the same and the high-energy index on average was slightly steeper.  
The difference of values in the high-energy index could perhaps be explained by the difference in spectral bandwidth between 
the two instruments.  The time-averaged $E_{peak}$ however, is on average harder, most likely due to the selection of the 
brightest bursts in that catalog.  This also affects the $E_{break}$, which is considerably harder at over 200 keV, and naturally 
the \citet{Kaneko06} catalog reports larger average photon and energy fluxes and much larger spreads in flux due to the 
increased sensitivity compared to GBM.

In conclusion, we have presented a systematic analysis of GRBs detected by GBM during its first two years of operation. This 
catalog contains four basic photon model fits to each burst, using two different selection criteria to facilitate an accurate estimate 
of the spectral properties of these GRBs.  We have described subsets of the full results in the form of data cuts based on 
parameter uncertainty, as well as employing model comparison techniques to select the most statistically preferred model for
each GRB.  The analysis of each GRB was performed as objectively as possible, in an attempt to minimize biased systematic 
errors inherent in subjective analysis.  The methods we have described treat all bursts equally, and we have presented the 
ensemble observed spectral properties of GBM GRBs.  Certainly there are avenues of investigation that require more 
detailed work and analysis or perhaps a different methodology.  This catalog should be treated as a starting point for future 
research on interesting bursts and ideas.  As has been the case in previous GRB spectral catalogs, we hope this catalog will be 
of great assistance and importance to the search for the physical properties of GRBs and other related studies.

%% Acknowledgments
\section{Acknowledgments}
AG acknowledges the support of the Graduate Student Researchers Program funded by NASA.  SMB acknowledges support of 
the European Union Marie Curie Reintegration Grant within the 7th Program under contract number 
PERG04-GA-2008-239176.  SF acknowledges the support of the Irish Research Council for Science, Engineering, and 
Technology, co-funded by Marie Curie Actions under FP7.  The GBM project is supported by the German Bundesministeriums 
f\"ur Wirtschaft und Technologie (BMWi)  via the Deutsches Zentrum f\"ur LuftÐ und Raumfahrt (DLR) under the contract 
numbers 50~QV~0301 and 50~OG~0502

%%Appendix
\appendix
\section{Spectral Catalog Data Files (SCATs)}
The spectral fit results are stored in a series of Flexible Image Transport System (FITS) data files, available online from NASA's 
High Energy Astrophysics Science Archive Research Center (HEASARC)\footnote{\url{http://heasarc.gsfc.nasa.gov/W3Browse/
fermi/fermigbrst.html}}. This data product provides different types of spectral fits to be included the official spectral catalog of GBM 
bursts.  The type is coded into the filename.  There are two types of spectrum categories:
\begin{enumerate}
\item Peak flux (`pflx') - a single spectrum over the time range of the burst's peak flux
\item Fluence (`flnc') - a single spectrum over the entire burst
\end{enumerate}
The time ranges for each category are included with the spectra.  The spectra are fit with a number of models, whose 
complexity often depends on the signal-to-noise ratio of the spectrum.  The current set is:
\begin{enumerate}
\item Power law (`plaw'),
\item Comptonized (exponentially attenuated power law; `comp')
\item Band (`band') 
\item Smoothly broken Power Law (`sbpl')
\end{enumerate}
The set of detectors used for the fit is not coded into the filename, and may not be the same as the set of detectors that 
triggered. The file naming convention is thus: 
\verb+glg_scat_all_bnyymmddfff_type_modl_vxx.fit+, with 
\verb+yymmdd+ = the date; 
\verb+fff+  = fraction of a day, in thousandths; 
\verb+xx+  = version number of the file; 
\verb+type+ = `pflx' or `flnc', the spectral categories; and 
\verb+modl+ = `plaw', `comp', `band', or `sbpl', the spectral model used. We have chosen to designate burst triggers by the 
fraction of a day in thousandths, because that time (86.4 s.) is less than the time for a complete trigger data readout, as currently 
defined in the GBM Flight Software (300 s.). Therefore, we will not confuse two successive triggers by name.

FITS files are generally self-documenting, consisting of header and data pairs, where the \verb+ASCII+ headers are human 
readable and are followed by the data, as described in the header format statements, in either \verb+ASCII+ or binary format. 
The first (or, `primary') header describes the file's content in general, and includes information that applies globally. 
The data section for the primary header is an optional image, and the GBM data files omit this.  An example primary header is 
as follows:
\begin{verbatim}
SIMPLE  =                    T /Written by IDL:  Mon Feb 14 13:36:27 2011
BITPIX  =                    8 /
NAXIS   =                    0 /
EXTEND  =                    T /File contains extensions
DATE    = `2011-02-14T19:36:28' /
FILETYPE= `SPECTRAL FITS'      /Unique FITS file type name
CREATOR = `rmfit 3.4rc1'      /Software/version creating file
ORIGIN  = `GIOC    '           /Name of organization
TELESCOP= `GLAST   '           /Name of mission
INSTRUME= `GBM     '           /Name of instrument
OBSERVER= `Meegan  '           /Name of instrument PI
MJDREFI =              51910.0 /MJD date of reference epoch, int part
MJDREFF = 7.428703703703703E-04 /MJD date of reference epoch, frac part
TIMESYS = `TT      '           /Time system
TIMEUNIT= `s       '           /Time unit used in TSTART, TSTOP and TRIGTIME
OBJECT  = `GRB081101167'       /
DATE-OBS= `2008-11-01T04:00:38' /Date of start of observation
DATE-END= `2008-11-01T04:00:39' /Date of start of observation
TSTART  =    247204838.5854399 /Observation start time, relative to MJDREF
TSTOP   =    247204839.6094399 /Observation end time, relative to MJDREF
TRIGTIME=    247204840.6334400 /Trigger time, relative to MJDREF
FILENAME= `glg_scat_all_bn081101167_flnc_band_v02.fit' /Name of FITS file
COMMENT This file consists of time-sequenced spectral fit parameters
CHECKSUM= `ZTAQbT4QZTAQbT3Q'   / HDU checksum updated 2011-02-14T19:36:28
DATASUM = `        0'         / data unit checksum updated 2011-02-14T19:36:28
END
\end{verbatim}

Here, we find that the event named `GRB081101167' (\verb+OBJECT+), triggered at 247204840.6334400 s Fermi Mission 
Elapsed Time (\verb+TRIGTIME+), or roughly 4 hours, 0 minutes and 38 seconds on November 1, 2008 UTC.  In addition to 
this, the standard header information gives the name of the file itself, the creation time and program version, as well as where 
the file originated and details of the mission and instrument with which it is associated. Finally, \verb+EXTEND+ is true, so the 
primary header is followed by one or more extensions. One cannot assume that the order of these extensions will be preserved, 
however they have identifiers that will allow readers to access the correct one. The original GBM files will always follow the 
order as given here. 

The first data extension describes the detectors used and the selections made, both in time and energy (channel), for the 
spectral fit. The unique name (\verb+EXTNAME+) for this extension and its structure is `DETECTOR DATA':
\begin{verbatim}
XTENSION= 'BINTABLE'           /Written by IDL:  Mon Feb 14 13:36:27 2011
BITPIX  =                    8 /
NAXIS   =                    2 /Binary table
NAXIS1  =                  244 /Number of bytes per row
NAXIS2  =                    3 /Number of rows
PCOUNT  =                 6156 /Random parameter count
GCOUNT  =                    1 /Group count
TFIELDS =                   12 /Number of columns
TFORM1  = `20A     '           /Character string
TTYPE1  = `INSTRMNT'           /Instrument name for this detector
TFORM2  = `20A     '           /Character string
TTYPE2  = `DETNAM'           /Detector number; if one of several available
TFORM3  = `20A     '           /Character string
TTYPE3  = `DATATYPE'           /Data type used for this analysis
TFORM4  = `20A     '           /Character string
TTYPE4  = `DETSTAT '           /Was this detector INCLUDED or OMITTED
TFORM5  = `60A     '           /Character string
TTYPE5  = `DATAFILE'           /File name for this dataset
TFORM6  = `60A     '           /Character string
TTYPE6  = `RSPFILE '           /Response file name for this dataset
TFORM7  = `60A     '           /Character string
TTYPE7  = `FIT_INT '           /Fit intervals
TFORM8  = `1J      '           /Integer*4 (long integer)
TTYPE8  = `CHANNUM '           /Total number of energy channels for this detecto
TFORM9  = `2J      '           /Integer*4 (long integer)
TTYPE9  = `FITCHAN '           /Channels selected in fitting this detector
TFORM10  = `1PE(129)'           / Real*4 (floating point), variable length
TUNIT10  = `keV     '           /
TTYPE10  = `E_EDGES '           /Energy edges for each selected detector
TFORM11 = `1PE(128)'           / Real*4 (floating point), variable length
TUNIT11 = `Photon cm^-2 s^-1 keV^-1' /
TTYPE11 = `PHTCNTS '           /Array of photon counts data
TFORM12 = `1PE(128)'           / Real*4 (floating point), variable length
TUNIT12 = `Photon cm^-2 s^-1 keV^-1' /
TTYPE12 = `PHTMODL '           /Array of photon model data
TFORM13 = `1PE(128)'           / Real*4 (floating point), variable length
TUNIT13 = `Photon cm^-2 s^-1 keV^-1' /
TTYPE13 = `PHTERRS '           /Array of errors in photon counts data
DATE    = `2011-02-14T19:36:28' /
EXTNAME = `DETECTOR DATA'      /Name of this binary table extension
ORIGIN  = `GIOC    '           /Name of organization
TELESCOP= `GLAST   '           /Name of mission
INSTRUME= `GBM     '           /Name of instrument
OBSERVER= `Meegan  '           /Name of instrument PI
MJDREFI =              51910.0 /MJD date of reference epoch, int part
MJDREFF = 7.428703703703703E-04 /MJD date of reference epoch, frac part
TIMESYS = `TT      '           /Time system
TIMEUNIT= `s       '           /Time unit used in TSTART, TSTOP and TRIGTIME
DATE-OBS= `2008-11-01T04:00:38' /Date of start of observation
DATE-END= `2008-11-01T04:00:39' /Date of start of observation
TSTART  =    247204838.5854399 /Observation start time, relative to MJDREF
TSTOP   =    247204839.6094399 /Observation end time, relative to MJDREF
TRIGTIME=    247204840.6334400 /Trigger time (s) relative to MJDREF
NUMFITS =                    1 /Number of spectral fits in the data
CHECKSUM= `H7PaJ5NYH5NaH5NW'   / HDU checksum updated 2011-02-14T19:36:28
DATASUM = `1048916985'         / data unit checksum updated 2011-02-14T19:36:28
END
\end{verbatim}
Since \verb+NAXIS2+ is 3, three detectors were used for this fit; the data must be read in the \verb+DETECTOR+ column to 
determine which ones. Descriptive ASCII text data identifies which of the GBM detectors were chosen (\verb+DETECTOR+; e.g. 
`BGO\_00'), which data type was selected (\verb+DATATYPE+; e.g. `TTE'), the original GBM data file name (\verb+DATAFILE+: 
e.g. `glg\_tte\_b0\_bn081101167\_v01.fit'), and the time and energy selection intervals (\verb+FIT_INT+: e.g. `-2.048: -1.024 s, 
284.65: 37989. keV, channels  3: 123'). Detector-specific binary data consist of the PHA channel energy edges 
(\verb+E_EDGES+), the deconvolved `photon' counts (\verb+PHTCNTS+) and uncertainties (\verb+PHTERRS+), used to 
reconstruct a plot of the deconvolved fit in photon space, rather than counts space, where the observed data were actually 
fitted, using the counts model (\verb+PHTMODL+). The number of energy edges is one more than the number of PHA 
channels, since they represent bin boundaries; or,  physically, they are the energy loss thresholds in the detector, along with 
one more edge to complete the last bin.

Finally, there is an extension (\verb+EXTNAME+ = `FIT PARAMS') that describes the spectral fit in detail:
\begin{verbatim}
XTENSION= 'BINTABLE'           /Written by IDL:  Mon Feb 14 13:36:27 2011
BITPIX  =                    8 /
NAXIS   =                    2 /Binary table
NAXIS1  =                  110 /Number of bytes per row
NAXIS2  =                    1 /Number of rows
PCOUNT  =                   96 /Random parameter count
GCOUNT  =                    1 /Group count
TFIELDS =                   15 /Number of columns
TFORM1  = `2D      '           /Real*8 (double precision)
TTYPE1  = `TIMEBIN '           /Start and stop times relative to trigger
TFORM2  = `3E      '           /Real*4 (floating point)
TTYPE2  = `PARAM0  '           /Band's GRB, Epeak: Amplitude
ASYMER0 =                    0 /Corresponding PARAM contains (value, +/- error)
TFORM3  = `3E      '           /Real*4 (floating point)
TTYPE3  = `PARAM1  '           /Band's GRB, Epeak: Epeak
ASYMER1 =                    0 /Corresponding PARAM contains (value, +/- error)
TFORM4  = `3E      '           /Real*4 (floating point)
TTYPE4  = `PARAM2  '           /Band's GRB, Epeak: alpha
ASYMER2 =                    0 /Corresponding PARAM contains (value, +/- error)
TFORM5  =`3E      '           /Real*4 (floating point)
TTYPE5  = `PARAM3  '           /Band's GRB, Epeak: beta
ASYMER3 =                    0 /Corresponding PARAM contains (value, +/- error)
TFORM6  = `2E      '           /Real*4 (floating point)
TTYPE6  = `PHTFLUX '           /Photon Flux (ph/s-cm^2) std energy (8-1000)
TFORM7  = `2E      '           /Real*4 (floating point)
TTYPE7  = `PHTFLNC '           /Photon Fluence (ph/cm^2) std energy (8-1000)
TFORM8  = `2E      '           /Real*4 (floating point)
TTYPE8  = `NRGFLUX '           /Energy Flux (erg/s-cm^2) std energy (8-1000)
TFORM9  = `2E      '           /Real*4 (floating point)
TTYPE9  = `NRGFLNC '           /Energy Fluence (erg/cm^2) std energy (8-1000)
TFORM10 = `2E      '           /Real*4 (floating point)
TTYPE10 = `REDCHSQ '           /Reduced Chi^2 (1) and fitting statistic (2)
TFORM11 = `1I      '           /Integer*2 (short integer)
TTYPE11 = `CHSQDOF '           /Degrees of Freedom
TFORM12 = `2E      '           /Real*4 (floating point)
TTYPE12 = `PHTFLUXB'           /Photon Flux (ph/s-cm^2) BATSE energy (50-300)
TFORM13 = `2E      '           /Real*4 (floating point)
TTYPE13 = `DURFLNC '           /Photon Fluence (ph-cm^2) for durations (user)
TFORM14 = `2E      '           /Real*4 (floating point)
TTYPE14 = `NRGFLNCB'           /Energy Fluence (erg-cm^2) BATSE energy (50-300)
TFORM15 = `16E     '           /Real*4 (floating point)
TTYPE15 = `COVARMAT'           /Covariance matrix for the fit (#free params ^2)
TDIM15  = `(4, 4)  '            /COVARMAT array dimensions
DATE    = `2011-02-14T19:36:28' /
ORIGIN  = `GIOC    '           /Name of organization
TELESCOP= `GLAST   '           /Name of mission
INSTRUME= `GBM     '           /Name of instrument
OBSERVER= `Meegan  '           /Name of instrument PI
MJDREFI =              51910.0 /MJD date of reference epoch, int part
MJDREFF = 7.428703703703703E-04 /MJD date of reference epoch, frac part
TIMESYS = `TT      '           /Time system
TIMEUNIT= `s       '           /Time unit used in TSTART, TSTOP and TRIGTIME
OBJECT  = `GRB081101167'       /
DATE-OBS= `2008-11-01T04:00:38' /Date of start of observation
DATE-END= `2008-11-01T04:00:39' /Date of start of observation
TSTART  =    247204838.5854399 /Observation start time, relative to MJDREF
TSTOP   =    247204839.6094399 /Observation end time, relative to MJDREF
TRIGTIME=    247204840.6334400 /Trigger time (s) relative to MJDREF
NUMFITS =                    1 /Number of spectral fits in the data
N_PARAM =                    4 /Total number of fit parameters (PARAMn)
FLU_LOW =              10.0000 /Lower limit of flux/fluence integration (keV)
FLU_HIGH=              1000.00 /Upper limit of flux/fluence integration (keV)
STATISTC= `Castor C-STAT'      /Indicates merit function used for fitting
EXTNAME = `FIT PARAMS'         /Name of this binary table extension
CHECKSUM= `Vc2jVc2hVc2hVc2h'   / HDU checksum updated 2011-02-14T19:36:28
DATASUM = `3846818573'         / data unit checksum updated 2011-02-14T19:36:28
END
\end{verbatim}
The start and end times of the spectral accumulation are found in the `TIMEBIN' array. This and the following columns are 
designed for several spectra results to be stored in a single file, as would be the case for time-resolved spectroscopy; for the 
purposes of the present work, there are only the results for a single fit. The following set of columns, labeled with `PARAMn', 
where `n' is an integer, contain both a fit parameter and the associated uncertainty.  The comment for the column contains the 
name of the function fitted as well as the parameter name, separated by a colon. So, for instance, the data column `PARAM1' 
contains the fitted Band GRB function $E_{\rm peak}$ parameter and uncertainty.  If the corresponding `ASYMERn' keyword is 
0 then the `PARAMn' columns contain symmetric errors, and if the keyword is 1 then the `PARAMn' columns contain the 
asymmetric errors.  All files created for this catalog contain symmetric errors only.  

The last parameter column (`PARAMn') is followed by nine standard analysis products, in turn followed by the covariance 
matrix for the fit. The analysis products are fluxes and fluence values, calculated over several different energy ranges, 
calculated by integrating the fitted photon model at high energy resolution. For the columns labelled, `PHTFLUX', `PHTFLNC', 
`NRGFLUX' and `NRGFLNC', the standard energy range for integration is given in the `FLU\_LOW' and `FLU\_HIGH' keyword 
values; in this case, 10 to 1000 keV. The columns with names beginning with `NRG' indicate that the corresponding value has 
been multiplied by the photon energy before integrating. The uncertainties associated with these values have 
been determined by numerically evaluating the partial derivative with respect to each free parameter in the fit and multiplying 
these into the covariance matrix (found in the column `COVARMAT', as a flattened row vector with $n \times n$ elements, $n$ 
being the total number of parameters as indicated by the `N\_PARAM' and `TDIM15' keywords. Rows and columns of fixed 
parameters (i.e.: those that have been frozen at a particular value during the fit) are set to zero in the matrix, as reconstructed by 
transforming the vector into an $n \times n$ matrix. Finally, the reduced `goodness-of-fit' test statistic value is given in the 
`REDCHSQ' column, as well as the reduced figure of merit.  The reduced values of these statistics are computed by dividing the 
actual statistic by the number of degrees of freedom, stored in `CHSQDOF'. These names are historical, since {\em rmfit} now 
allows both $\chi^2$ and $-2 \log$ likelihood fitting. The name of the figure of merit used in the fitting process is provided by the 
`STATISTC' keyword; which is the `Castor C-STAT' function, in the example.

%%References

\onecolumn
\clearpage

%% Figure 1

%% Figure 2

\clearpage

%% Figure 3

\clearpage

%% Figure 4

%% Figure 5

%% Figure 6

%% Figure 7

%% Figure 8

%% Figure 9

\clearpage

%% Figure 10

%% Figure 11

\clearpage

%% Figure 12

%% Figure 13

%% Figure 14

%% Figure 15

%% Figure 16

%% Figure 17

%% Figure 18

%% Figure 19

%% Figure 20

%% Figure 21

%% Figure 22

\clearpage

%% Figure 23

%% Figure 24

\clearpage

%% Table 1
\begin{deluxetable}{ c  c  c  c }
\tablecolumns{4}
\tablewidth{0pt}
\tabletypesize{\scriptsize}
\tablecaption{BEST GRB models \label{BestTable}}
\startdata
	\hline
	\bf PL & \bf SBPL & \bf BAND & \bf COMP \\ \hline

	\multicolumn{4}{ c }{\bf Fluence Spectra} \\ \hline 
	
	112 (23\%) & 68 (14\%) & 75 (15\%) & 232 (48\%) \\ \hline
	
	\multicolumn{4}{ c }{\bf Peak Flux Spectra} \\ \hline
	
	213 (44\%) & 51 (10\%) & 69 (14\%) & 154 (32\%) \\ \hline	

\enddata
\end{deluxetable}

%% Table 2
\begin{deluxetable}{ c  c  c  c  c  c  c }
\tablecolumns{7}
\tablewidth{0pt}
\tabletypesize{\scriptsize}
\tablecaption{Sample mean and standard deviation of the parameter distributions \label{ParamTable}}
\startdata
	\hline
	\multirow{2}{*}{\bf Model} & \bf Low-Energy & \bf High-Energy & \multirow{2}{*}{$\bf E_{peak} (keV)$} & \multirow{2}{*}{$\bf
	 E_{break} (keV)$} & \bf Photon Flux & \bf Energy Flux \\
 	& \bf Index & \bf Index & & & \bf (ph $\bf s^{-1} \ cm^{-2}$) & \bf ($\bf 10^{-7} \ erg \ s^{-1} \ cm^{-2}$) \\ \hline

	\multicolumn{7}{ c }{\bf Fluence Spectra} \\ \hline 
	
	PL & $-1.54^{+0.18}_{-0.25}$ & - & - & - & $2.93^{+3.49}_{-1.36}$ & $3.96^{+7.72}_{-1.97}$ \\

	COMP & $-0.90^{+0.42}_{-0.38}$ & - & $223.75^{+483.80}_{-123.80}$ & - & $2.82^{+3.49}_{-1.34}$ & $4.01^{+10.54}_	
	{-2.34}$ \\

	SBPL & $-1.08^{+0.40}_{-0.46}$ & $-2.29^{+0.47}_{-0.65}$ & $221.42^{+432.04}_{-129.12}$ & $129.74^{+290.08}_
	{-74.39}$ & $2.93^{+3.78}_{-1.54}$ & $4.48^{+11.4}_{-1.96}$ \\

	BAND & $-0.82^{+0.42}_{-0.38}$ & $-2.17^{+0.36}_{-0.47}$ & $185.58^{+428.94}_{-81.01}$ & $175.50^{+664.50}_	
	{-100.64}$ & $2.92^{+3.77}_{-1.39}$ & $4.48^{+11.05}_{-2.57}$ \\

	BEST & $-1.05^{+0.44}_{-0.45}$ & $-2.25^{+0.34}_{-0.73}$ & $204.75^{+359.36}_{-121.07}$ & $122.71^{+240.41}_	
	{-80.36}$ & $2.92^{+3.96}_{-1.31}$ & $4.03^{+9.38}_{-2.13}$\\ \hline

	\multicolumn{7}{ c }{\bf Peak Flux Spectra} \\ \hline 
	
	PL & $-1.54^{+0.16}_{-0.24}$ & - & - & - & $4.34^{+4.33}_{-2.04}$ & $5.85^{+7.08}_{-2.91}$\\

	COMP & $-0.81^{+0.44}_{-0.43}$ & - & $215.03^{+340.3}_{-113.00}$ & - & $4.67^{+7.84}_{-2.45}$ & $7.29^{+19.52}_	
	{-4.5}$\\

	SBPL & $-1.05^{+0.38}_{-0.49}$ & $-2.27^{+0.48}_{-0.53}$ & $217.88^{+395.88}_{-111.72}$ & $147.16^{+243.46}_	
	{-80.94}$ & $5.09^{+9.18}_{-2.68}$ & $8.33^{+21.87}_{-5.00}$\\
	
	BAND & $-0.75^{+0.41}_{-0.40}$ & $-2.16^{+0.40}_{-0.50}$ &$194.49^{+313.39}_{-100.38}$ & $372.96^{+5086.2}_	
	{-230.02}$ & $5.04^{+9.09}_{-2.70}$ & $8.91^{+22.34}_{-5.28}$ \\

	BEST & $-1.12^{+0.61}_{-0.50}$ & $-2.27^{+0.44}_{-0.50}$ & $223.12^{+351.55}_{-125.92}$ & $172.16^{+253.56}_	
	{-100.49}$ & $5.39^{+10.18}_{-2.87}$ & $8.35^{+22.61}_{-4.98}$ \\ \hline
	
\enddata
\end{deluxetable}

%% Table 3
\begin{deluxetable}{ c  c  c  c  c  c  c }
\tablecolumns{7}
\tablewidth{0pt}
\tabletypesize{\scriptsize}
\tablecaption{Comparison of the sample mean and standard deviation from different catalogs \label{BestComparison}}
\startdata
	\hline
	\multirow{2}{*}{\bf Dataset} & \bf Low-Energy & \bf High-Energy & \multirow{2}{*}{$\bf E_{peak}$} & \multirow{2}{*}{$\bf
	 E_{break}$} & \bf Photon Flux & \bf Energy Flux \\
 	& \bf Index & \bf Index & \bf (keV) & \bf (keV) & \bf (ph $\bf s^{-1} \ cm^{-2}$) & \bf ($\bf 10^{-7} \ erg \ s^{-1} \ cm^{-2}$) \\ 
	\hline

	\multicolumn{7}{ c }{\bf Fluence} \\ \hline
	This Catalog BEST & $-1.05^{+0.44}_{-0.45}$ & $-2.25^{+0.34}_{-0.73}$ & $205^{+359}_{-121}$ & $123^{+240}_{-80.4}$ 
	& $2.92^{+3.96}_{-1.31}$ & $4.03^{+9.38}_{-2.13}$\\
	
	\citet{Nava} & $-1.02^{+0.49}_{-0.56}$ & $-2.40^{+0.24}_{-0.45}$ & $190^{+336}_{-112}$ & - & - & $3.05^{+11.5}_{-1.54}$\\ 

	\citet{Kaneko06} & $-1.07^{+0.42}_{-0.36}$ & $-2.43^{+0.38}_{-0.59}$ & $260^{+233}_{-116}$ & $203^{+129}_{-80.0}$ & 
	$3.32^{+6.01}_{-2.04}$ & $8.56^{+16.0}_{-5.47}$\\ \hline

	\multicolumn{7}{ c }{\bf Peak Flux Spectra} \\ \hline 
	This Catalog BEST & $-1.12^{+0.61}_{-0.50}$ & $-2.27^{+0.44}_{-0.50}$ & $223^{+352}_{-126}$ & 
	$172^{+254}_{-100}$ & $5.39^{+10.18}_{-2.87}$ & $8.35^{+22.61}_{-4.98}$ \\
	
	\citet{Nava} & $-0.55^{+0.42}_{-0.37}$ & $-2.37^{+1.42}_{-0.61}$ & $241^{+387}_{-138}$ & - & - & $14.9^{+53.8}_{-10.7}$\\ 
	\hline 

\enddata
\end{deluxetable}

\end{document}